\documentclass[a4paper,11pt]{article}
\pdfoutput=1 

\usepackage{jheppub} 

\usepackage{hyperref}
\usepackage{cleveref}
\usepackage{supertabular}
\usepackage{mathrsfs}
\usepackage{array}
\usepackage{lipsum}
\usepackage{breqn}

\usepackage[T1]{fontenc} 
\newcommand{\mF}{\mathscr{F}}
\newcommand{\F}{\mathcal{F}}
\newcommand{\W}{\mathcal{W}}
\newcommand{\im}{\mathrm{Im}\,}

\newcommand{\ri}{\mathrm{i}}
\newcommand{\Ch}{\mathrm{Ch}}
\newcommand{\C}{\mathsf{C}}
\newcommand{\n}{\mathsf{n}}
\newcommand{\I}{\mathcal{I}}
\newcommand{\Wilson}[1]{Z_{ {W}_{#1}}}
\newcommand{\CHT}{\Ch_{\mu_2^t\mu_1^t}(Q_F,{q},{t})}
\newcommand{\CH}{\Ch_{\mu_1\mu_2}(Q_F,{t},{q})}
\def\FW{\mathcal{F}_{\mathbf{r}}}

\def\pq{{$(p,q)$}}

\title{\boldmath Topological strings and Wilson loops}

\author[a,b]{Min-xin Huang,}
\author[c]{Kimyeong Lee,}
\author[d]{Xin Wang}

\affiliation[a]{Interdisciplinary Center for Theoretical Study, University of Science and Technology of China,\\Hefei, Anhui 230026, China}
\affiliation[b]{Peng Huanwu Center for Fundamental Theory,\\Hefei, Anhui 230026, China}
\affiliation[c]{School of Physics, Korea Institute for Advanced Study,\\Hoegiro 85, Seoul 02455, Korea}
\affiliation[d]{Quantum Universe Center, Korea Institute for Advanced Study,\\Hoegiro 85, Seoul 02455, Korea}

\emailAdd{minxin@ustc.edu.cn}
\emailAdd{klee@kias.re.kr}
\emailAdd{wxin@kias.re.kr}

\preprint{\begin{flushright}USTC-ICTS/PCFT-22-09\\  KIAS-Q22003\end{flushright}}

\abstract{
We propose the refined topological string correspondence to the expectation values of half-BPS Wilson loop operators in 5d $\mathcal{N}=1$ gauge theory partition function on the Omega-deformed background $\mathbb{R}^4_{\epsilon_{1,2}}\times S^1$. We provide the refined topological vertex method and the refined holomorphic anomaly equation method in the topological string theory, from which we have exact computations on the 5d Wilson loops partition functions in both A- and B-models. Finally, with the exact results we have in B-model, we recover the quantum periods of local $\mathbb{P}^1\times\mathbb{P}^1$ model and local $\mathbb{P}^2$ model in the study of quantum geometry and we further give a refined generalization of A-period.}

\begin{document} 
\maketitle
\flushbottom
\section{Introduction}
\label{sec:intro}
The topological string theory proposed in \cite{Witten:1988xj} on Calabi-Yau threefolds has been intensively studied for decades, not only because of that they provided new techniques in computing mathematically well-defined topological invariants such as Gromov-Witten invariants, but also they predicted new kinds of topological invariants like Gopakumar-Vafa invariants. 

The amplitudes of refined topological strings on a non-compact Calabi-Yau threefold $X$ in the A-model capture the BPS spectra of M-theory compactified on $X$. The low energy theory obtained from geometric engineering \cite{Katz:1996fh,Katz:1997eq} is a 5d $\mathcal{N}=1$ supersymmetric quantum field theory (SQFT), which is a supersymmetric gauge theory or a non-Lagrangian theory with eight supercharges on Omega-deformed background $\mathbb{R}_{\epsilon_{1,2}}^4\times S^1$.
The BPS particles are understood in \cite{Gopakumar:1998ii,Gopakumar:1998jq} as M2-branes winding on holomorphic 2-cycles $C\in H_2(X,\mathbb{Z})$, where the BPS particles carry non-trivial spins $(j_L,j_R)$ in the representation $SU(2)_L\times SU(2)_R=SO(4)$, which is the little group of massive particles in $\mathbb{R}^4\times S^1$.

In the resulting gauge theory, an important physical observable is the Nekrasov's instanton partition function \cite{Nekrasov:2002qd,Nekrasov:2003rj} in the Coulomb branch, which takes the form $Z_{\mathrm{BPS}}=Z_{\mathrm{pert}}Z_{\mathrm{inst}}$. $Z_{\mathrm{pert}}$ is the perturbative partition function which is made of contributions from vector-multiplets and hyper-multiplets. $Z_{\mathrm{inst}}$ is the instanton contribution which takes the form $Z_{\mathrm{inst}}=1+\sum_{k=1}^{\infty}Z_k\mathfrak{q}^k$. Here $\mathfrak{q}=e^{-{8\pi^2\beta}/{g^2_{\mathrm{5d}}}}$ is the instanton counting parameter defined from the 5d gauge coupling $g_{\mathrm{5d}}$ and the radius $\beta$ of the Euclidean time circle $S^1$. $Z_k$ is the $k$-instanton partition function which could be computed from ADHM quantum mechanics as a Witten index
\begin{align}
    Z_k=\mathrm{Tr}_{\mathcal{H}_k}\left[(-1)^Fe^{-\beta \{Q,Q^{\dagger}\} }e^{-\epsilon_1(J_1+J_R)}e^{-\epsilon_2(J_2+J_R)}e^{-\alpha\cdot \Pi}e^{-m\cdot H} \right].
\end{align}
Here $\mathcal{H}_k$ is the $k$-instanton Hilbert space of 5d gauge theory on $\mathbb{R}^4$, $J_{1,2}$ are the Cartan generators of the $SO(4)$ Lorentz symmetry group on $\mathbb{R}^4$, $J_R$ is the Cartan of the $SU(2)_R$ R-symmetry. $\Pi$ and $H$ are the gauge and the flavor charges respectively. $\alpha$ is the chemical potential for the electric charges in the Coulomb branch, $m$ is the chemical potential for all other flavor symmetries. The ADHM construction has been used to calculate the more generic 5d instanton partition functions in \cite{Kim:2011mv,Rodriguez-Gomez:2013dpa,Bergman:2013ala,Hwang:2014uwa} and 6d elliptic genera in \cite{Haghighat:2013gba,Kim:2014dza,Haghighat:2014vxa,Gadde:2015tra,Kim:2015fxa,Kim:2016foj,Kim:2018gjo}.

In this paper, we are interested in another important physical observable, the correlation function in the presence of a half-BPS Wilson loop operator 
\begin{align}
    {W}_{\mathbf{r}}=\mathrm{Tr}_{\mathbf{r}}\mathcal{T}\exp\left(i\oint_{S^1} dt(A_0(t)-\varphi(t))\right)
\end{align}
in the representation $\mathbf{r}$, winding around the time circle $S^1$. For more notations and explanations, we refer the readers to Section \ref{sec:Wilsonloop}. The Wilson loop operator in the gauge theory can be generated by the worldline of a heavy stationary quark fixed at the origin of the space, which corresponds to a 1d line defect particle in the 5d theory. Inspired from the holographic Wilson loop for $\mathcal{N}=4$ super Yang-Mills theory in 4d \cite{Gomis:2006sb}, the brane configuration in type-II string theory for the line defect was proposed in \cite{Tong:2014cha} as a defect D4-brane perpendicular to the D4-branes in the $\mathbb{R}^4$ directions. Such a brane configuration suggest a ADHM quantum mechanics construction  for instanton partition functions with line defect,  and the expectation values of the half-BPS Wilson loop operators in the 5d gauge theories were computed from ADHM quantum mechanics in \cite{Gaiotto:2014ina,Nekrasov:2015wsu,Kim:2016qqs,Assel:2018rcw,Gaiotto:2015una,Haouzi:2020yxy} and the analogous 6d Wilson surface version were studied in \cite{Chen:2007ir,Agarwal:2018tso,Chen:2020jla,Chen:2021ivd,Chen:2021rek} as well. In particular, the 1d line defects play an important role in the study of quantum Seiberg-Witten curve and the non-perturbative Dyson-Schwinger equation \cite{Nekrasov:2015wsu,Haouzi:2020yxy}. More recently, the blowup equation for the instanton contributions with 5d/6d Wilson loops/surfaces were suggested in \cite{Kim:2021gyj}, and the Wilson loop expectation values for many theories which may not admit ADHM descriptions were predicted there.

In M-theory, the heavy stationary quark can be realized as particles from M2-branes winding on additional {\it stationary} {\it non-compact} holomorphic 2-cycles \cite{Kim:2021gyj}. In general, these additional curve classes can always be decomposed to non-decomposable curve classes $\C_i$, which are called {\it primitive curves}. Now we denote the background $(X,\{\C_i\})$ as a collection of the original Calabi-Yau manifold and the additional primitive curves, which can be embedded in a new Calabi-Yau manifold $\widehat{X}$. For the simplest case when $X$ is local $\mathbb{P}^1\times\mathbb{P}^1$, we show in Section \ref{sec:3.2} that $\widehat{X}$ can be understood as the $n$-points blow-ups of $X$. For a single primitive curve $\C$ insertion, the generated Wilson loop is in the non-decomposable representation $\mathbf{r}_{\C}$. It was conjectured in \cite{Kim:2021gyj} that the expectation value of the Wilson loop operator in the non-decomposable representation $\mathbf{r}_{\C}$ has a BPS expansion
\begin{align}\label{eq:1}
    \langle W_{\mathbf{r}_{\C}}\rangle=\F_{\mathrm{BPS},\{\C\}}=
    \sum_{C\in H_2(X,\mathbb{Z})}\sum_{j_L,j_R}(-1)^{2j_L+2j_R}\widetilde{N}_{j_L,j_R}^{C}{\chi_{j_L}(\epsilon_-)\chi_{j_R}(\epsilon_+)}e^{-t_{C}}
\end{align}
in term with non-negative integer       Wilson loop BPS invariants $\widetilde{N}_{j_L,j_R}^{C}$, where $\epsilon_{\pm}=\frac{1}{2}(\epsilon_1\pm \epsilon_2)$ and $\chi_j$ is the $SU(2)$ character with highest weight $j$. The right hand side of equation \eqref{eq:1} is nothing but the single heavy quark BPS sector $ \F_{\mathrm{BPS},\{\C\}}$ of M2-branes winding on the curve classes $\C+C$ in $\widehat{X}$ without the momentum factor ${2\sinh(\epsilon_1/2)\cdot 2\sinh(\epsilon_2/2)}$. For two primitive curves $\C_1,\C_2$ insertion, the resulting Wilson loop is in the representation of tensor product of $\mathbf{r}_1\otimes\mathbf{r}_2$, we find the expectation value has a BPS expansion
\begin{align}\label{eq:2}
    \langle W_{\mathbf{r}_1\otimes \mathbf{r}_2}\rangle =\F_{\mathrm{BPS},\{\C_1\}}\F_{\mathrm{BPS},\{\C_2\}}+{2\sinh(\epsilon_1/2)\cdot 2\sinh(\epsilon_2/2)}\cdot\F_{\mathrm{BPS},\{\C_1,\C_2\}},
\end{align}
each BPS sector in \eqref{eq:2} takes the BPS expansion as \eqref{eq:1} with the positive BPS invariants. 
More details and more generic discussion can be found in Section \ref{eq:Wr_op}.

The BPS states of the 5d gauge theory from M-theory compactification on $X$ are captured by topological strings on the non-compact Calabi-Yau threefold $X$. Thanks to the string-gauge theory duality to three dimensional Chern-Simons theory \cite{Witten:1992fb}, the large $N$ Chern-Simons theory leads to topological vertex calculation to the partition function in the A-model. The topological vertex was later generalized to the refined case \cite{Iqbal:2007ii,Aganagic:2012hs}, where the partition functions computed from the refined topological vertex were proved in \cite{Taki:2007dh} to be the same as the calculation from localization method \cite{Nekrasov:2002qd,Tachikawa:2004ur,Gottsche:2006bm}. In the B-model, from the worldsheet description of topological strings, there exists a holomorphic anomaly equation \cite{Bershadsky:1993cx} which gives the direct integration method to solve the topological string amplitudes \cite{Huang:2006si,Huang:2006hq,Grimm:2007tm,Haghighat:2008gw}. By combining the {\it gap condition} proposed in \cite{Huang:2006si,Huang:2006hq}, the direct integration method provide a very powerful tool to compute the string amplitudes. Later, the refined version of the holomorphic anomaly equation was proposed in \cite{Huang:2010kf,Krefl:2010fm} which was used in \cite{Huang:2010kf,Krefl:2010fm,Huang:2011qx,Huang:2013yta,Klemm:2015iya} to calculate the refined topological string amplitudes.

Recently, based on geometric descriptions and inspirations from the blowup equation for gauge theories \cite{Nakajima:2003pg,Nakajima:2003uh,Nakajima:2005fg,Nakajima:2009qjc}, the blowup equation for refined topological strings was proposed in \cite{Huang:2017mis}, which was further generalized to generic 5d/6d SQFTs and the blowup equation play an important role to compute the partition functions and elliptic genera for 5d/6d SQFTs \cite{Huang:2017mis,Gu:2018gmy,Gu:2019dan,Gu:2019pqj,Gu:2020fem,Kim:2019uqw,Kim:2020hhh,Duan:2021ges}. The generalization of the blowup equation to Wilson loop expectation values can be found in \cite{Nakajima:2003uh,Nakajima:2009qjc,Kim:2021gyj}. In particular, in \cite{Kim:2021gyj}, the blowup equation for Wilson loops in generic gauge theories and non-Lagrangian theories was suggested, and the Wilson loop BPS invariants for many 5d/6d theories were calculated there.

Even though the blowup formula for Wilson loop expectation values provides a very efficient method to calculate the BPS invariants of Wilson loops, the refined topological vertex approach and the refined holomorphic anomaly equation for 5d Wilson loop expectation values are still missing. 
Our goal in this paper is to establish these two approaches. In Section \ref{sec:SU2}, we explicitly calculated the Wilson loop expectation values in the representations $\mathbf{2}^{\otimes \n},\n=1,\cdots,5$ for $SU(2)$ theory by using the formalism of refined topological vertex. This method gives the exact all genus result for a given instanton number in the A-model. In Section \ref{sec:RHAE}, we proposed the refined holomorphic anomaly equation for Wilson loop amplitudes in the B-model. We observed that the refined holomorphic anomaly equation for the Wilson loop amplitudes is the same as the refined holomorphic equation for refined topological strings. This suggests that the Wilson loop free energies inherit many important properties of topological amplitudes. Indeed, as will be calculated in Section \ref{sec:RHAE}, we find the Wilson loop amplitudes are finite polynomials of the propagators and they are still weight zero quasi-modular forms as the conventional topological strings \cite{Aganagic:2006wq}.

The paper is organized as follows. In Section \ref{sec:2}, we review the basic concepts of topological strings on local Calabi-Yau threefolds. In Section \ref{sec:Wilsonloop}, we describe the M-theory understanding of Wilson loop operators insertion, we derive a BPS formula for the Wilson loop expectation values in generic representations. In Section \ref{sec:3.2}, we establish a refined topological vertex method to compute the Wilson loop expectation values, and we computed the Wilson loop expectation values for 5d pure $SU(2)$ theory in the representations $\mathbf{2}^{\otimes \n},\n=1,\cdots,5$. In Section \ref{sec:RHAE}, we review the direct integration method by using the refined holomorphic anomaly equation, and generalize it to the cases in the presence of Wilson loops in 5d. We explicitly test the direct integration method approach for local $\mathbb{P}^1\times \mathbb{P}^1$ and local $\mathbb{P}^2$ models in various representation. In Section \ref{sec:5}, we use the exact results from holomorphic anomaly equation, derived the quantum operators for local $\mathbb{P}^1\times \mathbb{P}^1$ and local $\mathbb{P}^2$ models studied in the quantum geometry and suggests the refined A-period and the refined version of the quantum operators. In Section \ref{sec:conclusion}, we summarize the result and discuss about the interesting future directions.

\section{Topological strings}\label{sec:2}
The topological string amplitudes $\F_g(t)$ in the A-twisted topological theory are defined as integrals over genus $g$ moduli space of worldsheet Riemann surfaces and are generating functions of the genus $g$ Gromov-Witten invariants. When the target space is a Calabi-Yau three-fold $X$, these generating functions are related to M-theory compactified on $\mathbb{R}^4\times S^1\times X$. The topological string amplitudes then compute the Gopakumar-Vafa invariants which
count the numbers of BPS particles coming from quantization of the moduli space of wrapped M2-branes on 2-cycles of $X$. These BPS particles carry $SU(2)_L\times SU(2)_R=SO(4)$ quantum numbers, where $SO(4)$ is the little group of massive particles in $\mathbb{R}^4\times S^1$. We can turn on a constant graviphoton field strength $G=F_1 dx^1\wedge dx^2-F_2 dx^3\wedge dx^4$, that the BPS particles carry non-trivial spins $(j_L,j_R)$ in representations of $SU(2)_L\times SU(2)_R$. The amplitudes then can be written from Schwinger integral \cite{Gopakumar:1998ii,Gopakumar:1998jq}, as
\begin{align}\label{eq:schwinger1}
    \F_{\mathrm{BPS}}=\sum_{C\in H_2(X,\mathbb{Z})}\sum_{j_L,j_R}N_{j_L,j_R}^C \int_{0}^{\infty}\frac{ds}{s}\frac{\mathrm{Tr}_{(j_L,j_R)}(-1)^{\sigma_L+\sigma_R}e^{-s m}e^{-2s e (\sigma_LF_++\sigma_RF_-)}}{2\sinh(seF_1/2)\cdot 2\sinh(-seF_2/2)},
\end{align}
where $\sigma_{L,R}$ is the Cartan of $SU(2)_{L,R}$, $m$ is the mass of the electric particle which is equal to the volume of 2-cycle $C$ that M2-brane is wrapped on, usually is denoted as $t_C=\int_C \omega$, where $\omega$ is the K\"ahler form of $X$. The number $N_{j_L,j_R}^C$ counts the number of BPS particles, when the Calabi-Yau three-fold $X$ is non-compact, it is an topological invariant which is a positive integer and is usually called refined BPS invariant. The numerator of \eqref{eq:schwinger1} in the integrand comes from the  summation of all eigenvalues of the harmonic operators in the Schwinger integral with respective two different graviphoton field strengths $F_{1},F_2$.

After performing the integral, the amplitude is then written as
\begin{align}
    \F_{\mathrm{BPS}}(t_C,\epsilon_1,\epsilon_2)=\sum_{\omega=1}^{\infty}\sum_{C\in H_2(X,\mathbb{Z})}\sum_{j_L,j_R}(-1)^{2j_L+2j_R}N_{j_L,j_R}^C\frac{\chi_{j_L}(\omega\epsilon_-)\chi_{j_R}(\omega\epsilon_+)}{\omega\cdot 2\sinh(\omega\epsilon_1/2)\cdot 2\sinh(\omega\epsilon_2/2)}e^{-\omega t_C},
\end{align}
where we defined $\epsilon_{1}=e F_{1},\epsilon_{2}=-e F_{2}$, $\epsilon_{\pm}=\frac{1}{2}(\epsilon_1\pm\epsilon_2)$, $\chi_j(\epsilon)=\frac{e^{(2j+1)\epsilon }-e^{-(2j+1)\epsilon}}{e^{\epsilon}-e^{-\epsilon}}$.

In the worldsheet description of topological string amplitudes, the parameters $\epsilon_{1,2}$ are identified as a refined version of the string coupling $g_s$. Genus expansion of the topological string amplitudes corresponding to small string coupling expansion
\begin{align}
    \F(t_C,\epsilon_1,\epsilon_2)=\sum_{n,g=0}^{\infty}(\epsilon_1+\epsilon_2)^{2n}(\epsilon_1\epsilon_2)^{g-1} \F^{(n,g)}(t_C),
\end{align}
where $\F^{(n,g)}(t_C)$ take the general form
\begin{align}
    \F^{(0,0)}(t_C)&=\int_X \omega\wedge\omega\wedge\omega+\sum_{C\in H_2(X,\mathbb{Z})}GW_{0,0}^{C}e^{-t_C},\\
    \F^{(0,1)}(t_C)&=-\frac{1}{24}\int_X\omega\wedge c_2(X)+\sum_{C\in H_2(X,\mathbb{Z})}GW_{0,1}^{C}e^{-t_C},\\
    \F^{(1,0)}(t_C)&=b_i^{(1,0)}t_i+\sum_{C\in H_2(X,\mathbb{Z})}GW_{1,0}^{C}e^{-t_C},,\\
    \F^{(n,g)}(t_C)&=\sum_{C\in H_2(X,\mathbb{Z})}GW_{n,g}^{C}e^{-t_C},\quad n+g\geq 2,
\end{align}
where $\omega$ is the K\"ahler form of $X$, $c_2(X)$ is the second Chern class of $X$, $GW_{n,g}^{C}$ is the refined version of Gromov-Witten invariant come from the coefficients of the genus expansion of $\F_{\mathrm{BPS}}(t_C,\epsilon_1,\epsilon_2)$. Here we denote $t_i$ the basis element of $t_C$, $b_i^{(1,0)}t_i$ is computed from the unstable maps of the relative Gromov-Witten invariants \cite{Bousseau:2020ckw}. The free energy $\F^{(n,g)}(t_C)$ is   easier to compute in the B-model method, where we will give a review in the following subsections. See \cite{Choi:2012jz} for a mathematical rigorous definition for the refined amplitudes.

\subsection{Local mirror symmetry}
We now focus on the topological strings on toric Calabi-Yau three-folds. A toric Calabi-Yau three-fold $X$ is a toric variety given by the quotient 
\begin{align}
    X=(\mathbb{C}^{k+3}-\mathcal{SR})/G,
\end{align}
where the group $G=(\mathbb{C}^*)^k$ and $\mathcal{SR}$ is the Stanley-Reisner ideal of $G$. The group action is specified by a matrix of charges $Q^{\alpha}_i$, $i=1,\cdots,k+3$, $\alpha=1,\cdots k$, satisfying the Calabi-Yau condition 
\begin{align}
   \sum_{i=1}^{k+3}Q_{i}^{\alpha}=0,
\end{align}
that act on the open subset as
\begin{align}
    G:\, x_i \mapsto \mu_{\alpha}^{Q_{i}^{\alpha}} x_i, \quad\quad i=1,\cdots,k+3,
\end{align}
where $\alpha=1,\cdots, k$, $\mu_{\alpha}\in \mathbb{C}^*$ and $Q_i^{\alpha}\in\mathbb{Z}$. One can define divisors $D_i\equiv\{x_i=0\}$, such that the charges $Q_i^{\alpha}$ are the intersection numbers of the divisors $D_i$ and the curve class $C^{\alpha}\in H^2(X,\mathbb{Z})$,
\begin{align}
    Q_i^{\alpha}=C^{\alpha}\cdot D_i.
\end{align}
All the data above can be encoded into a toric diagram, which is described from the rays $\nu^{(i)}=(1,\nu^{(i)}_1,\nu^{(i)}_2)$ with the constraints
\begin{align}
   \sum_{i=1}^{k+3} Q_i^{\alpha} \nu^{(i)}=0, \ \   \ \alpha=1,\cdots, k.
\end{align}

One can translate the language into the B-model geometry. Introducing the {\it{homogeneous coordinates}} $a_i$ with the identification
\begin{align}
    a_i \mapsto \lambda a_i, \quad \lambda \in \mathbb{C}^*.
\end{align}
Not all of the homogeneous coordinates are independent, they are constrained by 
\begin{align}\label{eq:consta}
    \prod_{i=1}^{k+3}a_i^{Q_i^{\alpha}}=z_{\alpha},\quad \alpha =1,\cdots,k.
\end{align}
Here $z_{\alpha}$'s are parameters in the complex structure moduli space of the B-model geometry.
The local B-model geometry is then defined by
\begin{align}\label{eq:MC}
    \omega^+ \omega^-=H=\sum_{i=1}^{k+3}a_i.
\end{align}
Since not all the homogeneous coordinates $a_i$ are independent. The homogeneity of the coordinates $a_i$ allows us to set one of $a_i$ to be one. Eliminating all the $a_i$ in \eqref{eq:MC} by using \eqref{eq:consta}, 
and choose other two as $e^x$ and $e^y$, then a proper choice always brings the mirror geometry into a form
\begin{align}\label{eq:mirrorcurve}
    H(e^x,e^y;z_{\alpha})=\sum_{i=1}^{k+3}a_i^*(z)e^{\nu^{(i)}_1x+\nu^{(i)}_2y}.
\end{align}
The geometry we describe here is called local Calabi-Yau \cite{Chiang:1999tz}. In particular, the period integrals over the holomorphic three-form of the B-model local Calabi-Yau is encoded in the periods of an algebraic curve
\begin{align}\label{eq:curve}
    H(e^x,e^y;z_{\alpha})=0,
\end{align}
which is a Riemann surface usually named as mirror curve.

The meromorphic one form $\lambda$ of the Riemann surface is induced from the holomorphic three-form of the Calabi-Yau three-fold and is    defined as 
\begin{align}
    \lambda=  y dx. 
\end{align}
The periods are defined as
\begin{align}
    \Pi^A_{i}=\oint_{A_{i}}\lambda,\quad\quad \Pi^B_{i}=\oint_{B_{i}}\lambda,
\end{align}
where $A_{i}$ and $B_{i}$ stand for the 2-cycles of the Riemann surface. $\Pi^A_{\alpha}$ and $\Pi^B_{\alpha}$ are called A- and B-period respectively. A simple way to compute the periods is from the Picard-Fuchs equations
\begin{align}
   \left( \prod_{Q_i^{\alpha}>0}\partial_{a_i}^{Q_i^{\alpha}}- \prod_{Q_i^{\alpha}<0}\partial_{a_i}^{Q_i^{\alpha}}\right)\Pi^{A,B}=0, \quad \alpha=1,\cdots,k.
\end{align}
The solutions are constructed by the Frobenius method. By using the fundamental period
\begin{align}
    \omega_0({z},{\rho})=\sum_{{n}}\frac{1}{\Gamma(Q_i^{\alpha}(n^{\alpha}+\rho^{\alpha})+1)}(z^{\alpha})^{n^{\alpha}+\rho^{\alpha}},
\end{align}
we have the solution
\begin{align}
    \Pi^A_{i}&=\partial_{\rho_i}\omega_0({z},{\rho})|_{{\rho}=0}=\log(z_i)+\mathcal{O}(z_i),
\end{align}
while $\Pi^B_{i}$ is given by a linear combination of the second order derivative of $\omega_0({z},{\rho}=0)$ with respective to $z$. We usually use the notation $t_i(z)\equiv \Pi^A_{i}({z})$ denote the K\"ahler parameter in the A-model and $t_i^D(z)\equiv \Pi^B_{i}({z})$ the dual parameters. The A-period here provide {\it{mirror maps}} between A-model K\"ahler parameters and B-model complex structure parameters. Since the 2-cycles in the period integral are not completely independent, there exists a function $\F^{(0,0)}$ which is called prepotential, such that
\begin{align}
   t_i^D({z})=\sum_{j=1}^{k} C_{ij} \frac{\partial \F^{(0,0)}}{\partial t_j},
\end{align}
where $C_{ij}$ is a subset of $Q_i^{\alpha}$ which is the intersection matrix of compact divisors and curves in the A-model Calabi-Yau manifold.

\section{Wilson loops}\label{sec:Wilsonloop}
In this section, we study partition functions of 5d $\mathcal{N}=1$ super Yang-Mills theory on $\mathbb{R}^4_{\epsilon_{1},\epsilon_{2}}\times S^1$, in the presence of Wilson loop operators winding around the time circle $S^1$. The insertion of a Wilson loop operator corresponding to adding a {\it{heavy stationary}} quark at the origin of the space $\mathbb{R}^4$, that can be obtained by the insertion a half-BPS non-local operator
\begin{align}\label{eq:Wr_op}
    {W}_{\mathbf{r}}=\mathrm{Tr}_{\mathbf{r}}\mathcal{T}\exp\left(i\oint_{S^1} dt(A_0(t)-\varphi(t))\right)
\end{align}
winding around the time circle $S^1$ in the path integral formalism. Here $\mathcal{T}$ is the time ordering, $\mathbf{r}$ is a representation of the gauge group, $A_0(t)=A_0(\vec{x}=0,t)$ is the zero component of the gauge field and $\varphi(t)=\varphi(\vec{x}=0,t)$ is a scalar field accompany with the gauge field to preserve half of the supersymmetries. 

\subsection{The $SU(N)$ gauge theory and Wilson loops}\label{sec:SUN_cod4}
\begin{figure}[]
\begin{center}
\includegraphics{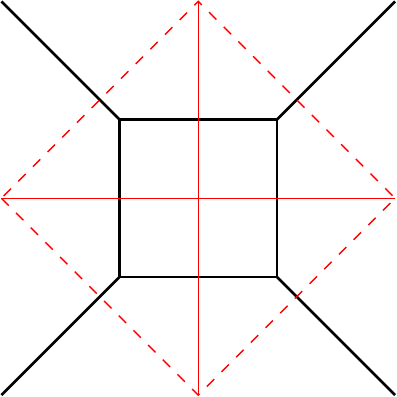}
\end{center}
\caption{{\pq}-brane web diagram (black) in the $x^{5,6}$ directions for pure $SU(2)$ gauge theory as a dual graph of $\mathbb{P}^2$ (red).}
\label{fig:su2brane1}
\end{figure}

In the following context, we focus on $SU(N)$ gauge theory with Chern-Simons level $\kappa$, sometimes also denoted as $SU(N)_{\kappa}$ in short. In the type IIB brane setups, a $SU(N)$ gauge theory is provided by $N$ D5-branes lying in the $x^{0,1,2,3,4,5}$ directions, which is stretched between two NS5-branes lying in the $x^{0,1,2,3,4,6}$ directions. The graph illustration of the type IIB brane configuration in the $x^{5,6}$ directions is called {\pq} five-brane diagram \cite{Aharony:1997bh}. An example of {\pq} five-brane diagram for $SU(2)_0$ theory can be found in Figure \ref{fig:su2brane1}. A Wilson line is then provided by inserting infinitely long fundamental strings in the $x^{0,6}$ directions, which is perpendicular to the D5-branes at $x^{5,6}$ plane. In \cite{Tong:2014cha}, following by \cite{Gomis:2006im}, the authors suggested to terminate the fundamental strings on a stack of defect D3-branes in $x^{0,7,8,9}$ directions, such that the fundamental strings are stationary along the $\mathbb{R}^4$ directions $x^{1,2,3,4}$, and then move the D3-branes to infinity. The insertion of D3-branes generates the insertion of {\it{fermionic}} heavy quarks in the lower dimensional quantum field theory. 

The brane configuration with $\n$ additional parallel defect D3-branes provides the partition function with 1d line defects in the ADHM construction
\begin{align}
    Z_{\mathrm{5d/1d}}^{(\n)}=Z^{\mathrm{pert}}_{\mathrm{5d}}\sum_{k=0}^{\infty}\mathfrak{q}^k\frac{1}{k!}\oint \left[\prod_{I=1}^k\frac{d \alpha_I}{2\pi \ri}\right] Z^{(k)}_{\mathrm{vec}}Z^{(k)}_{\mathrm{CS}}Z^{(k,\n)}_{\mathrm{defect}}.
\end{align}
where $Z^{(k)}_{\mathrm{CS}}=\prod_{I=1}^{k}e^{\kappa \alpha_I}$ and \footnote{Here we use the notations $\alpha_{IJ}\equiv\alpha_I-\alpha_J$ and $\sinh(\pm a+ b)\equiv\sinh(a+b)\cdot \sinh(-a+b)$.}
\begin{align}
    Z^{(k)}_{\mathrm{vec}}=\frac{\prod_{I\neq J }2\sinh\left(\frac{\alpha_{IJ}}{2}\right)\cdot\prod_{I, J=1 }^{k}2\sinh\left(\frac{\alpha_{IJ}+2\epsilon_+}{2}\right)}{\prod_{I=1 }^{k}\prod_{i=1 }^{N}2\sinh\left(\frac{\pm(\alpha_{I}-\phi_i)+\epsilon_+}{2}\right)\prod_{I, J=1 }^{k}2\sinh\left(\frac{\alpha_{IJ}+\epsilon_1}{2}\right)\cdot 2\sinh\left(\frac{\alpha_{IJ}+\epsilon_2}{2}\right)},
\end{align}
with $\phi_1+\cdots \phi_N=0$.
Here $Z^{(k,\n)}_{\mathrm{defect}}$ is the defect contribution which takes the expression
\begin{align}
    Z^{(k,n)}_{\mathrm{defect}}=\prod_{l=1}^{\n}\prod_{i=1}^{N}2\sinh(x_l-\phi_i)\prod_{I=1}^k\prod_{l=1}^{\n}\frac{\sinh(\pm(\alpha_I-x_l)+\epsilon_-)}{\sinh(\pm(\alpha_I-x_l)-\epsilon_+)},
\end{align}
where $x_l,\,l=1,\cdots \n,$ are the fugacities for the defect branes. The contour integral can be computed from JK-residues \cite{MR1318878} where the JK-poles are classified by $N+\n$ Young tableaux. See \cite{Hwang:2014uwa} for a recent discussion. 

As described in \cite{Assel:2018rcw}, the partition function $Z_{W_{\mathbf{r}}}$ of Wilson loop operator is computed from the line defect partition function $ Z_{\mathrm{5d/1d}}^{(\n)}$. In particular, for pure $SU(2)$ case, we have the expectation value
\begin{align}\label{eq:WilsonD3}
    \langle W_{\mathbf{2}^{\otimes \n}}\rangle =\frac{Z_{\mathrm{5d/1d}}^{(\n)}}{Z_{\mathrm{5d/1d}}^{(0)}}=(-1)^{\n}\oint \prod_{l=1}^{\n}\frac{d X_l}{2\pi i X_l}\frac{Z_{\mathrm{5d/1d}}^{(\n)}}{Z_{\mathrm{5d/1d}}^{(0)}},
\end{align}
where we have defined $X_l=e^{-x_l}$. $Z_{\mathrm{5d/1d}}^{(0)}$ is the partition function without defect. The expectation value $\langle W_{\mathbf{2}^{\otimes \n}}\rangle$ is expanded with instanton numbers $k$ as
\begin{align}
    \langle W_{\mathbf{2}^{\otimes \n}}\rangle=\langle W_{\mathbf{2}^{\otimes \n}}^{(0)}\rangle+\sum_{k=1}^{\infty}\mathfrak{q}^k\langle W_{\mathbf{2}^{\otimes \n}}^{(k)}\rangle,
\end{align}
where $\langle W_{\mathbf{2}^{\otimes \n}}^{(0)}\rangle$ is the zero-instanton or perturbative part
\begin{align}
    \langle W_{\mathbf{2}^{\otimes \n}}^{(0)}\rangle=(e^{\phi}+e^{-\phi})^{\n}, \quad \phi\equiv\phi_1=-\phi_2,
\end{align}
which is exactly the character of the representation $\mathbf{2}^{\otimes \n}$ and $\langle W_{\mathbf{2}^{\otimes \n}}^{(k)}\rangle$ is the $k$-instanton part. We will use the result from \eqref{eq:WilsonD3} to verify our calculations from a different approach in Section \ref{sec:SU2}. 

\subsection{Wilson loops from primitive non-compact 2-cycles}\label{sec:Wilson_general}
In M-theory, as explained in \cite{Kim:2021gyj}, the heavy quarks are generated by M2-branes winding on additional non-compact 2-cycles with infinite volumes, where the additional 2-cycles can be obtained from the blowups of the original Calabi-Yau three-fold $X$. However, these particles are not stationary on $\mathbb{R}^4$, they have an additional center of mass term or momentum term
\begin{align}
    \frac{1}{2\sinh(\epsilon_1/2)\cdot 2\sinh(\epsilon_2/2)},
\end{align}
which should be absorbed to the masses of these particles. Define $\C_l$ the primitive non-compact curves which is not decomposable, such that any other non-compact curve can be written as a linear combination of primitive non-compact curves with non-negative integer coefficients. By absorbing the momentum term, we define the effective mass of the additional stationary heavy particle as
\begin{align}
    M_{l}\equiv\frac{e^{-t_{\C_l}}}{2\sinh(\epsilon_1/2)\cdot 2\sinh(\epsilon_2/2)}.
\end{align}
From the Schwinger integral approach, the amplitude with a primitive curve $\C$ have an additional term which takes the expression
\begin{align}
    &\F^{\mathrm{Wilson}}_{\mathrm{BPS}}(t_{\C},t_C,\epsilon_1,\epsilon_2)\nonumber\\
    &=\sum_{\omega=1}^{\infty}\sum_{d=1}^{\infty}\sum_{C\in H_2(X,\mathbb{Z})}\sum_{j_L,j_R}(-1)^{2j_L+2j_R}N_{j_L,j_R}^{C,d\cdot\C}\frac{\chi_{j_L}(\omega\epsilon_-)\chi_{j_R}(\omega\epsilon_+)}{\omega\cdot 2\sinh(\omega\epsilon_1/2) \cdot 2\sinh(\omega\epsilon_2/2)}e^{-\omega t_{C}-\omega d\cdot t_{\C}}.
\end{align}
When the volume of the primitive curve is extremely large, the only surviving degrees are those with $\omega d\leq 1$, then the BPS partition function is reduced to 
\begin{align}
   & Z^{\mathrm{Wilson}}_{\mathrm{BPS}}=\exp\left(\F_{\mathrm{BPS}}(t_C,\epsilon_1,\epsilon_2)+\F^{\mathrm{Wilson}}_{\mathrm{BPS}}(t_{\C},t_C,\epsilon_1,\epsilon_2)\right)\nonumber\\
   &\sim Z_{\mathrm{BPS}}(t_C,\epsilon_1,\epsilon_2)\left(1+\sum_{C\in H_2(X,\mathbb{Z})}\sum_{j_L,j_R}(-1)^{2j_L+2j_R}N_{j_L,j_R}^{C,\C}{\chi_{j_L}(\epsilon_-)\chi_{j_R}(\epsilon_+)}e^{-t_{C}}M\right),
\end{align}
where
\begin{align}
    Z_{\mathrm{BPS}}(t_C,\epsilon_1,\epsilon_2)=\exp\left(\F_{\mathrm{BPS}}(t_C,\epsilon_1,\epsilon_2)\right)
\end{align}
is the BPS partition function of $X$.
We expect that the partition function $\Wilson{\mathbf{r}}$ in the presence of the Wilson loop operator $W_{\mathbf{r}}$ is the coefficient of $M$, then the expectation value of the Wilson loop operator has a BPS expansion
\begin{align}\label{eq:Wr}
    \langle W_{\mathbf{r}}\rangle \equiv \frac{\Wilson{\mathbf{r}}}{Z_{\mathrm{BPS}}}=\sum_{C\in H_2(X,\mathbb{Z})}\sum_{j_L,j_R}(-1)^{2j_L+2j_R}\widetilde{N}_{j_L,j_R}^{C}{\chi_{j_L}(\epsilon_-)\chi_{j_R}(\epsilon_+)}e^{-t_{C}},
\end{align}
where we use $\widetilde{N}_{j_L,j_R}^{C}\equiv N_{j_L,j_R}^{\C,C}$, which are always positive integers, for the BPS invariants of Wilson loops with representation $\mathbf{r}$ generated by the primitive curve $\C$. The non-decomposable property of the primitive curve indicates the representation $\mathrm{r}$ is also a non-decomposable representation. The BPS expansion \eqref{eq:Wr} for non-decomposable representation was first proposed in \cite{Kim:2021gyj} and was used to extract the BPS invariants of Wilson loop operators from the blowup equations.

For the case with $2$ primitive curves $\C_1,\C_2$, the amplitude has additional contributions
\begin{align}
    &\F^{\mathrm{Wilson}}_{\mathrm{BPS}}(t_{\C},t_C,\epsilon_1,\epsilon_2)=\F_{\mathrm{BPS},\{\C_1\}}M_1+\F_{\mathrm{BPS},\{\C_2\}}M_2+\I\cdot \F_{\mathrm{BPS},\{\C_1,\C_2\}}M_1M_2+\mathcal{O}(M_{1,2}^2),
\end{align}
where we define
\begin{align}
    \I\equiv 2\sinh(\epsilon_1/2)\cdot2\sinh(\epsilon_2/2),
\end{align}
and
\begin{align}
    \F_{\mathrm{BPS},\{\C_1,\cdots,\C_l\}}=\sum_{C\in H_2(X,\mathbb{Z})}\sum_{j_L,j_R}(-1)^{2j_L+2j_R}N_{j_L,j_R}^{C,\C_1,\cdots,\C_l}{\chi_{j_L}(\epsilon_-)\chi_{j_R}(\epsilon_+)}e^{-t_{C}}
\end{align}
as the amplitude of the curve class $C+\C_1+\cdots+\C_l$ for any $C\in H_2(X,\mathbb{Z})$.

At the large volume limit of the primitive curves, the partition function $\Wilson{\mathbf{r}_1\otimes \mathbf{r}_2}$ of the Wilson loop operator in the representation $\mathbf{r}_1\otimes \mathbf{r}_2$ can be obtained by picking up the coefficient of $M_1M_2$ in the partition function with primitive curves $\C_1,\C_2$, we then have the BPS expansion
\begin{align}\label{eq:wilson2}
    \langle W_{\mathbf{r}_1\otimes \mathbf{r}_2}\rangle &\equiv \frac{\Wilson{\mathbf{r}_1\otimes\mathbf{r}_2}}{Z_{\mathrm{BPS}}}=\F_{\mathrm{BPS},\{\C_1\}}\F_{\mathrm{BPS},\{\C_2\}}+\I\cdot\F_{\mathrm{BPS},\{\C_1,\C_2\}}.
\end{align}
From equation \eqref{eq:wilson2}, the $SU(2)_{L}\times SU(2)_R$ symmetry is not broken, thus we expect that the expectation value of the Wilson loop operator for the tensor product of two non-decomposable representations $\mathbf{r}_1$ and $\mathbf{r}_2$ can be still expanded as
\begin{align}
    \langle W_{\mathbf{r}_1\otimes \mathbf{r}_2}\rangle =\sum_{C\in H_2(X,\mathbb{Z})}\sum_{j_L,j_R}(-1)^{2j_L+2j_R}\widetilde{N}_{j_L,j_R}^{C}{\chi_{j_L}(\epsilon_-)\chi_{j_R}(\epsilon_+)}e^{-t_{C}}.
\end{align}
However, due to the existence of the momentum term $\I$ in \eqref{eq:wilson2}, $\widetilde{N}_{j_L,j_R}^{C}$ for decomposable representations would not always be positive.

One can generalize to the case with $\n$ primitive curves $\C_l,l=1,\cdots,\n,$,
From the Schwinger integral approach, the amplitude with $\n$ primitive curves $\C_l,l=1,\cdots,\n,$ take the expression
\begin{align}
    &\F^{\mathrm{Wilson}}_{\mathrm{BPS}}(t_{\C},t_C,\epsilon_1,\epsilon_2)=\sum_{l=1}^{\n}\mathcal{I}^{l-1}\cdot\sum_{\mathcal{S}_l}\sum_{\{C_1,\cdots \C_l\}\in\mathcal{S}_l}\F_{\mathrm{BPS},\{C_1,\cdots \C_l\}}\prod_{i=1}^lM_{i}+\mathcal{O}(M^{\n+1}),
\end{align}
where $\mathcal{S}_l$ is a subset of  $\{\C_1,\cdots,\C_{\n}\}$ with $l$ primitive curves. 
At the large volume limit of the primitive curves, we have the expectation value for the Wilson loop operator of generic tensor product of non-decomposable representations as
\begin{align}\label{eq:wilson_n}
    \langle W_{\mathbf{r}_1\otimes \cdots\otimes\mathbf{r}_{\n}}\rangle=\sum_{k=1}^{\n} \sum_{{\mathcal{S}^{(k)} }}\sum_{\{\C_{l_{i,1}},\cdots,\C_{l_{i,\n_i}}\}\in\mathcal{S}_{n_i}}\I^{\n-k}\cdot\prod_{i=1}^{k}\F_{\mathrm{BPS},\{\C_{l_{i,1}},\cdots,\C_{l_{i,\n_i}}\}}
\end{align}
by summing up all subsets $\mathcal{S}_{n_i}\in\mathcal{S}=\left\{\mathcal{S}_{n_1},\cdots,\mathcal{S}_{n_k}|\bigcup_{i=1}^{k} \mathcal{S}_{n_i}^{(k)}=\{\C_{1},\cdots,\C_{\n}\}\right\}.$ We claim that \eqref{eq:wilson_n} provides the BPS expansion for Wilson loop expectation value in arbitrary representation and the BPS invariants ${N}_{j_L,j_R}^{C,\{\C_{l_{i,1}},\cdots,\C_{l_{i,\n_i}}\}}$ are given by each BPS sector
\begin{align}\label{eq:BPSsector_expand}
    \F_{\mathrm{BPS},\{\C_{l_{i,1}},\cdots,\C_{l_{i,\n_i}}\}}=\sum_{C\in H_2(X,\mathbb{Z})}\sum_{j_L,j_R}(-1)^{2j_L+2j_R}{N}_{j_L,j_R}^{C,\{\C_{l_{i,1}},\cdots,\C_{l_{i,\n_i}}\}}{\chi_{j_L}(\epsilon_-)\chi_{j_R}(\epsilon_+)}e^{-t_{C}}.
\end{align} 
The expectation value \eqref{eq:wilson_n} can also be recast to the pseudo-BPS expansion
\begin{align}\label{eq:Wilson_n2}
    \langle W_{\mathbf{r}_1\otimes \cdots\otimes\mathbf{r}_{\n}}\rangle=\sum_{C\in H_2(X,\mathbb{Z})}\sum_{j_L,j_R}(-1)^{2j_L+2j_R}\widetilde{N}_{j_L,j_R}^{C}{\chi_{j_L}(\epsilon_-)\chi_{j_R}(\epsilon_+)}e^{-t_{C}},
\end{align}
with the {\it pseudo-BPS invariants} $\widetilde{N}_{j_L,j_R}^{C}$. When the representation is the non-decomposable representation, the pseudo-BPS invariant is indeed the BPS invariant which is always a positive integer. When the representation is decomposable, the pseudo-BPS invariant is not a proper BPS invariant which could be negative. Examples with negative integers can be found in Table \ref{tb:P1P1_z4_BPS} and Table \ref{tb:P1P1_z5_BPS} for $SU(2)$ gauge theory. 

\paragraph{Remarks} In general, the Wilson loop expectation of tensor product of representations are not equal to the product of Wilson loop expectation values in each individual representations. However, this is true in the NS limit as a consequence from the BPS expansion \eqref{eq:wilson_n}. It is clear to see that in the NS limit $\epsilon_2\rightarrow 0$, we have $\I\rightarrow 0$, such that
\begin{align}\label{eq:wilson_split}
    \left.\langle W_{\mathbf{r}_1\otimes \cdots\otimes\mathbf{r}_{\n}}\rangle \right|_{\epsilon_2\rightarrow0}=\prod_{l=1}^{\n}\left.\F_{\mathrm{BPS},\{\C_{l}\}}\right|_{\epsilon_2\rightarrow0}=\prod_{l=1}^{\n}\left.\langle W_{\mathbf{r}_l}\rangle\right|_{\epsilon_2\rightarrow0}.
\end{align}
This property can also be recovered from the refined holomorphic anomaly equation in Section \ref{sec:HAE_Wilson}.

\subsection{Example: pure $SU(2)$ gauge theory}\label{sec:SU2}
\begin{figure}[]
\begin{center}
\includegraphics{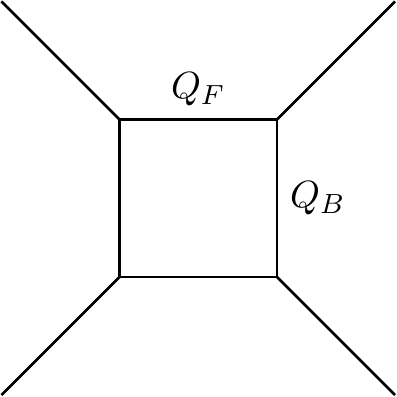}
\end{center}
\caption{{\pq}-brane web diagram in the $x^{5,6}$ directions for pure $SU(2)$ gauge theory.}
\label{fig:su2brane}
\end{figure}

In this section, we study the Wilson loop partition function for pure 5d $\mathcal{N}=1$ $SU(2)$ gauge theory with zero theta angle. Geometrically, this theory can be obtained from M-theory compactified on local $\mathbb{P}^1\times \mathbb{P}^1$, where the partition function can be computed from the {\pq}-brane web diagrams via refined topological vertex \cite{Iqbal:2007ii}. In the type IIB configuration, the 5d $SU(2)$ theory can be described by two D5-branes spanned over $x^{0,1,2,3,4,5}$ directions and two NS5-branes spanned over $x^{0,1,2,3,4,6}$ directions. Then the {\pq}-brane web diagram describes the {\pq} 5-brane configurations in the $x^{5,6}$ direction, as illustrated in Figure \ref{fig:su2brane}.

To generate Wilson loops, we need to add stationary heavy quarks. An easy way to obtain the heavy quarks is to insert fundamental flavors and then take the heavy mass limit. In the {\pq} 5-brane configuration, the fundamental matters are provided by additional D5-branes ending on 7-branes at infinity, the Wilson loops in the 5d gauge theory are provides by fundamental strings or {\pq}-strings ending on the additional D5-branes. One can then move the D5-branes to infinity to make the particles heavy. Now the {\pq}-strings are part of the background of the theory, where the supersymmetry is broken by half. We can now treat the heavy flavor matters as Wilson loop particles, however, the D5-branes we inserted are spanned over the direction $x^{0,1,2,3,4,5}$, the {\pq} strings ending on them are movable along $x^{1,2,3,4}$ thus are not stationary. The non-stationary property generate a momentum factor 
\begin{align}
 \mathcal{I}^{-1}=\frac{1}{2\sinh(\epsilon_1/2)\cdot 2\sinh(\epsilon_2/2)}=\frac{\sqrt{q_1q_2}}{(1-q_1)(1-q_2)}   
\end{align}
 that has to be absorbed into the flavor masses as described in Section \ref{sec:Wilson_general}.  The heavy quarks obtained from this setups are bosonic, which is different from the fermionic approach in Section \ref{sec:SUN_cod4}. However, the Wilson loop expectation values obtained in both method should be the same since the Wilson loop doesn't care about the spins of the source particles.

The partition function of pure $SU(2)$ theory can be computed using the refined topological vertex method \cite{Iqbal:2007ii} from the brane web diagram as shown in Figure \ref{fig:su2brane}, which is the graph dual of the toric diagram of local $\mathbb{P}^1\times\mathbb{P}^1$. The volume of two $\mathbb{P}^1$'s are described by the K\"ahler parameters $Q_F=e^{-2\phi}$ and $Q_B=\mathfrak{q} e^{-2\phi}$ respectively. Here $\phi$ is the Coulomb parameter of the gauge group $SU(2)$, $\mathfrak{q}=e^{-m_0}$ is the instanton counting parameter. Following the notation in \cite{Bao:2013pwa}, the partition function takes the expression \footnote{In this subsection, we use the standard notation $q=q_1=e^{\epsilon_1},t=\frac{1}{q_2}=e^{-\epsilon_2}$ in the refined topological vertex calculations. Note that $\mathcal{I}=(q_1q_2)^{-1/2}(1-q_1)(1-q_2)=-(qt)^{-1/2}(1-q)(1-t)$ in $t,q$ variables and the appearance of the minus sign.}
\begin{align}\label{eq:Z_A1_Nf0}
    Z^{SU(2)}=\sum_{\mu_1,\mu_2}Q_B^{|\mu_1|+|\mu_2|}{t}^{||\mu_1^t||^2}{q}^{||\mu_2||^2}\frac{\tilde{Z}_{\mu_1}({t},{q})\tilde{Z}_{\mu_2}({t},{q})\tilde{Z}_{\mu_1^t}({q},{t})\tilde{Z}_{\mu_2^t}({q},{t})}{\mathcal{R}_{\mu_2^t\mu_1}(Q_F\sqrt{\frac{q}{t}})\mathcal{R}_{\mu_2^t\mu_1}(Q_F\sqrt{\frac{t}{q}})},
\end{align}
where 
\begin{align}
    \tilde{Z}_{\mu}({t},{q})&=\prod_{(i,j)\in \mu}\left(1-t^{\mu_j^t-i+1}q^{\mu_i-j}\right)^{-1},\\
    \mathcal{R}_{\mu_1\mu_2}(Q;t,q)&=\prod_{i,j=1}^{\infty}\left(1-Qt^{i-\frac{1}{2}-\mu_{1,j}}q^{j-\frac{1}{2}-\mu_{2,i}}\right).
\end{align}
 It was proved in \cite{Iqbal:2007ii,Taki:2007dh} that the partition function \eqref{eq:Z_A1_Nf0} can be alternatively written as Nekrasov's partition function from instanton counting
\begin{align}
    Z^{SU(2)}(Q_B,Q_F;t,q)&=Z_{\mathrm{pert}}^{SU(2)}(Q_F;t,q)Z_{\mathrm{inst}}^{SU(2)}(Q_B,Q_F;t,q),\\
    Z_{\mathrm{pert}}^{SU(2)}(Q_F;t,q)&=\prod_{i,j=1}^{\infty}\left(1-Q_{F}t^{i-1}q^j\right)^{-1}\left(1-Q_{F}t^{i}q^{j-1}\right)^{-1},\\
    Z_{\mathrm{inst}}^{SU(2)}(Q_B,Q_F;t,q)&=\sum_{\mu_1,\mu_2}\left(\mathfrak{q}\sqrt{\frac{q}{t}}\right)^{|\mu_1|+|\mu_2|}\cdot\frac{1}{\prod_{i,j=1}^2N_{\mu_i\mu_j}(Q_{ij};t,q)}
\end{align}
where $Q_{11}=Q_{22}=1,Q_{12}=Q_{21}^{-1}=e^{-2\phi}$, 
and 
 \begin{align}
    N_{\mu_1\mu_2}(Q;t,q)=\prod_{(i,j)\in\mu_1}\left(1-Qt^{-\mu^t_{2,j}+i-1}q^{-\mu_{1,i}+j}\right)\cdot\prod_{(i,j)\in\mu_2}\left(1-Qt^{\mu^t_{1,j}-i}q^{\mu_{2,i}-j+1}\right).
\end{align}

\begin{figure}[]
\begin{center}
\includegraphics{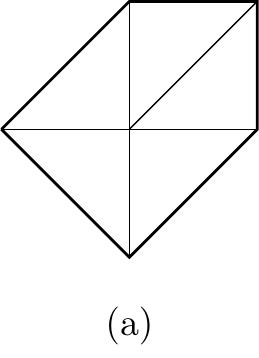}
\hspace{2cm}
\includegraphics{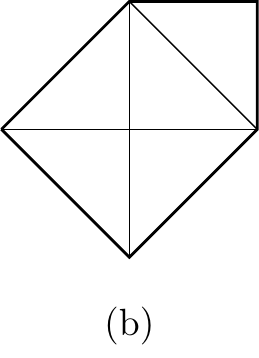}
\end{center}
\caption{Toric diagrams of local $\mathbb{P}^1\times\mathbb{P}^1$ blowup at one point. (a) is a star triangulation of the polygon commonly used in the mirror symmetry. (b) is a flop transition of (a)}
\label{fig:toric_P1P1_blowup}
\end{figure}

\begin{figure}[]
\begin{center}
\includegraphics{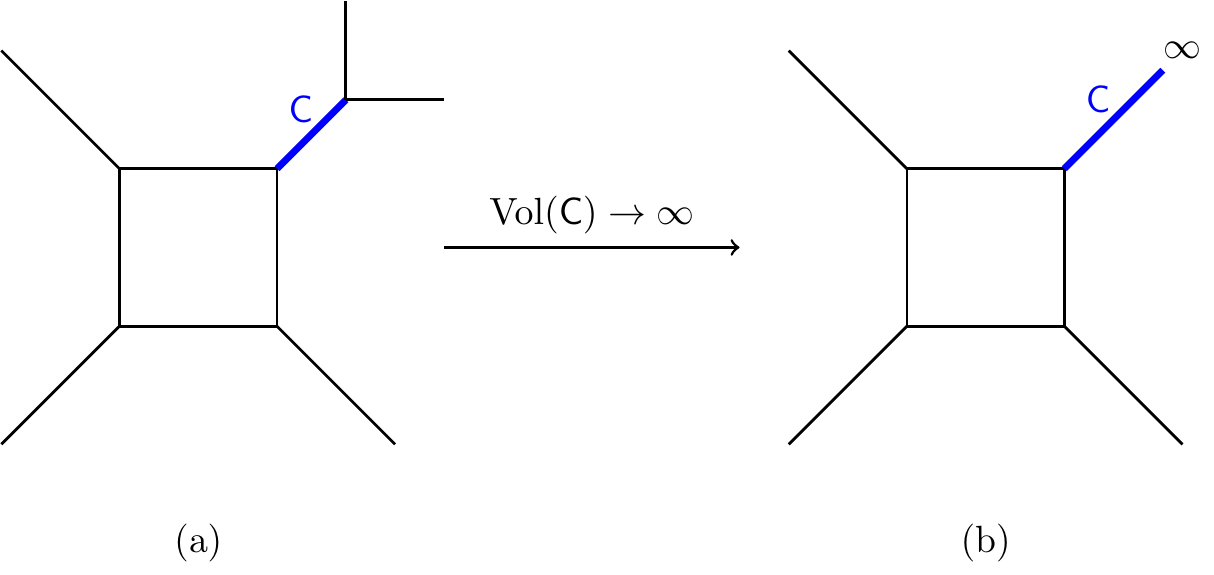}
\end{center}
\caption{Brane diagrams in the $x^{5,6}$ directions. (a) The brane diagram for $SU(2)$ gauge theory with one flavor. $\C$ is the curve class related to the edge in the diagram. (b) The brane diagram with an infinitely long curve $\C$.}
\label{fig:su2braneC}
\end{figure}
\paragraph{Wilson loop in the fundamental representation}
Now we want to add a fundamental flavor by blowing up the geometry $\mathbb{P}^1\times\mathbb{P}^1$ once at a point. In the standard calculation of mirror symmetry, e.g. in \cite{Chiang:1999tz,Huang:2013yta}, we usually chose a Calabi-Yau phase that that the the polygon has a star triangulation as in Figure \ref{fig:toric_P1P1_blowup}(a). However, to make the degree of the additional curve always positive, one needs to do a flop transition such that we get a different Calabi-Yau phase described from the triangulation as in Figure \ref{fig:toric_P1P1_blowup}(b). In this phase, the resulting geometry has an additional curve class $\C$ as described in the dual graph Figure \ref{fig:su2braneC}(a). Denote $Q_1=e^{\phi-m}$ the K\"ahler parameter related to the curve $\C$, where $m$ is the mass of the fundamental flavor, then from the refined topological vertex method \cite{Iqbal:2007ii}, the partition function of $SU(2)$ theory with one fundamental matter is \cite{Bao:2013pwa}
\begin{align}\label{eq:Z_A1_Nf1}
    Z^{SU(2),N_f=1}=Z_{\mathrm{pert}}^{SU(2)}\sum_{\mu_1,\mu_2}\left(\mathfrak{q}\sqrt{\frac{q}{t}}\right)^{|\mu_1|+|\mu_2|}\cdot\frac{\mathcal{R}_{\mu_1^t\emptyset}(Q_1)\mathcal{R}_{\mu_2^t\emptyset}(Q_1Q_F)}{\prod_{i,j=1}^2N_{\mu_i\mu_j}(Q_{ij};t,q)}
\end{align}
by summing all the partitions $\mu_1,\mu_2$. Noticing that \footnote{Here we use the notation $\mathrm{PE}[\pm f(x)]=\exp\left[\pm\sum\limits_{\omega=1}^{\infty}\frac{1}{\omega}\cdot f(x^{\omega})\right]$}
\begin{equation}
\begin{split}
    \mathcal{R}_{\mu^t\emptyset}(Q;t,q)&=\mathcal{R}_{\emptyset\mu^t}(Q;q,t)=\prod_{i,j=1}^{\infty}\left(1-Q{t}^{i-\frac{1}{2}-\mu_{j}^t}{q}^{j-\frac{1}{2}}\right)\\
    &=\mathrm{PE}\left[-\frac{Q{t}^{\frac{1}{2}}}{1-{t}} \left(\sum_{i=1}^{\infty}{q}^{i-\frac{1}{2}}+\sum_{i=1}^{\mu_{1}}{q}^{i-\frac{1}{2}}({t}^{-\mu_{i}^t}-1)\right)\right]\\
    &=\mathrm{PE}\left[-\frac{Q({t}{q})^{\frac{1}{2}}}{(1-{t})(1-{q})}-Q \sum_{(i,j)\in\mu^t}{t}^{\frac{1}{2}-j}{q}^{i-\frac{1}{2}}\right],\\
    &=\mathrm{PE}\left[-\frac{Q({t}{q})^{\frac{1}{2}}}{(1-{t})(1-{q})}-Q \sum_{(i,j)\in\mu}{t}^{\frac{1}{2}-i}{q}^{j-\frac{1}{2}}\right],
\end{split}
\end{equation}
we have that comparing with pure water $SU(2)$ case, the extra factors appear in \eqref{eq:Z_A1_Nf1} can be written as
\begin{equation}
\begin{split}
   \mathcal{R}_{\mu_1^t\emptyset}(Q_1)\mathcal{R}_{\mu_2^t\emptyset}(Q_1Q_F)=&\prod_{i,j=1}^{\infty}{\left(1-Q_1{t}^{i-\frac{1}{2}-\mu_{1,i}^t}{q}^{j-\frac{1}{2}}\right)\left(1-Q_1 Q_F{t}^{i-\frac{1}{2}-\mu_{2,i}^t}{q}^{j-\frac{1}{2}}\right)}\\
   =\,&\mathrm{PE}\left[-\frac{e^{-m}({t}{q})^{\frac{1}{2}}}{(1-{t})(1-{q})}\Ch_{\mu_1\mu_2}(Q_F,{t},{q})\right],
\end{split}
\end{equation}
where $\Ch_{\mu_1\mu_2}(Q_F,{t},{q})$ is the equivariant Chern character defined in \cite{Losev:2003py,Shadchin:2004yx} which have the expression 
\begin{align}
    \Ch_{\mu_1\mu_2}(Q_F,{t},{q})=e^a+e^{-a}+\frac{(1-{t})(1-{q})}{({t}{q})^{\frac{1}{2}}}\left(e^a\sum_{(i,j)\in\mu_1} {t}^{\frac{1}{2}-i}{q}^{j-\frac{1}{2}}+e^{-a}\sum_{(i,j)\in\mu_2}{t}^{\frac{1}{2}-i}{q}^{j-\frac{1}{2}}\right).
\end{align}

In the massive limit $m\rightarrow \infty$, only the leading term of $e^{-m}$ is dominated. The $(1,1)$-string lived on the $(1,1)$ five-brane then provides the Wilson line. Together with the center of mass term, the coefficient of effective mass $M=-\frac{e^{-m}({t}{q})^{\frac{1}{2}}}{(1-{t})(1-{q})}$ gives the Wilson loop $\Wilson{\mathbf{2}}$ in the fundamental representation.

Our result here indicates that the partition function of $SU(2)$ with one heavy fundamental flavor has the following expansion
\begin{align}\label{eq:ZSU2Nf1}
    Z^{SU(2),N_f=1}=Z^{SU(2)}+\Wilson{\mathbf{2}}M+\mathcal{O}(M^2),
\end{align}
where
\begin{align}
    \Wilson{\mathbf{2}}=Z_{\mathrm{pert}}^{SU(2)}\sum_{\mu_1,\mu_2}\left(\mathfrak{q}\sqrt{\frac{q}{t}}\right)^{|\mu_1|+|\mu_2|}\cdot\frac{\Ch_{\mu_1\mu_2}(Q_F,{t},{q})}{\prod_{i,j=1}^2N_{\mu_i\mu_j}(Q_{ij};t,q)}
\end{align}
On the other hand, the partition function $Z^{SU(2),N_f=1}$ has the BPS expansion
\begin{align}
    Z^{SU(2),N_f=1}=\exp\left[\sum_{\omega=1}^{\infty}\sum_{{d},d_m,j_L,j_R} \frac{1}{\omega}N^{({d},d_m)}_{j_L,j_R}\frac{\chi_{j_L}({\omega}\epsilon_-)\chi_{j_R}({\omega}\epsilon_+)}{({t}^{-\omega/2}-{t}^{\omega/2})({q}^{\omega/2}-{q}^{-\omega/2})} e^{-\omega (d_1m_0+{d_2} {\phi})}e^{-d_mm}\right],
\end{align}
where $m_0=-\log \mathfrak{q}$. By comparing with the expansion \eqref{eq:ZSU2Nf1}, we immediately have the conclusion that the Wilson loop expectation value $\langle{W}_{\mathbf{2}}\rangle$ has a BPS expansion 
\begin{align}
   \langle{W}_{\mathbf{2}}\rangle\equiv \frac{\Wilson{\mathbf{2}}}{Z^{SU(2)}}=\sum_{d_1=0}^{\infty}\sum_{d_2=-1}^{\infty}\sum_{j_L,j_R} N^{{d},d_m=1}_{j_L,j_R}{\chi_{j_L}(\epsilon_-)\chi_{j_R}(\epsilon_+)}e^{-\omega (d_1m_0+{d_2} {\phi})},
\end{align}
while the expansion coefficients $N^{{d},d_m=1}_{j_L,j_R}$ count the BPS states with degree $m+(d_1m_0+{d_2} {\phi})$ in the blown up geometry of $\mathbb{P}^1\times \mathbb{P}^1$ with the phase in Figure \ref{fig:toric_P1P1_blowup}(b). Here the degree of Coulomb parameter $\phi$ start with negative charge $-1$ which comes from negative charge carried by the primitive curve. At perturbative and one-instanton level, we have
\begin{align}
    \label{eq:Wislon_2_0}
    \langle W_{\mathbf{2}}^{(0)}\rangle&=e^{\phi}+e^{-\phi},\\
    \langle W_{\mathbf{2}}^{(1)}\rangle&=-\frac{q_1q_2(e^{\phi}+e^{-\phi})}{(1-q_1q_2e^{2\phi})(1-q_1q_2e^{-2\phi})} \nonumber \\
     &=  \frac{ e^{-\phi}(1+e^{-2\phi})}{(1-q_1q_2e^{-2\phi})(1-(q_1q_2)^{-1}e^{-2\phi})},
\end{align}
which agrees with the result in Section \ref{sec:SUN_cod4} and result in \cite{Kim:2021gyj} from the blowup equations.
The instanton contributions come with the positive coefficients for the Fourier expansion in $e^{-\phi}$.
The two-instanton result is listed in Appendix \ref{appendix:2-inst}. We list few BPS invariants of the instanton part of $\langle W_{\mathbf{2}}\rangle$ in Table \ref{tb:P1P1_z1_BPS}. The perturbative part only contributes spin $(0,0)$ particle states, which can be read from equation \eqref{eq:Wislon_2_0} directly.
\begin{table}
\centering
 \begin{tabular}{|c|>{\centering\arraybackslash}m{5.4cm}||c|>{\centering\arraybackslash}m{5.4cm}|} \hline 
$ \mathbf{d} $ & $ \oplus \widetilde{N}_{j_l, j_r}^{\mathbf{d}} (j_l, j_r) $ & $ \mathbf{d} $ & $ \oplus \widetilde{N}_{j_l, j_r}^{\mathbf{d}} (j_l, j_r) $ \\ \hline
$(1, 1)$&$(0,0)$&$(1, 3)$&$(0,1)$\\ \hline 
 $(1, 5)$&$(0,2)$&$(1, 7)$&$(0,3)$\\ \hline 
 $(1, 9)$&$(0,4)$&$(2, 5)$&$(0,2)$\\ \hline 
 $(2, 7)$&$(0,2)\oplus2(0,3)\oplus(1/2,7/2)$&$(2, 9)$&$(0,2)\oplus2(0,3)\oplus3(0,4)\oplus(1/2,7/2)\oplus2(1/2,9/2)\oplus(1,5)$\\ \hline 
 $(3, 7)$&$(0,3)$&$(3, 9)$&$(0,2)\oplus2(0,3)\oplus3(0,4)\oplus(1/2,7/2)\oplus2(1/2,9/2)\oplus(1,5)$\\ \hline 
 \end{tabular}
\caption{BPS Spectrum of $SU(2)$ Wilson loop expectation value in the representation $\mathbf{2}$ for the curve class $d_1m_0+d_2\phi$ with $ d_1 =1,2, 3, $ and $ d_2 \leq 9 $. The $d_1=0$ part can be read from \eqref{eq:Wislon_2_0}.}
\label{tb:P1P1_z1_BPS}
\end{table}

\begin{figure}[h]

\begin{center}
\includegraphics{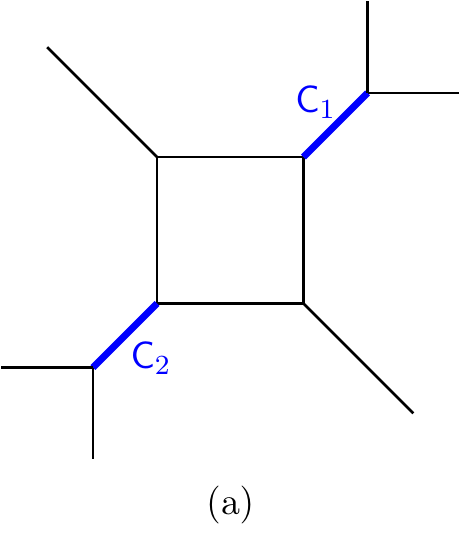}
\hspace{2.2cm}
\includegraphics{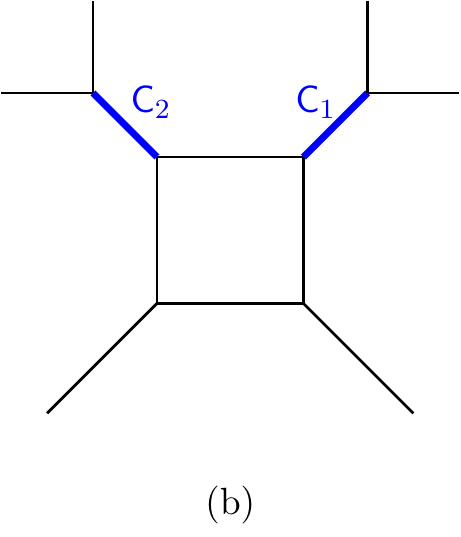}
\end{center}
\caption{Brane diagrams for $SU(2)$ gauge theory with two flavors, by inserting curves $\C_{1,2}$ in different positions.}
\label{fig:su2brane_Nf2}
\end{figure}

\paragraph{Wilson loop in the representation $\mathbf{2}\otimes \mathbf{2}$}
For the tensor product of two fundamental representations $\mathbf{2}\otimes \mathbf{2}$, we can introduce two fundamental flavor matters to generate it. In the brane diagram as illustrated in Figure \ref{fig:su2brane_Nf2}, we have two independent ways to add two fundamental flavor matters.

The partition function of diagrams (a) and (b) in Figure \ref{fig:su2brane_Nf2} are computed in \cite{Bao:2013pwa} as
\begin{align}
    &Z^{SU(2),N_f=2}_{\text{(a)}}=Z_{\mathrm{pert}}^{SU(2)}\sum_{\mu_1,\mu_2}\left(\mathfrak{q}\sqrt{\frac{q}{t}}\right)^{|\mu_1|+|\mu_2|}\cdot\frac{1}{\prod_{i,j=1}^2N_{\mu_i\mu_j}(Q_{ij};t,q)}\nonumber\\
   &\quad\quad\quad\quad\quad\quad\quad\quad\quad\quad \quad\quad \times \mathcal{R}_{\mu_1^t\emptyset}(Q_1)\mathcal{R}_{\mu_2^t\emptyset}(Q_1Q_F)\mathcal{R}_{\emptyset\mu_2}(Q_2)\mathcal{R}_{\emptyset\mu_1}(Q_2Q_F)\nonumber\\
   & \quad=Z^{SU(2)}+\Wilson{\mathbf{2}}(M_1+M_2)+Z_{{W}^{\text{(a)}}_{\mathbf{2}\otimes\mathbf{2}}} M_1M_2+\cdots,
\end{align}
\begin{align}
    &Z^{SU(2),N_f=2}_{\text{(b)}}=Z_{\mathrm{pert}}^{SU(2)}\sum_{\mu_1,\mu_2}\left(\mathfrak{q}\sqrt{\frac{q}{t}}\right)^{|\mu_1|+|\mu_2|}\cdot\frac{1}{\prod_{i,j=1}^2N_{\mu_i\mu_j}(Q_{ij};t,q)}\nonumber\\
   &\quad\quad\quad\quad\quad \times \mathcal{R}_{\mu_1^t\emptyset}(Q_1)\mathcal{R}_{\mu_2^t\emptyset}(Q_1Q_F)\mathcal{R}_{\mu_1^t\emptyset}(Q_2)\mathcal{R}_{\mu_2^t\emptyset}(Q_2Q_F)\mathcal{R}_{\emptyset\emptyset}(Q_1Q_2Q_B\sqrt{q/t})\nonumber\\
   & \quad=Z^{SU(2)}+\Wilson{\mathbf{2}}(M_1+M_2)+Z_{{W}^{\text{(b)}}_{\mathbf{2}\otimes\mathbf{2}}}M_1M_2+\cdots,
\end{align}
where we have
\begin{align}\label{eq:W22a}
    Z_{{W}^{\text{(a)}}_{\mathbf{2}\otimes\mathbf{2}}}=Z_{\mathrm{pert}}^{SU(2)}\sum_{\mu_1,\mu_2}\left(\mathfrak{q}\sqrt{\frac{q}{t}}\right)^{|\mu_1|+|\mu_2|}\cdot\frac{\Ch_{\mu_1\mu_2}(Q_F,{t},{q})\Ch_{\mu_2^t\mu_1^t}(Q_F,{q},{t})}{\prod_{i,j=1}^2N_{\mu_i\mu_j}(Q_{ij};t,q)}
\end{align}
and
\begin{align}\label{eq:W22b}
    Z_{{W}^{\text{(b)}}_{\mathbf{2}\otimes\mathbf{2}}}=\,&Z_{\mathrm{pert}}^{SU(2)}\sum_{\mu_1,\mu_2}\left(\mathfrak{q}\sqrt{\frac{q}{t}}\right)^{|\mu_1|+|\mu_2|}\cdot\frac{1}{\prod_{i,j=1}^2N_{\mu_i\mu_j}(Q_{ij};t,q)}\nonumber\\
    &\times\left(\Ch_{\mu_1\mu_2}(Q_F,{t},{q})^2+(1-q)(1-1/t) \frak{q}\right)
\end{align}
Even though the expression of \eqref{eq:W22a} and \eqref{eq:W22b} are quite different, we observe that they are identical
\begin{align}
    Z_{{W}^{\text{(a)}}_{\mathbf{2}\otimes\mathbf{2}}}=Z_{{W}^{\text{(b)}}_{\mathbf{2}\otimes\mathbf{2}}}
\end{align}
by comparing them order by order.

At perturbative and one-instanton level, we have
\begin{align}
    \label{eq:Wislon_z2_0}
    \langle W_{\mathbf{2}\otimes\mathbf{2}}^{(0)}\rangle=\,&(e^{\phi}+e^{-\phi})^2, \\
    \langle W_{\mathbf{2}\otimes \mathbf{2}}^{(1)}\rangle=\,& -\frac{2q_1q_2(e^{\phi}+e^{-\phi})^2}{(1-q_1q_2e^{2\phi})(1-q_1q_2e^{-2\phi})} 
     +\frac{(1-q_1)(1-q_2)(1+q_1q_2)}{(1-q_1q_2e^{2\phi})(1-q_1q_2e^{-2\phi})} \nonumber \\
     =\, & \frac{2 (1+e^{-2\phi})^2}{(1-q_1q_2e^{-2\phi})(1-(q_1q_2)^{-1}e^{-2\phi})} 
     -\frac{{\I}\cdot ((q_1q_1)^{\frac{1}{2}}+(q_1q_2)^{-\frac{1}{2}}) e^{-2\phi}}{(1-q_1q_2e^{-2\phi})(1-(q_1q_2)^{-1}e^{-2\phi})}\label{eq:Wislon_z2_1},
\end{align}
which agrees with the result in Section \ref{sec:SUN_cod4} and result in \cite{Kim:2021gyj} from the blowup equations. The two-instanton result is listed in Appendix \ref{appendix:2-inst}. The expectation value $\langle W_{\mathbf{2}\otimes \mathbf{2}}\rangle$ has an BPS expansion according to \eqref{eq:wilson_n}
\begin{align}\label{eq:BPS_P1P1_z2}
    \langle W_{\mathbf{2}\otimes \mathbf{2}}\rangle=\F_{\mathrm{BPS},\{\C\}}^2+\mathcal{I}\cdot\F_{\mathrm{BPS},\{\C^2\}},
\end{align}
where the BPS sector $\F_{\mathrm{BPS},\{\C\}}=\langle W_{ \mathbf{2}}\rangle$ is generated by a single primitive curve and $\F_{\mathrm{BPS},\{\C^2\}}$ is generated by two primitive curves. At one-instanton, the BPS expansion \eqref{eq:BPS_P1P1_z2} becomes
\begin{align}
    \langle W^{(1)}_{\mathbf{2}\otimes \mathbf{2}}\rangle=2\F_{\mathrm{BPS},\{\C\}}^{(0)}\F_{\mathrm{BPS},\{\C\}}^{(1)}+\mathcal{I}\cdot\F_{\mathrm{BPS},\{\C^2\}}^{(1)},
\end{align}
where we used the notation $\F_{\mathrm{BPS},\{\cdot\}}^{(k)}$ to denote the $k$-instanton part of the BPS sector $\F_{\mathrm{BPS},\{\cdot\}}$ and we find
\begin{align}\label{eq:Wislon_z2BPS_1}
    \F_{\mathrm{BPS},\{\C^2\}}^{(1)}=-\frac{ ((q_1q_1)^{\frac{1}{2}}+(q_1q_2)^{-\frac{1}{2}}) e^{-2\phi}}{(1-q_1q_2e^{-2\phi})(1-(q_1q_2)^{-1}e^{-2\phi})}.
\end{align}
By comparing the Fourier expansion of the single instanton contribution \eqref{eq:Wislon_z2BPS_1} in $e^{-\phi}$, the sign $(-1)^{2j_L+2j_R}$ for the internal half-spins lead the non-negative integers in the BPS invariants $N^{C,\{\C_1,\C_2\}}_{j_L,j_R}$ according to \eqref{eq:BPSsector_expand}.  
 We list few BPS invariants of the BPS sector $\F_{\mathrm{BPS},\{\C^2\}}$ in Table \ref{tb:BPS_P1P1_z2_BPS} and the pseudo-BPS invariants of the instanton part of $\langle W_{\mathbf{2}\otimes \mathbf{2}}\rangle$ in Table \ref{tb:P1P1_z2_BPS}. 
\begin{table}
\centering
 \begin{tabular}{|c|>{\centering\arraybackslash}m{5.4cm}||c|>{\centering\arraybackslash}m{5.4cm}|} \hline
$ \mathbf{d} $ & $ \oplus \widetilde{N}_{j_l, j_r}^{\mathbf{d}} (j_l, j_r) $ & $ \mathbf{d} $ & $ \oplus \widetilde{N}_{j_l, j_r}^{\mathbf{d}} (j_l, j_r) $ \\ \hline
$(1, 0)$&$2(0,0)$&$(1, 2)$&$(0,0)\oplus(0,1)\oplus(1/2,1/2)$\\ \hline 
 $(1, 4)$&$(0,1)\oplus(0,2)\oplus(1/2,3/2)$&$(1, 6)$&$(0,2)\oplus(0,3)\oplus(1/2,5/2)$\\ \hline 
 $(1, 8)$&$(0,3)\oplus(0,4)\oplus(1/2,7/2)$&$(2, 2)$&$(0,0)$\\ \hline 
 $(2, 4)$&$(0,1)\oplus(0,2)\oplus(1/2,3/2)$&$(2, 6)$&$(0,0)\oplus3(0,2)\oplus2(0,3)\oplus(1/2,3/2)\oplus2(1/2,5/2)\oplus(1/2,7/2)\oplus(1,3)$\\ \hline 
 $(2, 8)$&$(0,1)\oplus2(0,2)\oplus5(0,3)\oplus4(0,4)\oplus(1/2,3/2)\oplus2(1/2,5/2)\oplus5(1/2,7/2)\oplus2(1/2,9/2)\oplus(1,3)\oplus2(1,4)\oplus(1,5)\oplus(3/2,9/2)$&$(3, 6)$&$(0,2)\oplus(0,3)\oplus(1/2,5/2)$\\ \hline 
 $(3, 8)$&\multicolumn{3}{>{\centering\arraybackslash}m{12cm}|}{$(0,1)\oplus2(0,2)\oplus5(0,3)\oplus4(0,4)\oplus(1/2,3/2)\oplus2(1/2,5/2)\oplus5(1/2,7/2)\oplus2(1/2,9/2)\oplus(1,3)\oplus2(1,4)\oplus(1,5)\oplus(3/2,9/2)$}\\ \hline 
 \end{tabular}
\caption{Pseudo-BPS Spectrum of $SU(2)$ Wilson loop expectation value in the representation $\mathbf{2}\otimes\mathbf{2}$ for the curve class $d_1m_0+d_2\phi$ with $ d_1 =1,2, 3, $ and $ d_2 \leq 8 $. The $d_1=0$ part can be read from \eqref{eq:Wislon_z2_0}.}
\label{tb:P1P1_z2_BPS}
\end{table}

\begin{table}
\centering
 \begin{tabular}{|c|>{\centering\arraybackslash}m{5.4cm}||c|>{\centering\arraybackslash}m{5.4cm}|} \hline
$ \mathbf{d} $ & $ \oplus {N}_{j_l, j_r}^{\mathbf{d},\{\C^{2}\}} (j_l, j_r) $ & $ \mathbf{d} $ & $ \oplus {N}_{j_l, j_r}^{\mathbf{d},\{\C^{2}\}} (j_l, j_r) $ \\ \hline
$(1, 2)$&$(0,1/2)$&$(1, 4)$&$(0,3/2)$\\ \hline 
 $(1, 6)$&$(0,5/2)$&$(1, 8)$&$(0,7/2)$\\ \hline 
 $(2, 4)$&$(0,3/2)$&$(2, 6)$&$(0,3/2)\oplus3(0,5/2)\oplus(1/2,3)$\\ \hline 
 $(2, 8)$&$(0,3/2)\oplus3(0,5/2)\oplus5(0,7/2)\oplus(1/2,3)\oplus3(1/2,4)\oplus(1,9/2)$&$(3, 6)$&$(0,5/2)$\\ \hline 
 $(3, 8)$&\multicolumn{3}{>{\centering\arraybackslash}m{12cm}|}{$(0,3/2)\oplus3(0,5/2)\oplus5(0,7/2)\oplus(1/2,3)\oplus3(1/2,4)\oplus(1,9/2)$}\\ \hline 
 \end{tabular}
\caption{BPS Spectrum of the BPS sector $\F_{\mathrm{BPS},\{\C^{2}\}}$ for the curve class $d_1m_0+d_2\phi$ with $ d_1\leq 3 $ and $ d_2 \leq 8 $. }
\label{tb:BPS_P1P1_z2_BPS}
\end{table}

\begin{figure}[]

\begin{center}
\includegraphics{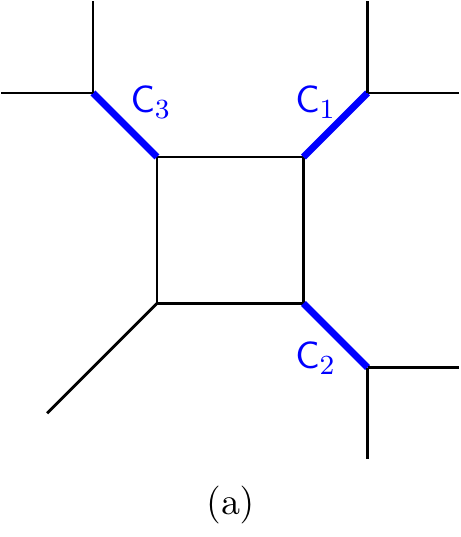}
\hspace{2.2cm}
\includegraphics{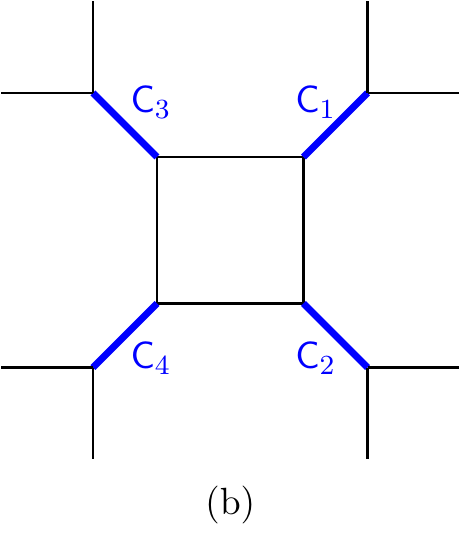}
\end{center}

\caption{Brane diagrams for $SU(2)$ gauge theory with (a) three flavors and (b) four flavors. In the extremely massive limit, the Wilson loops for SU(2) theory are introduced by inserting curves $\C_i,i=1,\cdots 4$.}
\label{fig:su2brane_Nf}
\end{figure}
\paragraph{Wilson loops in the representations $\mathbf{2}\otimes \mathbf{2}\otimes \mathbf{2}$ and $\mathbf{2}^{\otimes 4}$}
Wilson loops with more higher representations can be obtained by adding more flavors. For example, the brane diagrams with three and four fundamental flavors are illustrated in Figure \ref{fig:su2brane_Nf}, where we have the partition functions
\begin{align}
    &Z^{SU(2),N_f=3}=Z_{\mathrm{pert}}^{SU(2)}\sum_{\mu_1,\mu_2}\left(\mathfrak{q}\sqrt{\frac{q}{t}}\right)^{|\mu_1|+|\mu_2|}\cdot\frac{1}{\prod_{i,j=1}^2N_{\mu_i\mu_j}(Q_{ij};t,q)}\nonumber\\
   &\quad\quad\quad\quad\quad\quad\quad \times \mathcal{R}_{\mu_1^t\emptyset}(Q_{1})\mathcal{R}_{\mu_2^t\emptyset}(Q_1Q_F)\mathcal{R}_{\mu_1^t\emptyset}(Q_3)\mathcal{R}_{\mu_2^t\emptyset}(Q_3Q_F)\nonumber\\
   &\quad\quad\quad\quad\quad\quad\quad \times \mathcal{R}_{\emptyset\mu_2}(Q_2)\mathcal{R}_{\emptyset\mu_1}(Q_2Q_F)\mathcal{R}_{\emptyset\emptyset}(Q_1Q_3Q_B\sqrt{q/t})\nonumber\\
   & \quad=Z^{SU(2)}+\Wilson{\mathbf{2}\otimes\mathbf{2}\otimes\mathbf{2}}M_1M_2M_3+\cdots,\\
\end{align}
\begin{align} 
    &Z^{SU(2),N_f=4}=Z_{\mathrm{pert}}^{SU(2)}\sum_{\mu_1,\mu_2}\left(\mathfrak{q}\sqrt{\frac{q}{t}}\right)^{|\mu_1|+|\mu_2|}\cdot\frac{1}{\prod_{i,j=1}^2N_{\mu_i\mu_j}(Q_{ij};t,q)}\nonumber\\
   &\quad\quad\quad\quad\quad\quad\quad \times \mathcal{R}_{\mu_1^t\emptyset}(Q_{1})\mathcal{R}_{\mu_2^t\emptyset}(Q_1Q_F)\mathcal{R}_{\mu_1^t\emptyset}(Q_3)\mathcal{R}_{\mu_2^t\emptyset}(Q_3Q_F)\mathcal{R}_{\emptyset\emptyset}(Q_1Q_3Q_B\sqrt{q/t})\nonumber\\
   &\quad\quad\quad\quad\quad\quad\quad \times \mathcal{R}_{\emptyset\mu_2}(Q_2)\mathcal{R}_{\emptyset\mu_1}(Q_2Q_F)\mathcal{R}_{\emptyset\mu_2}(Q_4)\mathcal{R}_{\emptyset\mu_1}(Q_4Q_F)\mathcal{R}_{\emptyset\emptyset}(Q_2Q_4Q_B\sqrt{t/q})\nonumber\\
   & \quad=Z^{SU(2)}+\Wilson{\mathbf{2}^{\otimes 4}}M_1M_2M_3M_4+\cdots.
\end{align}
Then we find the Wilson loop partition functions are
\begin{align}
    \Wilson{\mathbf{2}\otimes\mathbf{2}\otimes\mathbf{2}}=\,&Z_{\mathrm{pert}}^{SU(2)}\sum_{\mu_1,\mu_2}\left(\mathfrak{q}\sqrt{\frac{q}{t}}\right)^{|\mu_1|+|\mu_2|}\cdot\frac{1}{\prod_{i,j=1}^2N_{\mu_i\mu_j}(Q_{ij};t,q)}\nonumber\\
    &\times \Ch_{\mu_2^t\mu_1^t}(Q_F,{q},{t})\left(\Ch_{\mu_1\mu_2}(Q_F,{t},{q})^2+(1-q)(1-1/t) \frak{q}\right),\\
    \Wilson{\mathbf{2}^{\otimes 4}}=\,&Z_{\mathrm{pert}}^{SU(2)}\sum_{\mu_1,\mu_2}\left(\mathfrak{q}\sqrt{\frac{q}{t}}\right)^{|\mu_1|+|\mu_2|}\cdot\frac{1}{\prod_{i,j=1}^2N_{\mu_i\mu_j}(Q_{ij};t,q)}\nonumber\\
    &\times \left[\Ch_{\mu_2^t\mu_1^t}(Q_F,{q},{t})^2\left(\Ch_{\mu_1\mu_2}(Q_F,{t},{q})^2 +(1-q)(1-1/t) \frak{q}\right)\right.\nonumber\\
    &\quad\quad\quad\quad\quad\quad\quad\quad\quad\quad +(1-t)(1-1/q) \frak{q}\Ch_{\mu_1\mu_2}(Q_F,{t},{q})^2\nonumber\\
    &\quad\quad\quad\quad\quad\quad\quad\quad\quad\quad \left.+(t^{1/2}-t^{-1/2})^2(q^{1/2}-q^{-1/2})^2\mathfrak{q}^2\right],
\end{align}
from which we get the perturbative and one-instanton Wilson loop expectation values
\begin{align}
    \label{eq:Wislon_z3_0}
    \langle W_{\mathbf{2}\otimes\mathbf{2}\otimes\mathbf{2}}^{(0)}\rangle=\,&(e^{\phi}+e^{-\phi})^3,\\
    \label{eq:Wislon_z3_1}
    \langle W_{\mathbf{2}\otimes\mathbf{2}\otimes\mathbf{2}}^{(1)}\rangle=\,&-\frac{3q_1q_2(e^{\phi}+e^{-\phi})^3}{(1-q_1q_2e^{2\phi})(1-q_1q_2e^{-2\phi})}\nonumber\\
    &-\frac{(1-q_1) (1-q_2) \left((1-q_1) (1-q_2)-3(1+q_1 q_2)\right)(e^{\phi}+e^{-\phi})}{(1-q_1q_2e^{2\phi})(1-q_1q_2e^{-2\phi})}
\end{align}
and
\begin{align}
    \label{eq:Wislon_z4_0}
    \langle W_{\mathbf{2}^{\otimes 4}}^{(0)}\rangle=\,&(e^{\phi}+e^{-\phi})^4,\\
     \langle W_{\mathbf{2}^{\otimes 4}}^{(1)}\rangle=\,&-\frac{4q_1q_2(e^{\phi}+e^{-\phi})^4}{(1-q_1q_2e^{2\phi})(1-q_1q_2e^{-2\phi})}\nonumber\\
     &+\frac{2(1-q_1) (1-q_2) \left(3(1+q_1q_2)-2(1-q_1)(1-q_2)\right)(e^{\phi}+e^{-\phi})^2}{(1-q_1q_2e^{2\phi})(1-q_1q_2e^{-2\phi})}\nonumber\\
     &+\frac{(1-q_1)^3 (1-q_2)^3 (1+q_1 q_2)}{q_1q_2(1-q_1q_2e^{2\phi})(1-q_1q_2e^{-2\phi})},
\end{align}
which agree with the results in Section \ref{sec:SUN_cod4}. Note that the one-instanton result for the representation $\mathbf{2}\otimes\mathbf{2}\otimes\mathbf{2}$ agrees with the result in \cite{Kim:2021gyj} from the blowup equations, but the one-instanton expectation value $\langle W_{\mathbf{2}^{\otimes 4}}^{(1)}\rangle$ is different from the result in \cite{Kim:2021gyj} by a Coulomb parameter independent term, which is not captured by the blowup equation properly. The two-instanton result is listed in Appendix \ref{appendix:2-inst}. 
The expectation value $\langle W_{\mathbf{2}\otimes \mathbf{2}\otimes \mathbf{2}}\rangle$ and $\langle W_{\mathbf{2}^{\otimes 4}}\rangle$ have the BPS expansions 
\begin{align}
    \langle W_{\mathbf{2}\otimes \mathbf{2}\otimes \mathbf{2}}\rangle=\F_{\mathrm{BPS},\{\C\}}^3+3\mathcal{I}\cdot\F_{\mathrm{BPS},\{\C\}}\F_{\mathrm{BPS},\{\C^2\}}+\mathcal{I}^2\cdot\F_{\mathrm{BPS},\{\C^3\}},
\end{align}
and
\begin{align}
    \langle W_{\mathbf{2}^{\otimes 4}}\rangle=\F_{\mathrm{BPS},\{\C\}}^4&+6\mathcal{I}\cdot\F_{\mathrm{BPS},\{\C\}}^2\F_{\mathrm{BPS},\{\C^2\}}+3\mathcal{I}^2\cdot\F_{\mathrm{BPS},\{\C^2\}}^2\nonumber\\
    &+4\mathcal{I}^2\cdot\F_{\mathrm{BPS},\{\C\}}\F_{\mathrm{BPS},\{\C^3\}}+\mathcal{I}^3\cdot\F_{\mathrm{BPS},\{\C^4\}},
\end{align}
where the BPS sector $\F_{\mathrm{BPS},\{\C^3\}}$ is generated by three primitive curves and $\F_{\mathrm{BPS},\{\C^4\}}$ is generated by four primitive curves. We list few BPS invariants of the BPS sector $\F_{\mathrm{BPS},\{\C^3\}}$ and $\F_{\mathrm{BPS},\{\C^4\}}$ in Table \ref{tb:BPS_P1P1_z3_BPS} and Table \ref{tb:BPS_P1P1_z4_BPS} respectively. We also list the pseudo-BPS invariants of the instanton part of $\langle W_{\mathbf{2}\otimes \mathbf{2}\otimes \mathbf{2}}\rangle$ and $\langle W_{\mathbf{2}^{\otimes 4}}\rangle$ in Table \ref{tb:P1P1_z3_BPS} and \ref{tb:P1P1_z4_BPS} respectively.
\begin{table}
\centering
 \begin{tabular}{|c|>{\centering\arraybackslash}m{5.4cm}||c|>{\centering\arraybackslash}m{5.4cm}|} \hline
$ \mathbf{d} $ & $ \oplus \widetilde{N}_{j_l, j_r}^{\mathbf{d}} (j_l, j_r) $ & $ \mathbf{d} $ & $ \oplus \widetilde{N}_{j_l, j_r}^{\mathbf{d}} (j_l, j_r) $ \\ \hline
$(1, -1)$&$3(0,0)$&$(1, 1)$&$5(0,0)\oplus(0,1)\oplus(1/2,1/2)\oplus(1,0)$\\ \hline 
 $(1, 3)$&$(0,0)\oplus3(0,1)\oplus(0,2)\oplus(1/2,1/2)\oplus(1/2,3/2)\oplus(1,1)$&$(1, 5)$&$(0,1)\oplus3(0,2)\oplus(0,3)\oplus(1/2,3/2)\oplus(1/2,5/2)\oplus(1,2)$\\ \hline 
 $(2, 1)$&$3(0,0)$&$(2, 3)$&$(0,0)\oplus3(0,1)\oplus(0,2)\oplus(1/2,1/2)\oplus(1/2,3/2)\oplus(1,1)$\\ \hline 
 $(2, 5)$&$(0,0)\oplus(0,1)\oplus8(0,2)\oplus2(0,3)\oplus(1/2,1/2)\oplus3(1/2,3/2)\oplus2(1/2,5/2)\oplus(1/2,7/2)\oplus(1,1)\oplus2(1,2)\oplus(1,3)\oplus(3/2,5/2)$&$(3, 3)$&$(0,0)$\\ \hline 
 $(3, 5)$&\multicolumn{3}{>{\centering\arraybackslash}m{12cm}|}{$(0,1)\oplus3(0,2)\oplus(0,3)\oplus(1/2,3/2)\oplus(1/2,5/2)\oplus(1,2)$}\\ \hline 
 \end{tabular}
\caption{Pseudo-BPS Spectrum of $SU(2)$ Wilson loop expectation value in the representation $\mathbf{2}\otimes\mathbf{2}\otimes\mathbf{2}$ for the curve class $d_1m_0+d_2\phi$ with $ d_1 =1,2, 3, $ and $ d_2 \leq 5 $. The $d_1=0$ part can be read from \eqref{eq:Wislon_z3_0}.}
\label{tb:P1P1_z3_BPS}
\end{table}

\begin{table}
\centering
 \begin{tabular}{|c|>{\centering\arraybackslash}m{5.4cm}||c|>{\centering\arraybackslash}m{5.4cm}|} \hline
$ \mathbf{d} $ & $ \oplus {N}_{j_l, j_r}^{\mathbf{d},\{\C^{3}\}} (j_l, j_r) $ & $ \mathbf{d} $ & $ \oplus {N}_{j_l, j_r}^{\mathbf{d},\{\C^{3}\}} (j_l, j_r) $ \\ \hline
$(1, 1)$&$(0,0)$&$(1, 3)$&$(0,1)$\\ \hline 
 $(1, 5)$&$(0,2)$&$(1, 7)$&$(0,3)$\\ \hline 
 $(2, 3)$&$(0,1)$&$(2, 5)$&$(0,1)\oplus4(0,2)\oplus(1/2,5/2)$\\ \hline 
 $(2, 7)$&$(0,1)\oplus4(0,2)\oplus8(0,3)\oplus(1/2,5/2)\oplus4(1/2,7/2)\oplus(1,4)$&$(3, 5)$&$(0,2)$\\ \hline 
 $(3, 7)$&\multicolumn{3}{>{\centering\arraybackslash}m{12cm}|}{$(0,1)\oplus4(0,2)\oplus8(0,3)\oplus(1/2,5/2)\oplus4(1/2,7/2)\oplus(1,4)$}\\ \hline 
 \end{tabular}
\caption{BPS Spectrum of the BPS sector $\F_{\mathrm{BPS},\{\C^{3}\}}$ for the curve class $d_1m_0+d_2\phi$ with $ d_1\leq 3 $ and $ d_2 \leq 7 $. }
\label{tb:BPS_P1P1_z3_BPS}
\end{table}

\begin{table}
\centering
 \begin{tabular}{|c|>{\centering\arraybackslash}m{5.4cm}||c|>{\centering\arraybackslash}m{5.4cm}|} \hline
$ \mathbf{d} $ & $ \oplus \widetilde{N}_{j_l, j_r}^{\mathbf{d}} (j_l, j_r) $ & $ \mathbf{d} $ & $ \oplus \widetilde{N}_{j_l, j_r}^{\mathbf{d}} (j_l, j_r) $ \\ \hline
$(1, -2)$&$4(0,0)$&$(1, 0)$&$14(0,0)\oplus2(0,1)\oplus-2(1/2,1/2)\oplus4(1,0)$\\ \hline 
 $(1, 2)$&$7(0,0)\oplus4(0,1)\oplus(0,2)\oplus4(1/2,1/2)\oplus(1/2,3/2)\oplus(1,0)\oplus(1,1)\oplus(3/2,1/2)$&$(1, 4)$&$(0,0)\oplus4(0,1)\oplus4(0,2)\oplus(0,3)\oplus(1/2,1/2)\oplus4(1/2,3/2)\oplus(1/2,5/2)\oplus(1,1)\oplus(1,2)\oplus(3/2,3/2)$\\ \hline 
 $(2, 0)$&$6(0,0)$&$(2, 2)$&$7(0,0)\oplus4(0,1)\oplus(0,2)\oplus4(1/2,1/2)\oplus(1/2,3/2)\oplus(1,0)\oplus(1,1)\oplus(3/2,1/2)$\\ \hline 
 $(2, 4)$&$3(0,0)\oplus7(0,1)\oplus10(0,2)\oplus2(0,3)\oplus2(1/2,1/2)\oplus10(1/2,3/2)\oplus2(1/2,5/2)\oplus(1/2,7/2)\oplus4(1,1)\oplus2(1,2)\oplus(1,3)\oplus(3/2,1/2)\oplus2(3/2,3/2)\oplus(3/2,5/2)\oplus(2,2)$&$(3, 2)$&$4(0,0)$\\ \hline 
 $(3, 4)$&\multicolumn{3}{>{\centering\arraybackslash}m{12cm}|}{$(0,0)\oplus4(0,1)\oplus4(0,2)\oplus(0,3)\oplus(1/2,1/2)\oplus4(1/2,3/2)\oplus(1/2,5/2)\oplus(1,1)\oplus(1,2)\oplus(3/2,3/2)$}\\ \hline 
 \end{tabular}
\caption{Pseudo-BPS Spectrum of $SU(2)$ Wilson loop expectation value in the representation $\mathbf{2}^{\otimes 4}$ for the curve class $d_1m_0+d_2\phi$ with $ d_1 =1,2, 3, $ and $ d_2 \leq 4 $. The $d_1=0$ part can be read from \eqref{eq:Wislon_z4_0}.}
\label{tb:P1P1_z4_BPS}
\end{table}

\begin{table}
\centering
 \begin{tabular}{|c|>{\centering\arraybackslash}m{5.4cm}||c|>{\centering\arraybackslash}m{5.4cm}|} \hline
$ \mathbf{d} $ & $ \oplus {N}_{j_l, j_r}^{\mathbf{d},\{\C^{4}\}} (j_l, j_r) $ & $ \mathbf{d} $ & $ \oplus {N}_{j_l, j_r}^{\mathbf{d},\{\C^{4}\}} (j_l, j_r) $ \\ \hline
$(1, 2)$&$(0,1/2)$&$(1, 4)$&$(0,3/2)$\\ \hline 
 $(1, 6)$&$(0,5/2)$&$(1, 8)$&$(0,7/2)$\\ \hline 
 $(2, 2)$&$(0,1/2)$&$(2, 4)$&$(0,1/2)\oplus5(0,3/2)\oplus(1/2,2)$\\ \hline 
 $(2, 6)$&$(0,1/2)\oplus5(0,3/2)\oplus12(0,5/2)\oplus(1/2,2)\oplus5(1/2,3)\oplus(1,7/2)$&$(2, 8)$&$(0,1/2)\oplus5(0,3/2)\oplus12(0,5/2)\oplus20(0,7/2)\oplus(1/2,2)\oplus5(1/2,3)\oplus12(1/2,4)\oplus(1,7/2)\oplus5(1,9/2)\oplus(3/2,5)$\\ \hline 
 $(3, 4)$&$(0,3/2)$&$(3, 6)$&$(0,1/2)\oplus5(0,3/2)\oplus12(0,5/2)\oplus(1/2,2)\oplus5(1/2,3)\oplus(1,7/2)$\\ \hline 
 $(3, 8)$&\multicolumn{3}{>{\centering\arraybackslash}m{12cm}|}{$5(0,1/2)\oplus17(0,3/2)\oplus37(0,5/2)\oplus47(0,7/2)\oplus4(0,9/2)\oplus(1/2,1)\oplus6(1/2,2)\oplus21(1/2,3)\oplus37(1/2,4)\oplus(1/2,5)\oplus(1,5/2)\oplus6(1,7/2)\oplus17(1,9/2)\oplus(3/2,4)\oplus5(3/2,5)\oplus(2,11/2)$}\\ \hline 
 \end{tabular}
\caption{BPS Spectrum of the BPS sector $\F_{\mathrm{BPS},\{\C^{4}\}}$ for the curve class $d_1m_0+d_2\phi$ with $ d_1\leq 3 $ and $ d_2 \leq 8 $. }
\label{tb:BPS_P1P1_z4_BPS}
\end{table}

\begin{figure}[h]
\begin{center}
\includegraphics{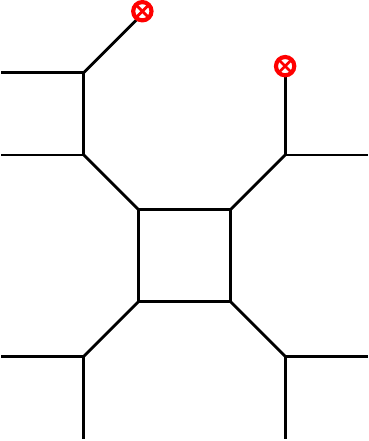}
\hspace{3cm}
\includegraphics{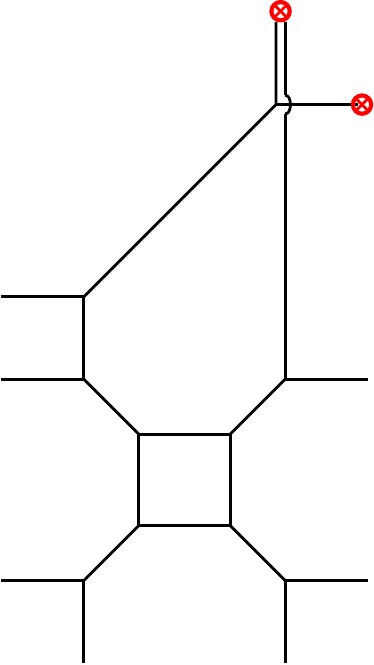}
\end{center}
\caption{Brane diagrams for $SU(2)$ gauge theory with five flavors, where the red points stand for the 7-branes spanned over $x^{0,1,2,3,4,7,8,9}$ directions.}
\label{fig:su2brane_Nf_5}
\end{figure}
\begin{figure}[h]
\begin{center}
\includegraphics{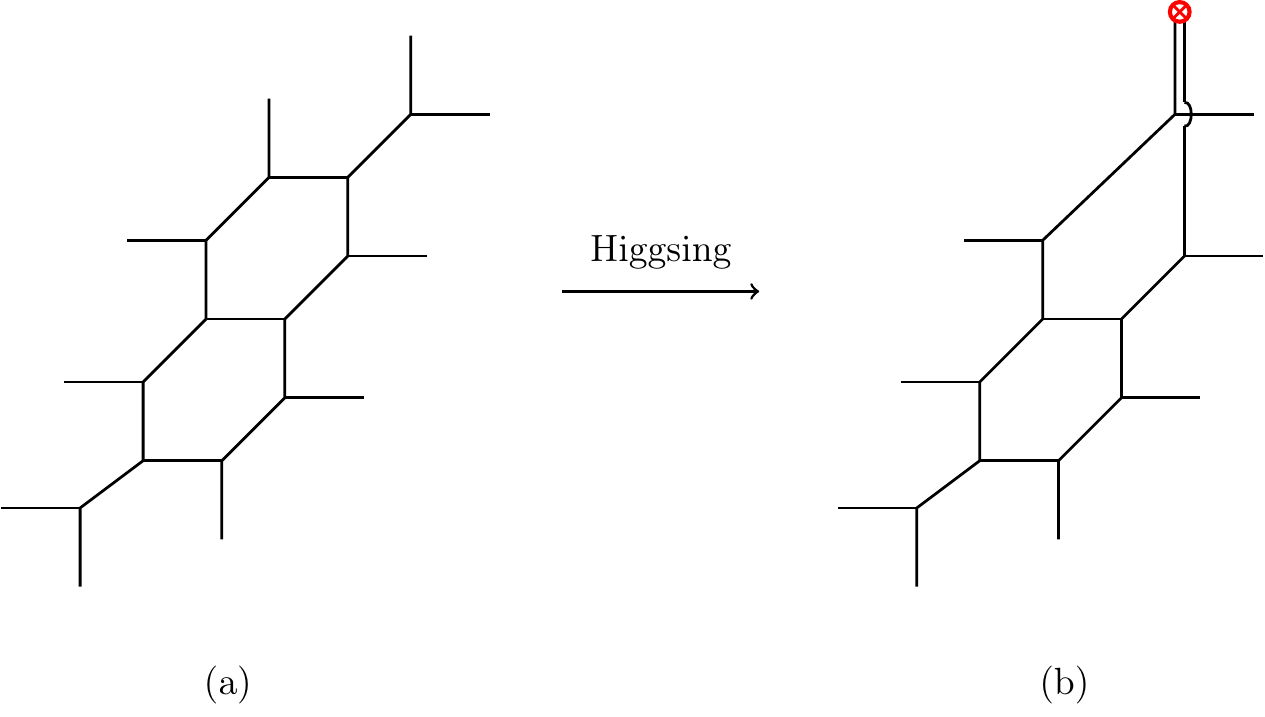}
\end{center}
\caption{Higgsing from (a) $SU(3),N_f=6$ theory  to (b) $SU(2),N_f=5$ theory.}
\label{fig:strip}
\end{figure}

\paragraph{Wilson loop in the representation $\mathbf{2}^{\otimes 5}$}
In the last, we consider a more non-trivial example, the Wilson loop with representation $\mathbf{2}^{\otimes 5}$, which can be obtained from $SU(2)$ theory with 5 flavors, described by the brane diagram in Figure \ref{fig:su2brane_Nf_5}(a). The red points in Figure \ref{fig:su2brane_Nf_5} are 7-branes spanned over $x^{0,1,2,3,4,7,8,9}$ directions that the 5-branes are ending. When the 7-branes are moved to infinity, the 7-branes across 5-branes and generate new 5-brane configuration in Figure \ref{fig:su2brane_Nf_5}(b). To compute the partition function of brane diagram \ref{fig:su2brane_Nf_5}(b), a simplest way is to start with the brane diagram \ref{fig:strip}(a), which is the brane diagram of $SU(3)$ theory with 6 fundamental flavors, and then Higgs to the $SU(2)$ theory with 5 fundamental flavors as described in Figure \ref{fig:strip}(b). Similar Higgsing process can also be found in \cite{Hayashi:2013qwa,Hayashi:2014wfa}.
The partition function of the brane diagram \ref{fig:strip}(a) can be compute by gluing the partition function of strips
\begin{align}
    &Z_{\phi\boldsymbol{\mu}}^{\text{strip}}(Q_{m_1},Q_{l_1},Q_{m_2},Q_{l_2},Q_{m_3};t,q)=\prod_{i=1}^3 q^{\frac{1}{2}||\mu_i||^2}\tilde{Z}_{\mu_i}(t,q)\times\mathcal{R}_{\emptyset \mu_1}(Q_{m_1})\nonumber\\
    &\quad\quad\quad\quad\times\mathcal{R}_{\emptyset \mu_2}(Q_{m_1}Q_{m_2}Q_{l_1})\mathcal{R}_{\emptyset \mu_2}(Q_{m_2})\mathcal{R}_{\mu_1^t\emptyset}(Q_{l_1})\mathcal{R}_{\mu_1^t\emptyset}(Q_{m_2}Q_{l_1}Q_{l_2})\mathcal{R}_{\mu_2^t\emptyset}(Q_{l_2})\nonumber\\
    &\quad\quad\quad\quad\times\frac{\mathcal{R}_{\emptyset\mu_3}(Q_{m_3})\mathcal{R}_{\emptyset\mu_3}(Q_{m_2}Q_{m_3}Q_{l_2})\mathcal{R}_{\emptyset\mu_3}(Q_{m_1}Q_{m_2}Q_{m_3}Q_{l_1}Q_{l_2})}{\mathcal{R}_{\mu_1^t\mu_2}(Q_{l_1}Q_{m_2}\sqrt{\frac{t}{q}})\mathcal{R}_{\mu_2^t\mu_3}(Q_{l_2}Q_{m_3}\sqrt{\frac{t}{q}})\mathcal{R}_{\mu_1^t\mu_3}(Q_{m_2}Q_{m_3}Q_{l_1}Q_{l_2}\sqrt{\frac{t}{q}})},
\end{align}
that we get
\begin{align}
    Z^{SU(3),N_f=6}=&\sum_{\boldsymbol{\mu}}(-Q_{b_1})^{\mu_1}(-Q_{b_2})^{\mu_2}(-Q_{b_3})^{\mu_3}\times Z_{\phi\boldsymbol{\mu}}^{\text{strip}}(Q_{m_1},Q_{l_1},Q_{m_2},Q_{l_2},Q_{m_3};t,q)\nonumber\\
    &\quad\times Z_{\phi\boldsymbol{\mu}^t}^{\text{strip}}(Q_{m_1}^{\prime},Q_{l_1}^{\prime},Q_{m_2}^{\prime},Q_{l_2}^{\prime},Q_{m_3}^{\prime};q,t) \mathcal{R}_{\emptyset\emptyset}(Q_{b_1}Q_{m_1})\mathcal{R}_{\emptyset\emptyset}(Q_{b_3}Q_{m_1}^{\prime}).
\end{align}
The parameters here are not completely independent, they satisfy relations
\begin{align}
    Q_{l_1}Q_{m_2}=Q_{l_2}^{\prime}Q_{m_3}^{\prime},\,\, Q_{l_2}Q_{m_3}=Q_{l_1}^{\prime}Q_{m_2}^{\prime},\,\,Q_{m_3}Q_{b_3}=Q_{b_2}Q_{m_2}^{\prime},\,\, Q_{m_3}^{\prime}Q_{b_1}=Q_{m_2}Q_{b_2}.
\end{align}
To get the partition function of $SU(2)$ theory with $5$ flavors, we Higgs the theory by turning the parameters
\begin{align}
    Q_{m_1}^{\prime}=Q_{b_3}=\sqrt{\frac{t}{q}},
\end{align}
in the partition function $Z^{SU(3),N_f=6}$. Then the term
\begin{align}
    \mathcal{R}_{\emptyset\mu_3^t}(Q_{m_1}^{\prime};q,t)=\mathcal{R}_{\mu_3^t\emptyset}(Q_{m_1}^{\prime};t,q)=\prod_{i,j=1}^{\infty}(1-Q_{m_1}^{\prime}t^{i-\frac{1}{2}}q^{j-\frac{1}{2}})\cdot\prod_{(i,j)\in\mu_3}(1-Q_{m_1}^{\prime}t^{\frac{1}{2}-i}q^{j-\frac{1}{2}}),
\end{align}
is non-zero only for $\mu_3=\emptyset$. When $\mu_3=\emptyset$, it is an irrelevant term and should be factored out from the partition function. 
By further using the notation read from the brane diagram
\begin{align}
    &\quad Q_F=Q_{l_1}Q_{m_2}=Q_{l_2}^{\prime}Q_{m_3}^{\prime}, \quad Q_B=Q_{b_1}Q_{m_3}^{\prime}=Q_{b_2}Q_{m_2}, \nonumber\\
    &  Q_1=Q_{m_1},\quad Q_2=\frac{1}{Q_{m_2}},\quad Q_3=Q_{l_2},\,\,Q_4=\frac{1}{Q_{m_3}^{\prime}},\quad Q_5=Q_{m_2}^{\prime},
\end{align}
we have the partition function of $SU(2)$ theory with 5 fundamental flavors as
\begin{align}
    Z^{SU(2),N_f=5}=&\sum_{\boldsymbol{\mu}}(-Q_B Q_4)^{|\mu_1|}(-Q_BQ_2)^{|\mu_2|}\times \prod_{i=1}^2 t^{\frac{1}{2}||\mu_i^t||^2}q^{\frac{1}{2}||\mu_i||^2}\tilde{Z}_{\mu_i}(t,q)\tilde{Z}_{\mu_i^t}(q,t)\nonumber\\
    &\times \mathcal{R}_{\emptyset\mu_1}(Q_1)\mathcal{R}_{\emptyset\mu_2}(Q_FQ_1)\mathcal{R}_{\emptyset\mu_2}(Q_2^{-1})\mathcal{R}_{\mu_1^t\emptyset}(Q_FQ_2)\mathcal{R}_{\mu_2^t\emptyset}(Q_3)\mathcal{R}_{\mu_1^t\emptyset}(Q_FQ_3)\nonumber\\
    &\times \mathcal{R}_{\mu_1^t\emptyset}(Q_4^{-1})\mathcal{R}_{\emptyset\mu_2}(Q_FQ_4)\mathcal{R}_{\mu_2^t\emptyset}(Q_5)\mathcal{R}_{\mu_1^t\emptyset}(Q_FQ_5)\mathcal{R}_{\emptyset\emptyset}(Q_BQ_1Q_4\sqrt{\frac{t}{q}})\nonumber\\
    &\times\mathcal{R}_{\emptyset\emptyset}(Q_BQ_3Q_5\sqrt{\frac{q}{t}})\mathcal{R}_{\emptyset\emptyset}(Q_BQ_FQ_2Q_3Q_4Q_5\sqrt{\frac{q}{t}})\mathcal{R}_{\emptyset\emptyset}(Q_BQ_2Q_5\sqrt{\frac{q}{t}})\nonumber\\
    &\times {\mathcal{R}_{\emptyset\emptyset}(Q_BQ_2Q_3\sqrt{\frac{q}{t}})\mathcal{R}_{\emptyset\emptyset}(Q_BQ_FQ_1Q_2Q_3Q_5\sqrt{\frac{q}{t}})}\mathcal{R}_{\mu_1^t\mu_2}^{-1}(Q_F\sqrt{\frac{t}{q}})\nonumber\\
    &\times\mathcal{R}_{\mu_1^t\mu_2}^{-1}(Q_F\sqrt{\frac{q}{t}})\mathcal{R}_{\mu_2^t\emptyset}^{-1}(Q_BQ_2Q_3Q_5\frac{q}{t})\mathcal{R}_{\mu_1^t\emptyset}^{-1}(Q_FQ_BQ_2Q_3Q_5\frac{q}{t}).
\end{align}
In the last, we need to do flop transitions to get the partition function in the phase with heavy flavor masses. Notice that
\begin{align}
    \mathcal{R}_{\lambda^t\nu}(Q)=(-Q)^{|\nu|+|\lambda|}t^{\frac{1}{2}(-||\lambda^t||^2+||\nu^t||^2)}q^{\frac{1}{2}(||\lambda||^2-||\nu||^2)}N_{\lambda\nu}(Q^{-1}\sqrt{\frac{t}{q}};t,q)\mathcal{R}_{\emptyset\emptyset}(Q).
\end{align}
After flopping $Q\rightarrow Q^{-1}$, the function $\mathcal{R}_{\emptyset\emptyset}(Q)$ is analytically continuously transform to $\mathcal{R}_{\emptyset\emptyset}(Q^{-1})$ according to the integral representation of the generalized MacMahon function, so we have
\begin{align}
    \mathcal{R}_{\lambda^t\nu}(Q)\rightarrow(-Q)^{|\nu|+|\lambda|}t^{\frac{1}{2}(-||\lambda^t||^2+||\nu^t||^2)}q^{\frac{1}{2}(||\lambda||^2-||\nu||^2)}\mathcal{R}_{\nu^t\lambda}(Q^{-1}).
\end{align}
After flop transitions by applying analytic continuation, we have the $SU(2),N_f=5$ partition function
\begin{align}\label{eq:partitionfunction_SU2_Nf5}
    Z^{SU(2),N_f=5}=\,&Z_{\mathrm{pert}}^{SU(2)}\sum_{\mu_1,\mu_2}\left(\mathfrak{q}\sqrt{\frac{q}{t}}\right)^{|\mu_1|+|\mu_2|}\cdot\frac{1}{\prod_{i,j=1}^2N_{\mu_i\mu_j}(Q_{ij};t,q)}\nonumber\\
    &\times \mathcal{R}_{\emptyset\mu_2}(Q_1)\mathcal{R}_{\emptyset\mu_1}(Q_FQ_1)\mathcal{R}_{\mu_1^t\emptyset}(Q_2)\mathcal{R}_{\mu_2^t\emptyset}(Q_FQ_2)\mathcal{R}_{\mu_1^t\emptyset}(Q_3)\mathcal{R}_{\mu_2^t\emptyset}(Q_FQ_3)\nonumber\\
    &\times \mathcal{R}_{\emptyset\mu_2}(Q_4)\mathcal{R}_{\emptyset\mu_1}(Q_FQ_4)\mathcal{R}_{\mu_1^t\emptyset}(Q_5)\mathcal{R}_{\mu_2^t\emptyset}(Q_FQ_5)\mathcal{R}_{\emptyset\emptyset}(Q_BQ_1Q_4\sqrt{\frac{t}{q}})\nonumber\\
    &\times\mathcal{R}_{\emptyset\emptyset}(Q_BQ_3Q_5\sqrt{\frac{q}{t}})\mathcal{R}_{\emptyset\emptyset}(Q_BQ_FQ_2Q_3Q_4Q_5\sqrt{\frac{q}{t}})\mathcal{R}_{\emptyset\emptyset}(Q_BQ_2Q_5\sqrt{\frac{q}{t}})\nonumber\\
    &\times {\mathcal{R}_{\emptyset\emptyset}(Q_BQ_2Q_3\sqrt{\frac{q}{t}})\mathcal{R}_{\emptyset\emptyset}(Q_BQ_FQ_1Q_2Q_3Q_5\sqrt{\frac{q}{t}})}\nonumber\\
    &\times\mathcal{R}_{\mu_1^t\emptyset}^{-1}(Q_BQ_2Q_3Q_5\frac{q}{t})\mathcal{R}_{\mu_2^t\emptyset}^{-1}(Q_FQ_BQ_2Q_3Q_5\frac{q}{t}),
\end{align}
where the coefficients of $M_1M_2M_3M_4M_5=\left(-\frac{Q_F^{1/2}({t}{q})^{\frac{1}{2}}}{(1-{t})(1-{q})}\right)^5Q_1Q_2Q_3Q_4Q_5$ in \eqref{eq:partitionfunction_SU2_Nf5} gives the partition function for Wilson loop in the representation $\mathbf{2}^{\otimes 5}$
\begin{align}
    \Wilson{\mathbf{2}^{\otimes 5}}=\,&Z_{\mathrm{pert}}^{SU(2)}\sum_{\mu_1,\mu_2}\left(\mathfrak{q}\sqrt{\frac{q}{t}}\right)^{|\mu_1|+|\mu_2|}\cdot\frac{1}{\prod_{i,j=1}^2N_{\mu_i\mu_j}(Q_{ij};t,q)}\nonumber\\
    &\times \left[\CH^3\CHT^2+(1-t)(1-1/q)\CH^3\right.\nonumber\\
    &\quad\quad -\mathfrak{q}(1-q)^2(1-1/t)^2\CH\CHT^2 \nonumber\\
    &\quad\quad +3\mathfrak{q}(1-q)(1-1/t)\CH\CHT^2 \nonumber\\
    &\quad\quad +3\mathfrak{q}^2(1-q)^2(1-t^2)/t/q\CH\nonumber\\
    &\quad\quad \left.-(1-t)^3(1-q)^3/q/t^2\left(2\mathfrak{q}\CHT-\mathfrak{q}^2\CH\right)\right].
\end{align}
At perturbative and one-instanton level, we have
\begin{align}
    \label{eq:Wislon_z5_0}
    \langle W_{\mathbf{2}^{\otimes 5}}^{(0)}\rangle=\,&(e^{\phi}+e^{-\phi})^5,\\
    \langle W_{\mathbf{2}^{\otimes 5}}^{(1)}\rangle=\,&\frac{-5q_1q_2(e^{\phi}+e^{-\phi})^5}{(1-q_1q_2e^{2\phi})(1-q_1q_2e^{-2\phi})}\nonumber\\
    &+\frac{10(1-q_1)(1-q_2)\left((1+q_1q_2)-(1-q_1)(1-q_2)\right)(e^{\phi}+e^{-\phi})^3}{(1-q_1q_2e^{2\phi})(1-q_1q_2e^{-2\phi})}\nonumber\\
    &+\frac{(1-q_1)^3 (1-q_2)^3 \left(5(1+q_2q_2)-(1-q_1)(1-q_2)\right)(e^{\phi}+e^{-\phi})}{q_1q_2(1-q_1q_2e^{2\phi})(1-q_1q_2e^{-2\phi})},\\
\end{align}
which agrees with the result in Section \ref{sec:SUN_cod4}. The two-instanton result is listed in Appendix \ref{appendix:2-inst}. 
The expectation value $\langle W_{\mathbf{2}^{\otimes 5}}\rangle$ has an BPS expansion 
\begin{align}
    \langle W_{\mathbf{2}^{\otimes 5}}\rangle=\F_{\mathrm{BPS},\{\C\}}^5&+10\mathcal{I}\cdot\F_{\mathrm{BPS},\{\C\}}^3\F_{\mathrm{BPS},\{\C^2\}}+15\mathcal{I}^2\cdot\F_{\mathrm{BPS},\{\C\}}\F_{\mathrm{BPS},\{\C^2\}}^2\nonumber\\
    &+10\mathcal{I}^2\cdot\F_{\mathrm{BPS},\{\C\}}^2\F_{\mathrm{BPS},\{\C^3\}}+10\mathcal{I}^3\cdot\F_{\mathrm{BPS},\{\C^2\}}\F_{\mathrm{BPS},\{\C^3\}}\nonumber\\
    &+5\mathcal{I}^3\cdot\F_{\mathrm{BPS},\{\C\}}\F_{\mathrm{BPS},\{\C^4\}}+\mathcal{I}^4\cdot\F_{\mathrm{BPS},\{\C^5\}},
\end{align}
where the BPS sector $\F_{\mathrm{BPS},\{\C^5\}}$ is generated by five primitive curves. We list few BPS invariants of the BPS sector $\F_{\mathrm{BPS},\{\C^5\}}$ in Table \ref{tb:BPS_P1P1_z5_BPS} and the pseudo-BPS invariants of the instanton part of $\langle W_{\mathbf{2}^{\otimes 5}}\rangle$ in Table \ref{tb:P1P1_z5_BPS}.

\begin{table}
\centering
 \begin{tabular}{|c|>{\centering\arraybackslash}m{5.4cm}||c|>{\centering\arraybackslash}m{5.4cm}|} \hline
$ \mathbf{d} $ & $ \oplus \widetilde{N}_{j_l, j_r}^{\mathbf{d}} (j_l, j_r) $ & $ \mathbf{d} $ & $ \oplus \widetilde{N}_{j_l, j_r}^{\mathbf{d}} (j_l, j_r) $ \\ \hline
$(1, -3)$&$5(0,0)$&$(1, -1)$&$30(0,0)\oplus5(0,1)\oplus-10(1/2,1/2)\oplus10(1,0)$\\ \hline 
 $(1, 1)$&$35(0,0)\oplus9(0,1)\oplus(0,2)\oplus-6(1/2,1/2)\oplus(1/2,3/2)\oplus14(1,0)\oplus(1,1)\oplus(3/2,1/2)\oplus(2,0)$&$(1, 3)$&$9(0,0)\oplus10(0,1)\oplus5(0,2)\oplus(0,3)\oplus5(1/2,1/2)\oplus5(1/2,3/2)\oplus(1/2,5/2)\oplus(1,0)\oplus5(1,1)\oplus(1,2)\oplus(3/2,1/2)\oplus(3/2,3/2)\oplus(2,1)$\\ \hline 
 $(2, -1)$&$10(0,0)$&$(2, 1)$&$35(0,0)\oplus9(0,1)\oplus(0,2)\oplus-6(1/2,1/2)\oplus(1/2,3/2)\oplus14(1,0)\oplus(1,1)\oplus(3/2,1/2)\oplus(2,0)$\\ \hline 
 $(2, 3)$&$15(0,0)\oplus34(0,1)\oplus16(0,2)\oplus2(0,3)\oplus(1/2,3/2)\oplus2(1/2,5/2)\oplus(1/2,7/2)\oplus(1,0)\oplus22(1,1)\oplus2(1,2)\oplus(1,3)\oplus3(3/2,1/2)\oplus2(3/2,3/2)\oplus(3/2,5/2)\oplus(2,0)\oplus2(2,1)\oplus(2,2)\oplus(5/2,3/2)$&$(3, 1)$&$10(0,0)$\\ \hline 
 $(3, 3)$&\multicolumn{3}{>{\centering\arraybackslash}m{12cm}|}{$9(0,0)\oplus10(0,1)\oplus5(0,2)\oplus(0,3)\oplus5(1/2,1/2)\oplus5(1/2,3/2)\oplus(1/2,5/2)\oplus(1,0)\oplus5(1,1)\oplus(1,2)\oplus(3/2,1/2)\oplus(3/2,3/2)\oplus(2,1)$}\\ \hline 
 \end{tabular}
\caption{Pseudo-BPS Spectrum of $SU(2)$ Wilson loop expectation value in the representation $\mathbf{2}^{\otimes 5}$ for the curve class $d_1m_0+d_2\phi$ with $ d_1 =1,2, 3, $ and $ d_2 \leq 3 $. The $d_1=0$ part can be read from \eqref{eq:Wislon_z5_0}.}
\label{tb:P1P1_z5_BPS}
\end{table}

\begin{table}
\centering
 \begin{tabular}{|c|>{\centering\arraybackslash}m{5.4cm}||c|>{\centering\arraybackslash}m{5.4cm}|} \hline
$ \mathbf{d} $ & $ \oplus {N}_{j_l, j_r}^{\mathbf{d},\{\C^{5}\}} (j_l, j_r) $ & $ \mathbf{d} $ & $ \oplus {N}_{j_l, j_r}^{\mathbf{d},\{\C^{5}\}} (j_l, j_r) $ \\ \hline
$(1, 1)$&$(0,0)$&$(1, 3)$&$(0,1)$\\ \hline 
 $(1, 5)$&$(0,2)$&$(1, 7)$&$(0,3)$\\ \hline 
 $(2, 1)$&$(0,0)$&$(2, 3)$&$(0,0)\oplus6(0,1)\oplus(1/2,3/2)$\\ \hline 
 $(2, 5)$&$(0,0)\oplus6(0,1)\oplus17(0,2)\oplus(1/2,3/2)\oplus6(1/2,5/2)\oplus(1,3)$&$(2, 7)$&$(0,0)\oplus6(0,1)\oplus17(0,2)\oplus32(0,3)\oplus(1/2,3/2)\oplus6(1/2,5/2)\oplus17(1/2,7/2)\oplus(1,3)\oplus6(1,4)\oplus(3/2,9/2)$\\ \hline 
 $(3, 3)$&$(0,1)$&$(3, 5)$&$(0,0)\oplus6(0,1)\oplus17(0,2)\oplus(1/2,3/2)\oplus6(1/2,5/2)\oplus(1,3)$\\ \hline 
 $(3, 7)$&\multicolumn{3}{>{\centering\arraybackslash}m{12cm}|}{$5(0,0)\oplus23(0,1)\oplus58(0,2)\oplus84(0,3)\oplus5(0,4)\oplus(1/2,1/2)\oplus7(1/2,3/2)\oplus28(1/2,5/2)\oplus58(1/2,7/2)\oplus(1/2,9/2)\oplus(1,2)\oplus7(1,3)\oplus23(1,4)\oplus(3/2,7/2)\oplus6(3/2,9/2)\oplus(2,5)$}\\ \hline 
 \end{tabular}
\caption{BPS Spectrum of the BPS sector $\F_{\mathrm{BPS},\{\C^{5}\}}$ for the curve class $d_1m_0+d_2\phi$ with $ d_1\leq 3 $ and $ d_2 \leq 7 $. }
\label{tb:BPS_P1P1_z5_BPS}
\end{table}

\subsection{Wilson loops and Seiberg-Witten curves}\label{sec:3.2}
We are now preparing the B-model approach to the Wilson loop expectation values in topological string theory. Our key idea is to identify the Wilson loop expectation values with the B-model complex structure parameters, due to the correspondence of mirror curves and Seiberg-Witten curves.

The partition function with 1d line defects that we reviewed in Section \ref{sec:SUN_cod4} has a close connection to the (quantum) Seiberg-Witten curve. For simplicity, we consider the insertion of one defect D3-brane, with $\n=1$. Then it is known that for $SU(N)_{\kappa}$ theory the 5d/1d partition function has an expansion 
\begin{align}
    \frac{Z_{\mathrm{5d/1d}}}{Z_{\mathrm{5d}}}=X^{N/2}-H_1 X^{N/2-1}+H_2 X^{N/2-2}+\cdots +(-1)^{N-1}H_{N-1} X^{1-N/2}+(-1)^N X^{-N/2},
\end{align}
such that it satisfies a difference equation \cite{Nekrasov:1996cz,Nekrasov:2009rc} in the Nekrasov-Shatashvili (NS) limit $\epsilon_1\rightarrow \hbar,\epsilon_2\rightarrow 0$ 
\begin{align}\label{eq:q_SW}
    [Y+\mathfrak{q}e^{\frac{1}{2}\kappa\hbar}X^{\kappa}Y^{-1}]\Psi(x;\hbar)=\lim_{\substack{\epsilon_1\rightarrow \hbar,\\\epsilon_2\rightarrow 0}}\frac{Z_{\mathrm{5d/1d}}}{Z_{\mathrm{5d}}}\Psi(x;\hbar),
\end{align}
with $X=\exp(\hat{x}),Y=\exp(\hat{y})$.
Here $H_i$ is nothing but the Wilson loop expectation value in the NS limit
\begin{align}
    H_i=\lim_{\substack{\epsilon_1\rightarrow \hbar,\\\epsilon_2\rightarrow 0}}\langle{{{W}}_{{\mathbf{r}_i}}}\rangle
\end{align}
in the representation $\mathbf{r}_i$ whose highest weight is the $i$-th fundamental weight of the gauge group $SU(N)$.

\begin{figure}[h]
\begin{center}
\includegraphics{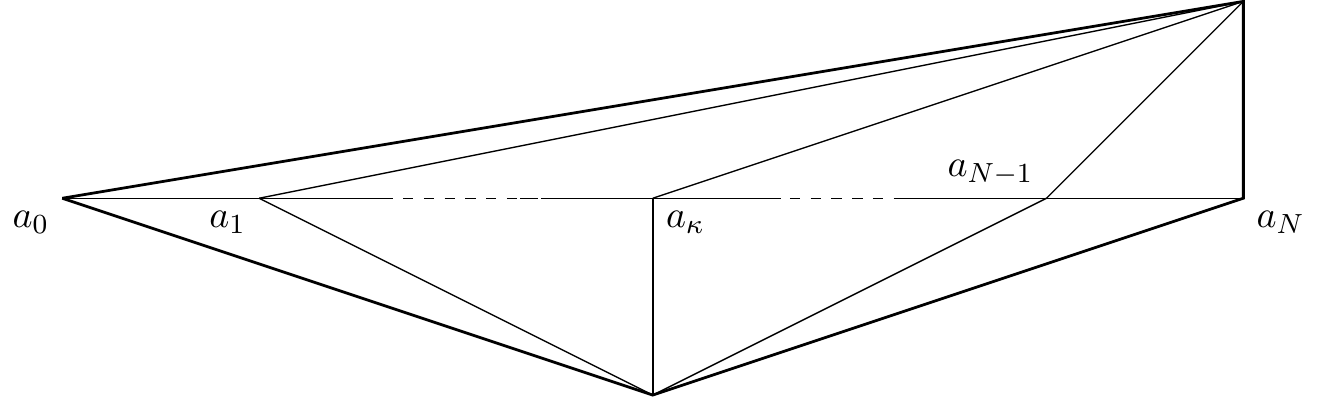}
\end{center}
\caption{Toric diagram for $Y^{N,\kappa}$ system. Here $a_i,i=0,\cdots N$ are complex structure parameters.}
\label{fig:toric_Ypq}
\end{figure}

Geometrically, the $SU(N)_{\kappa}$ gauge theory is described by M-theory compactified on Sasaki-Einstein manifold $Y^{N,\kappa}$ with the toric description in Figure \ref{fig:toric_Ypq}.
In the classical limit $\hbar\rightarrow 0$, the Seiberg-Witten curve \eqref{eq:q_SW} coincides with the classical mirror curve \eqref{eq:mirrorcurve} read from the toric diagram in Figure \ref{fig:toric_Ypq}, such that we have an identification of the parameters as
\begin{align}\label{eq:ai_W0}
    a_i=\lim_{\epsilon_{1,2}\rightarrow 0}\langle{{{W}}_{{\mathbf{r}_i}}}\rangle,\quad i=1,\cdots, N-1,
\end{align}
where $a_i$ can be treated as the homogeneous coordinates of the toric diagram or more precisely the complex structure parameters dual to the compact 4-cycles of the B-model geometry. See \cite{Grassi:2017qee,Grassi:2018bci} for related discussions. Equation \eqref{eq:ai_W0} means that in the classical limit $\epsilon_{1,2}\rightarrow 0$, the Wilson loop expectation values are provided by homogeneous coordinates related to the compact 4-cycles, which will be used in Section \ref{sec:RHAE} as an initial condition of the refined holomorphic anomaly equation.

\section{Refined holomorphic anomaly equations}\label{sec:RHAE}
In this section, we study the refined holomorphic anomaly equation for the Wilson loop amplitudes. We first review the refined holomorphic anomaly equation for the refined topological string amplitudes, and then generalized it to Wilson loop amplitudes. 
\subsection{Topological strings}\label{sec:HAE}
Define the small string coupling expansion of the free energy
\begin{align}
    \F=\sum_{n,g}(\epsilon_1+\epsilon_2)^{2n}(\epsilon_1\epsilon_2)^{g-1}\F^{(n,g)}.
\end{align}
The holomorphic anomaly equation derived in \cite{Bershadsky:1993cx} states that the anti-holomorphic derivative of $\F^{(n,g)}$ is not zero, but has boundary contributions in the moduli space. This anomaly suggests an anti-holomorphic completion $\mF^{(n,g)}$ satisfies a holomorphic anomaly equation \cite{Bershadsky:1993cx} .  
The refined holomorphic anomaly equations for topological strings have been proposed in \cite{Huang:2010kf,Krefl:2010fm} as
\begin{align}\label{eq:HAE0}
    \bar{\partial}_{\bar{i}}\mF^{(n,g)}=\frac{1}{2} \bar{C}_{\bar{i}}^{jk}\left( D_jD_k \mF^{(n,g-1)} +{\sum_{n^{\prime},g^{\prime}}}^{\prime}D_{j} \mF^{(n^{\prime},g^{\prime})}\cdot D_k \mF^{(n-n^{\prime},g-g^{\prime})} \right).
\end{align}
for $n+g>1$, where $\mF^{(n,g-1)}$ is the anti-holomorphic completion of $\F^{(n,g-1)}$. Here the prime in the summation means the omission of $(n^{\prime},g^{\prime})=(0,0)$ and $(n,g)$. $D_i$ is the covariant derivative defined in the geometry of the moduli space. When $n+g=1$, the free energies take the most general ansatz
\begin{align}
    \mF^{(1,0)}&=\frac{1}{24} \log \left(\Delta \prod_{i}z_i^{a_i}\right),\\
    \mF^{(0,1)}&=\frac{1}{2} \log \left(\Delta^a |g_{z_i\bar{z}_{\bar{j}}}^{-1}|\prod_{i}z_i^{b_i}\right).
\end{align}
where $\Delta$ is the discriminant of the mirror geometry, and the constant $a,a_i,b_j$ can be fixed by some initial data of the geometry or few BPS invariants. The metric $g_{z_i\bar{z}_{\bar{j}}}=\partial_{z_i}\bar{\partial}_{\bar{z}_{\bar{j}}}K$ is defined from the K\"ahler potential $K$ which can be re-expressed as
\begin{align}\label{eq:gij}
    g_{z_i\bar{z}_{\bar{j}}}=\frac{\partial^2 K}{\partial_{t_k}{\partial}_{\bar{t}_{\bar{l}}}}\cdot\frac{\partial t_{k}}{\partial{z_i}}\frac{\partial \bar{t}_{\bar{l}}}{\partial{\bar{z}_{\bar{j}}}}.
\end{align}

In the direct integration method of the holomorphic anomaly equation, it is useful to introduce the propagators $S^{jk}$ such that $\bar{C}_{\bar{i}}^{jk}=\bar{\partial}_{\bar{i}}S^{jk}$. Then we can re-express the anti-holomorphic derivative of $\mF^{(n,g)}$ as
\begin{align}
    \bar{\partial}_{\bar{i}}\mF^{(n,g)}=\bar{C}_{\bar{i}}^{jk}\frac{\partial \mF^{(n,g)}}{\partial S^{jk}}.
\end{align}
In the holomorphic limit $\im t\rightarrow \infty$, $\F^{(n,g)}=\lim_{\im t \rightarrow \infty}\mF^{(n,g)}$, the refined holomorphic anomaly equation \eqref{eq:HAE0} becomes the form
\begin{align}\label{eq:HAE}
    \frac{\partial \F^{(n,g)}}{\partial S^{ij}}=\frac{1}{2} \left( D_iD_j \F^{(n,g-1)} +{\sum_{n^{\prime},g^{\prime}}}^{\prime}D_{i} \F^{(n^{\prime},g^{\prime})}\cdot D_j \F^{(n-n^{\prime},g-g^{\prime})} \right), \quad n+g\geq 2.
\end{align}
In the local geometry we are interested in, the K\"ahler potential is a constant in the holomorphic limit, then the metric and the connection can be computed from the mirror maps
\begin{align}
    g_{z_i\bar{z}_{\bar{j}}}\rightarrow\frac{\partial t_{k}}{\partial{z_i}},\quad \Gamma^{i}_{ik}=\frac{\partial z_i}{\partial t_l}\frac{\partial^2t_l}{\partial z_j \partial z_k}.
\end{align}
The propagator $S^{ij}$ here satisfies two identities derived from special geometry as 
\begin{align}
    \label{eq:S_relation1}
    D_iS^{jk}=&-C_{imn}S^{im}S^{kn}+f_{i}^{jk},\\
    \label{eq:S_relation2}
    \Gamma_{ij}^k=&-C_{ijk}S^{kl}+\tilde{f}_{ij}^k,
\end{align}
where $f_{i}^{jk}$ and $\tilde{f}_{ij}^k$ are rational functions with respect to complex structure parameters, which can be chosen to make the expression of the propagator simpler. 
The genus one free energies in the holomorphic limit take the expression
\begin{align}\label{eq:F1_general}
    \F^{(1,0)}&=\frac{1}{24} \log \left(\Delta \prod_{i}z_i^{a_i}\right),\\
    \F^{(0,1)}&=\frac{1}{12} \log \left(\Delta^a \prod_{i}z_i^{b_i}\right)-\frac{1}{2}\log\left|\det\frac{\partial t_i}{\partial z_j}\right|.
\end{align}

Define the effective coupling $\tau_{ij}=C_{il}\frac{\partial t_{D,j}}{\partial t_l}$, the free energy $\F^{(n,g)}$ with $n+g>0$ are weight zero quasi-modular forms \cite{Aganagic:2006wq,Witten:1993ed}, under the modular transformation of the effective coupling
\begin{equation}
\tau  \mapsto \gamma\tau=(A\tau+B)(C\tau+D)^{-1},\quad \gamma=\left(
\begin{array}{cc}
 A & B \\
 C & D \\
\end{array}
\right)\in Sp(2g;\mathbb{Z}).
\end{equation}
The complex moduli parameters $z_i$'s are modular functions, and the propagator $S^{ij}$ is proportional to the the weight two quasi-modular form $E^{ij}(\tau)$. It has the completion to a non-holomorphic modular form
\begin{align}
    \widetilde{E}^{ij}(\tau,\bar{\tau})=E^{ij}(\tau)+\left((\mathrm{Im}\,\tau)^{-1}\right)^{ij},
\end{align}
such that under modular transformation
\begin{align}
    \widetilde{E}^{ij}(\tau,\bar{\tau})\mapsto {(C\tau+D)^{i}}_k{(C\tau+D)^{j}}_l\widetilde{E}^{kl}(\tau,\bar{\tau}).
\end{align}
Then it is quite clear that to solve the free energy recursively from the refined holomorphic anomaly equation \eqref{eq:HAE}, one first integrates over the propagator $S^{ij}$, then up to a holomorphic function $f^{(n,g)}(z_i)$, the free energy is completely solved. The holomorphic function $f^{(n,g)}(z_i)$ is called holomorphic ambiguity, which can be fixed partially or completely from other boundary conditions. A frequently used condition is called {\it{gap condition}}, which is derived by integrating out the massless particles in the Schwinger loop computation
\begin{align}\label{eq:schwinger_gap}
    \F(\epsilon_1,\epsilon_2,t_c)=&\int_{0}^{\infty}\frac{ds}{s}\frac{\exp(-s t_c)}{4\sinh(s \epsilon_1/2)4\sinh(s \epsilon_2/2)}+\mathcal{O}(t_c^0)\nonumber\\
    &=\left[-\frac{1}{12}+\frac{1}{24}(\epsilon_1+\epsilon_2)^2(\epsilon_1\epsilon_2)^{-1}\right]\log(t_c)\nonumber\\
    &+\frac{1}{\epsilon_1\epsilon_2}\sum_{n_1,n_2}\frac{(2n_1+2n_2-3)!}{t_c^{2n_1+2n_2-2}}\hat{B}_{2n_1}\hat{B}_{2n_2}\epsilon_1^{2n_1}\epsilon_2^{2n_2}+\mathcal{O}(t_c^0),
\end{align}
where $\hat{B}_{n}=(1-2^{1-n})\frac{B_n}{n!}$ and $B_n$ denoting the Bernoulli numbers. Equation \eqref{eq:schwinger_gap} indicates that the free energy near the conifold point has a expansion
\begin{align}\label{eq:gap_condition}
    \F_c^{(n,g)}=\frac{\text{const}_{n,g}}{t_c^{2n+2g-2}}+\mathcal{O}(t_c^0),
\end{align}
such that there is no contribution between the lowest order $t_c^{2n+2g-2}$ and the zero order $t_c^0$, which gives a {\it gap}.

\subsection{Wilson loops}\label{sec:HAE_Wilson}
For the Wilson loop partition function, we can also define the free energy $\FW$ as a logarithm of the partition function, which has a genus expansion \begin{align}
    \FW=\log \Wilson{\mathbf{r}}=\sum_{n,g=0}^{\infty}(\epsilon_1+\epsilon_2)^{2n}(\epsilon_1\epsilon_2)^{g-1}\FW^{(n,g)}.
\end{align}
We propose that the free energy $\FW$ satisfies a refined holomorphic anomaly equation 
\begin{align}\label{eq:HAEW}
    \frac{\partial \FW^{(n,g)}}{\partial S^{ij}}=\frac{1}{2} \left( D_iD_j \FW^{(n,g-1)} +{\sum_{n^{\prime},g^{\prime}}}^{\prime}D_{i} \FW^{(n^{\prime},g^{\prime})}\cdot D_j \FW^{(n-n^{\prime},g-g^{\prime})} \right), \quad n+g\geq 2.
\end{align}
In order to solve the refined holomorphic anomaly equation, we need information about the low genus property of the Wilson loop free energy.

By separating the conventional refined topological string amplitude $\F$, we define the free energy of the Wilson loop expectation value which is called {\it Wilson loop amplitude} as
\begin{align}
    \W_{\mathbf{r}}=\log  \langle W_{\mathbf{r}}\rangle=\FW-\F=\sum_{n,g}(\epsilon_1+\epsilon_2)^{2n}(\epsilon_1\epsilon_2)^{g-1}\W_{\mathbf{r}}^{(n,g)}.
\end{align}
From the 5d gauge theory point of view, e.g the BPS expansion \eqref{eq:wilson_n} and \eqref{eq:Wilson_n2}, the Wilson loop expectation value does not have pole at $\epsilon_{1,2}=0$, this property indicates that the Wilson loop amplitudes is zero at genus $(n,g=0)$,
\begin{align}\label{eq:W10}
    \W_{\mathbf{r}}^{(n,g=0)}=0,
\end{align}
for any $n\geq 0$. So at genus zero and genus one level, the only non-vanishing Wilson loop amplitude is $\W_{\mathbf{r}_i}^{(0,1)}$. As analyzed from the Seiberg-Witten curve and mirror curve correspondence in Section \ref{sec:3.2}, we expect that genus one free energy for the representation $\mathbf{r}_i$ whose highest weight is the $i$-th fundamental weight of the gauge
group $SU(N)$ is written as the logarithm of the homogeneous coordinate $a_i$
\begin{align}\label{eq:W01}
    \W_{\mathbf{r}_i}^{(0,1)}=\log(a_i)=C_{ij}^{-1}\log(z_j),
\end{align}
where the last identity in \eqref{eq:W01} is derived from \eqref{eq:consta} and $C_{ij}$ is the intersection matrix. 
For any representations $\mathbf{r}_1$ and $\mathbf{r}_2$ of a given gauge group $G$, one can perform tensor product and direct sum operations for them. As a direct consequence of \eqref{eq:wilson_split}, one may expect that at genus one, the tensor product of two representations can be decomposed as the sum of two amplitudes at genus one
\begin{align}\label{eq:otimes}
    \W_{\mathbf{r}_1\otimes \mathbf{r}_2}^{(0,1)}=\W_{\mathbf{r}_1}^{(0,1)} +\W_{\mathbf{r}_2}^{(0,1)}. 
\end{align}
For the direct sum of two representations, the Wilson loop partition function should be additive, such that we conjecture
\begin{align}\label{eq:oplus}
    \exp(\W_{\mathbf{r}_1\oplus \mathbf{r}_2}^{(0,1)})=\exp(\W_{\mathbf{r}_1}^{(0,1)}) +\exp(\W_{\mathbf{r}_2}^{(0,1)}).
\end{align}
In particular, by using \eqref{eq:otimes}, \eqref{eq:oplus} and \eqref{eq:W01}, we have that for the tensor product representation of the fundamental weight $\mathbf{r}_{i_1}^{\otimes j_1}\otimes\cdots\otimes \mathbf{r}_{i_l}^{\otimes j_1}$, 
\begin{align}\label{eq:W01_additive}
    \W_{\mathbf{r}_{i_1}^{\otimes j_1}\otimes\cdots\otimes \mathbf{r}_{i_l}^{\otimes j_1}}^{(0,1)}=j_1\W_{\mathbf{r}_{i_1}}^{(0,1)}+\cdots +j_l \W_{\mathbf{r}_{i_l}}^{(0,1)}=\left(j_1C_{i_1k}^{-1}+\cdots +j_lC_{i_lk}^{-1}\right)\log(z_k).
\end{align}

From \eqref{eq:W10} and \eqref{eq:W01}, the genus one free energies $\FW^{(0,1)}$ and $\FW^{(1,0)}$ are completely fixed, one can follow a similar method as we described in Section \ref{sec:HAE} to solve the higher genus Wilson loop free energies via the holomorphic anomaly equation \eqref{eq:HAEW}.

Finally, an interesting conclusion can be easily made from the holomorphic anomaly equation \eqref{eq:HAEW}. It is known that the quantum Seiberg-Witten curve computes the free energy in the NS limit $\epsilon_2\rightarrow 0$. In the NS limit, only $\W^{(n,g=1)}$ components survives, it satisfies the holomorphic anomaly equation 
\begin{align}\label{eq:HAEW1}
    \frac{\partial \W_{\mathbf{r}}^{(n,1)}}{\partial S^{ij}}=\sum_{n^{\prime}=0}^{n-1}D_{i} \W_{\mathbf{r}}^{(n^{\prime},1)}\cdot D_j \F^{(n-n^{\prime},0)}.
\end{align}
The equation \eqref{eq:HAEW1} is linear in $\W_{\mathbf{r}}^{(n,1)}$ which means it is additive
\begin{align}\label{eq:Wn1_additive}
    \frac{\partial }{\partial S^{ij}}\left(\W_{\mathbf{r}_1}^{(n,1)}+\cdots +\W_{\mathbf{r}_l}^{(n,1)}\right)=\sum_{n^{\prime}=1}^{n-1}D_{i} \left(\W_{\mathbf{r}_1}^{(n^{\prime},1)}+\cdots +\W_{\mathbf{r}_l}^{(n^{\prime},1)}\right)\cdot D_j \F^{(n-n^{\prime},0)}.
\end{align}
Together with \eqref{eq:W01_additive} and \eqref{eq:Wn1_additive}, we conjecture that
\begin{align}\label{eq:additiveruleinNS}
    \W_{\mathbf{r}_{i_1}\otimes\cdots\otimes \mathbf{r}_{i_l}}^{(n,1)}= \W_{\mathbf{r}_1}^{(n,1)}+\cdots +\W_{\mathbf{r}_l}^{(n,1)},\quad n\geq 0,
\end{align}
which suggests that in the NS limit, the Wilson loop expectation value of tensor products is equal to the product of the Wilson loop expectation values of each representations. This indeed agrees with the conclusion we have made in Section \ref{sec:Wilson_general}.
\subsection{Local $\mathbb{P}^1\times \mathbb{P}^1$}\label{sec:P1P1}
In this subsection, we calculate the Wilson loop amplitudes for local $\mathbb{P}^1\times \mathbb{P}^1$ from the refined holomorphic anomaly equation. We first review the computation of the refined free energies of local $\mathbb{P}^1\times \mathbb{P}^1$ without Wilson loop, most of the contents here can be also found in \cite{Haghighat:2008gw,Huang:2010kf,Huang:2013yta}.

\begin{figure}[h]
\begin{center}
\includegraphics{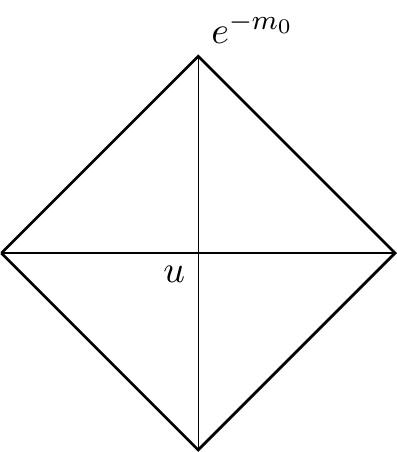}
\end{center}
\caption{Toric diagram for local $\mathbb{P}^1\times \mathbb{P}^1$. }
\label{fig:toric_P1P1}
\end{figure}

The toric diagram for local $\mathbb{P}^1\times \mathbb{P}^1$ is illustrated in Figure \ref{fig:toric_P1P1}. There are two complex structure parameters $z_1=\frac{1}{u^2}$ and $z_2=\frac{e^{-m_0}}{u^2}$, Where $u$ is the homogeneous coordinate related to the interior lattice point, and $m_0$ is a mass parameter related to a boundary point. 

The toric diagram gives the Picard-Fuchs operators
\begin{equation}
\begin{split}
     \mathcal{L}_1&=\Theta_1^2-2z_1(\Theta_1+\Theta_2) (1+2\Theta_1+2\Theta_2),\\
     \mathcal{L}_2&=\Theta_2^2-2z_2(\Theta_1+\Theta_2) (1+2\Theta_1+2\Theta_2),
\end{split}    
\end{equation}
where $\Theta_{i}\equiv z_{i}\frac{\partial}{\partial z_i}, i=1,2$. For simplicity, we turn off the mass parameter by setting $m_0=0$ such that $z=z_1=z_2$. The Picard-Fuchs operator for massless case is computed in \cite{Huang:2013yta} as
\begin{align}\label{eq:PF_P1P1}
    \mathcal{L}=z^2(1-16z)\partial_z^3+z(3-64z)\partial_z^2+(1-36z)\partial_z.
\end{align}
From \eqref{eq:PF_P1P1}, we can compute the mirror map
\begin{equation}\label{P1P1:mirror}
     -t(z)=\log (z)+4 z+18 z^2+\frac{400 z^3}{3}+1225 z^4+\frac{63504 z^5}{5} + \mathcal{O}(z^6),
\end{equation}
as a solution of the Picard-Fuchs equation
\begin{align}
    \mathcal{L}t(z)=0.
\end{align}
Even though the coefficients in the expansion \eqref{P1P1:mirror} are not always integers, by inverting the series, we find integral series \footnote{The integral Fourier expansion were also studied in \cite{Lian:1994zv} for some K3 surfaces and in \cite{Closset:2021lhd} for rank-one 4d $\mathcal{N}=2$ theories.}
\begin{align}\label{eq:P1P1_ztoQ}
    z(Q)=Q-4 Q^2+6 Q^3-16 Q^4-15 Q^5-216 Q^6+\mathcal{O}(Q^7),\quad Q=e^{-t}.
\end{align}
The genus genus one free energies are computed by fitting the ansatz \eqref{eq:F1_general} with few BPS invariants, we get
\begin{align}\label{eq:P1P1_F01}
    \F^{(0,1)}&=-\frac{1}{12}\log( z^7\Delta)-\frac{1}{2}\log \left|\frac{\partial t}{\partial z}\right|,\\
    \F^{(1,0)}&=\frac{1}{24}\log( z^{-2}\Delta),
\end{align}
where $\Delta=(1-16z)$ is the discriminant of the mirror geometry. 
To compute the higher genus free energies, we need to integrate over the propagator $S^{zz}$, where the propagator can be fixed by a proper choice holomorphic functions appear in \eqref{eq:S_relation1} and \eqref{eq:S_relation2}. Following the calculations in \cite{Haghighat:2008gw}, we fix the special geometry relations
\begin{align}
   & D_zS^{zz}=\partial_zS^{zz}+2\Gamma^{z}_{zz}S^{zz}=-C_{zzz}S^{zz}S^{zz}-\frac{z(1-12z)}{9(1-16z)},\\
   & \Gamma^{z}_{zz}=\frac{\partial z}{\partial t}\frac{\partial^2t}{\partial z^2}=-C_{zzz}S^{zz}-\frac{4(1-18z)}{3z(1-16z)}.
\end{align}
where $C_{zzz}$ is the Yukawa coupling
\begin{align}\label{eq:P1P1_Czzz}
    C_{zzz}=\left(\frac{\partial t}{\partial z}\right)^3\cdot\frac{\partial^3 \F^{(0,0)}}{\partial t^3}= -\frac{1}{z^3(1-16z)},
\end{align}
such that the propagator is
\begin{align}\label{eq:P1P1_Szz}
    S^{zz}=\frac{1}{C_{zzz}}\left(2\partial_z \F^{(0,1)}-\frac{1}{6z}\right)=z^3(1-16z){\partial_z \log \frac{\partial t}{\partial z}}+\frac{4}{3}z^2(1-18z).
\end{align}
The special geometry relations also indicate a derivative rule
\begin{align}\label{eq:P1P1_Srule}
    \partial_z S^{zz}=-\frac{1}{9z^3\Delta} (9 S^2 - 24 S z^2 + 432 S z^3 + z^4 - 12 z^5).
\end{align}
Substituting equation \eqref{eq:P1P1_Szz} in \eqref{eq:P1P1_Srule}, we have that the third order derivative of $t(z)$ can be reduced to lower derivatives
\begin{align}\label{eq:P1P1_trule}
    \partial_z^3t(z)=\frac{1}{z^2\Delta}[(64z^2-3z)\partial_z^2+(36z-1)\partial_z]t(z).
\end{align}
Since there is only one component of the propagator, we use $S$ to stand for $S^{zz}$ in short. By using equation \eqref{eq:P1P1_F01} and \eqref{eq:P1P1_ztoQ}, the propagator can be written as a series expansion of the A-model K\"ahler modulus as
\begin{align}\label{eq:StoQ}
    S=\frac{Q^2}{3}-\frac{20 Q^3}{3}+\frac{148 Q^4}{3}-\frac{680 Q^5}{3}+\frac{2198
   Q^6}{3}+O\left(Q^7\right).
\end{align}

The higher genus free energies are solved by integrating over the refined holomorphic anomaly equation \eqref{eq:HAE}. However, there are holomorphic ambiguities need to be fixed. At genus two, up to holomorphic ambiguities, we have
\begin{equation}\label{eq:Fg2} 
\begin{split}
    \F^{(0,2)}&=\frac{1}{288 z^6 \Delta^2}(60 S^3-48 S^2 z^2+480 S^2 z^3+13 S z^4-288 S z^5+1792 S z^6)+h^{(0,2)},\\
    \F^{(1,1)}&=\frac{1}{144 z^4 \Delta^2}(6 S^2-48 S^2 z-3 S z^2+40 S z^3-512 S z^4)+h^{(1,1)},\\
    \F^{(2,0)}&=\frac{S (-1+8 z)^2}{288 z^2 \Delta^2}+h^{(2,0)},
\end{split}
\end{equation}
where $h^{(n,g)}$ is the holomorphic ambiguity which is a rational function of $z$, which takes the ansatz
\begin{align}\label{eq:hng}
    h^{(n,g)}=\frac{{h^{\prime}}^{(n,g)}(z)}{\Delta^{2(n+g)-2}},
\end{align}
with ${h^{\prime}}^{(n,g)}(z)$ is a polynomial in $z$.
As was pointed out in \cite{Haghighat:2008gw,Huang:2010kf}, the holomorphic ambiguities for local $\mathbb{P}^1\times \mathbb{P}^1$ model can be completely fixed by considering the behavior of the free energies around conifold point and orbifold point. We have \footnote{The constant term in $\F^{(n,g)}$ is fixed by the constant map contribution at large volume limit, which is irrelevant from the Gromov-Witten invariants. We are more interested in the Gromov-Witten invariants, thus we set the constant term to be zero.} 
\begin{align}
    \F^{(0,2)}=\,&\frac{5 S^3}{24 \Delta ^2 z^6}+\frac{S^2 \left(21600 z^3-2160 z^2\right)}{12960 \Delta ^2
   z^6}+\frac{S }{12960 \Delta ^2z^6}\left(80640 z^6-12960 z^5+585 z^4\right)\nonumber\\
   &+\frac{1}{12960 \Delta ^2 z^6}\left(110592 z^9-24000 z^8+1884 z^7-55 z^6\right),\\
    \F^{(1,1)}=\,&\frac{S^2 }{2160 \Delta ^2 z^4}(90-720 z)+\frac{S }{2160 \Delta ^2 z^4}\left(-7680 z^4+600 z^3-45z^2\right)\nonumber\\
   &+\frac{1}{2160 \Delta^2 z^4}\left(-21504 z^7+2560 z^6-108 z^5+5 z^4\right),\\
    \F^{(2,0)}=\,&\frac{S \left(960 z^2-240 z+15\right)}{4320 \Delta ^2 z^2}+\frac{1}{4320 \Delta ^2 z^2}(10752 z^5-3200 z^4+164z^3-5 z^2).
\end{align}
The higher genus free energies are solved in a similar way. By computing the free energies to a higher enough genus, one can recover the BPS invariants as listed in \cite{Huang:2010kf,Huang:2013yta}. 
\paragraph{Wilson loop in representation $\mathbf{2}^{\otimes \n}$}
We now use holomorphic anomaly equation to solve the free energies of the Wilson loop partition function $\F_{\mathbf{2}^{\otimes \n}}^{(n,g)}=\W_{\mathbf{2}^{\otimes \n}}^{(n,g)}+\F^{(n,g)}$ for the representation $\mathbf{r}=\mathbf{2}^{\otimes \n}$. Following equation \eqref{eq:W01_additive}, noticing that the entry of the $1\times 1$ intersection matrix is ${-2}$, the genus one free energies are computed as
\begin{align}
    \W_{\mathbf{2}^{\otimes \n}}^{(1,0)}=0,\quad \W_{\mathbf{2}^{\otimes \n}}^{(0,1)}=-\frac{\n}{2}\log(z).
\end{align}
The higher genus results are computed via the refined holomorphic equation \eqref{eq:HAEW}. For the fundamental representation when $\n=1$, the Wilson loop free energies can be completely solved to arbitrary genus from the gap condition and the regularity at orbifold point. However, the regularity at orbifold point is broken down for higher representation, we observe that they satisfy a singular behavior
\begin{align}
    \W_{\mathbf{2}^{\otimes \n}}^{(n,g)}=\frac{a_{o,g-1}}{t_o^{2g-2}}+\cdots+\frac{a_{o,1}}{t_o^{2}}+\mathcal{O}(t_o^0),\quad \n>1,
\end{align}
where the unknown coefficients $a_{o,g-1}$ reduce the constraints on the holomorphic ambiguities, that we can not solve the free energies completely. Thus we use few (pseudo-)BPS invariants achieved in Section \ref{sec:SU2}, to fix the holomorphic ambiguities. We find at genus two, $\W_{\mathbf{r}}^{(2,0)}=0$,
\begin{align}\label{eq:P1P1W11}
   \W_{\mathbf{2}^{\otimes \n}}^{(1,1)}=\frac{\n}{72 z^2 \Delta}(3 S-24 S z-z^2+20 z^3),
\end{align}
\begin{align}
    \W_{\mathbf{2}}^{(0,2)}&=\frac{9 S^2+24 S z^3-z^4+16 z^5}{36 z^4 \Delta },\\
    \W_{\mathbf{2}\otimes \mathbf{2}}^{(0,2)}&=\frac{18 S^2+9 S z^2-96 S z^3-5 z^4+116 z^5-576 z^6}{36 z^4 \Delta },\\
    \W_{\mathbf{2}\otimes \mathbf{2}\otimes \mathbf{2}}^{(0,2)}&=\frac{9 S^2+9 S z^2-120 S z^3-4 z^4+100 z^5-576 z^6}{12 z^4 \Delta },\\
    \W_{\mathbf{2}^{\otimes 4}}^{(0,2)}&=\frac{18 S^2+27 S z^2-384 S z^3-11 z^4+284 z^5-1728 z^6}{18 z^4 \Delta },\\
    \W_{\mathbf{2}^{\otimes 5}}^{(0,2)}&=\frac{5 \left(9 S^2+18 S z^2-264 S z^3-7 z^4+184 z^5-1152 z^6\right)}{36 z^4 \Delta },\\
    &\quad\quad\vdots \nonumber
\end{align}
Similarly, by computing the Wilson loop free energies to higher enough genus, we solve the (pseudo-)BPS invariants by fitting the ansatz
\begin{align}
    \exp(\W_{\mathbf{r}})=\sum_{d=1}^{\infty}\sum_{j_L,j_R} (-1)^{2j_L+2j_R}\widetilde{N}^{d}_{j_L,j_R}{\chi_{j_L}(q_L)\chi_{j_R}(q_R)}Q^d,
\end{align}
which agrees with the result in Section \ref{sec:SU2} in the massless limit $m_0=0$.
We list the free energies in Appendix \ref{sec:C} and the (pseudo-)BPS spectrum in Appendix \ref{sec:BPS}. Notice that even though the direct integration method looks quite indirect to get the BPS invariants, an advantage of the results, e.g. \eqref{eq:P1P1W11}, is that they are {\it{exact}} that one can expand them to arbitrary degree of the complex structure parameters or the K\"ahler parameters. It is also convenient to get the behavior of the free energies around other points in the Calabi-Yau moduli space, e.g. conifold point and orbifold point \cite{Aganagic:2006wq}.   

\subsection{Local $\mathbb{P}^2$}\label{sec:P2}

\begin{figure}[h]
\begin{center}
\includegraphics{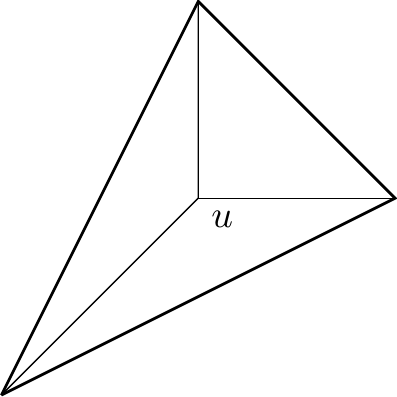}
\end{center}
\caption{Toric diagram for local $\mathbb{P}^2$. }
\label{fig:toric_P2}
\end{figure}

The toric diagram for local $\mathbb{P}^2$ is described in Figure \ref{fig:toric_P2}. There is only one complex structure parameter $z_1=\frac{1}{u^3}$, Where $u$ is the homogeneous coordinate related to the interior lattice point. 

The toric diagram gives the Picard-Fuchs operator
\begin{equation}
\begin{split}
     \mathcal{L}&=\Theta^3+3z(3\Theta+2)(3\Theta+1)\Theta,
\end{split}    
\end{equation}
where $\Theta\equiv z\frac{\partial}{\partial z}$. We can then compute the mirror map
\begin{equation}
     -t=\log(z)  -6 z+45 z^2-560 z^3+\frac{17325 z^4}{2}-\frac{756756 z^5}{5} + \mathcal{O}(z^6),
\end{equation}
and the inverse map
\begin{align}
   z(Q)= Q+6 Q^2+9 Q^3+56 Q^4-300 Q^5+3942 Q^6+ \mathcal{O}(Q^7).
\end{align}
The genus one free energies are
\begin{align}
    \F^{(0,1)}&=-\frac{1}{12}\log( z^7\Delta)-\frac{1}{2}\log \left(\frac{\partial t}{\partial z}\right),\\
    \F^{(1,0)}&=\frac{1}{24}\log( z^{-1}\Delta),
\end{align}
where $\Delta=1+27 z$ is the discriminant.
By fixing the ambiguities in equations \eqref{eq:S_relation1} and \eqref{eq:S_relation2} as
\begin{align}
   & D_zS^{zz}=\partial_zS^{zz}+2\Gamma^{z}_{zz}S^{zz}=-C_{zzz}S^{zz}S^{zz}-\frac{z}{12(1+27z)},\\
   & \Gamma^{z}_{zz}=\frac{\partial z}{\partial t}\frac{\partial^2t}{\partial z^2}=-C_{zzz}S^{zz}-\frac{7+216z}{6z(1+27z)},
\end{align}
where $C_{zzz}$ is the Yukawa coupling
\begin{align}
    C_{zzz}=\left(\frac{\partial t}{\partial z}\right)^3\cdot\frac{\partial^3 \F^{(0,0)}}{\partial t^3}= -\frac{1}{3z^3(1+27z)},
\end{align}
we can fix the propagator
\begin{align}
    S^{zz}=\frac{2\partial_z \F^{(0,1)}}{C_{zzz}}=3 z^3 (1 + 27 z){\partial_z \log \frac{\partial t}{\partial z}}+\frac{1}{2} z^2 (7 + 216 z).
\end{align}
With all the ingredients, we can solve the topological string amplitudes from the refined holomorphic equation. For the case without Wilson loop, as pointed out in \cite{Haghighat:2008gw,Huang:2010kf}, from the gap condition and regularity at orbifold point, the holomorphic anomaly can be completely fixed.

The local $\mathbb{P}^2$ model is corresponded to non-Lagrangian theory in 5d without a gauge group. Such that there is no notation of representations for the Wilson loop operators. We follow the same notation as introduced in \cite{Kim:2021gyj}, to use $[-1]$ stand for the fundamental representation comes from the insertion of a primitive curve, such that 
\begin{align}
    \W_{[-1]}^{(0,1)}=-\frac{1}{3}\log z,
\end{align}
and the $\n$-th tensor product of the representation is naturally given by
\begin{align}
    \W_{[-1]^{\otimes \n}}^{(0,1)}=-\frac{\n}{3}\log z.
\end{align}
The higher genus Wilson loop free energies are solved from the refined holomorphic anomaly equation. For representation $[-1]$ and $[-1]\otimes [-1]$, we observe that the gap condition and regularity at orbifold point still holds, such that we can solve the Wilson loop free energies to arbitrary genus.
For example, at genus $n+g=2$, we have
\begin{align}
   \W_{[-1]}^{(1,1)}&=\frac{2 S-z^2-54 z^3}{144 z^2 (1+27 z)},\\
    \W_{[-1]}^{(0,2)}&=\frac{2 S^2+S z^2-z^4-27 z^5}{36 z^4 (1+27 z)},\\
\end{align}
for the representation $[-1]$, and 
\begin{align}
    \W_{[-1]\otimes [-1]}^{(1,1)}&=\frac{2 S-z^2-54 z^3}{72 z^2 (1+27 z)},\\
    \W_{[-1]\otimes [-1]}^{(0,2)}&=\frac{2 S^2+3 S z^2+54 S z^3-2 z^4-54 z^5}{18 z^4 (1+27 z)},\\
\end{align}
for the representation $[-1]\otimes [-1]$. By computing the Wilson loop to higher and higher genus, we recover the refined (pseudo-)BPS invariants as listed in Table \ref{tab:P2_z1} and Table \ref{tab:P2_z2}. 
\section{(Refined) mirror map from Wilson loop}\label{sec:5}
The quantum mirror curves and quantum periods are studied in \cite{Nekrasov:2009rc,Aganagic:2011mi}. The basic idea to get the quantum mirror curve is by promoting the coordinates $x,y$ to operators $\hat{x},\hat{y}$ in the classical curve \eqref{eq:curve}, with the canonical commutation relation
\begin{align}
    [\hat{x},\hat{y}]=\hbar.
\end{align}
Then the curve \eqref{eq:curve} becomes an operator equation,
\begin{align}\label{eq:quantum:equation}
    \hat{H}(e^{\hat{x}},e^{\hat{y}};z_{\alpha},\hbar)\Psi(x;\hbar)=0.
\end{align}
Equation \eqref{eq:quantum:equation} can be treated as a quantum mechanics system with Hamiltonian $ \hat{H}$, and $\Psi(x;\hbar)$ is the wave function which can be solved from the standard WKB method by imposing 
\begin{align}
    \Psi(x;\hbar)=e^{-\frac{1}{\hbar}\int \partial_x S(x;\hbar) dx}, \quad S(x;\hbar)=S_0(x)+\sum_{i=1}^{\infty}S_{2i}(x)\hbar^{2i}.
\end{align}
Now the periods of the curve become quantum periods which are computed from the period integral over the quantum one-form $\partial S(x)$. Denote the quantum periods
\begin{align}\label{eq:q_period}
    \Pi_i({z};\hbar)=\Pi_{i}({z})+\sum_{j=1}^{\infty}\Pi_{i,2j}({z})\hbar^{2j},
\end{align}
for each $j>0$, there exists an differential operator $\mathcal{D}_{2j}$ with respect to $z$
\begin{align}\label{eq:q_period2}
    \Pi_{i,2j}({z})=\mathcal{D}_{2j}\Pi_{i}({z}).
\end{align}
Equation \eqref{eq:q_period} indicates that the quantum A-period or quantum mirror map $t(z;\hbar)$ has a small $\hbar$ expansion
\begin{align}\label{eq:q_mirror_map}
    t(z;\hbar)=[1+\sum_{i=1}^{\infty}\hbar^{2i}\mathcal{D}_{2i}]t(z),
\end{align}
which is generated by the quantum operators $\mathcal{D}_{2i}$, while the quantum B-period 
\begin{align}
    t_D(z;\hbar)=C\frac{\partial F^{\mathrm{NS}}(t(z;\hbar);\hbar)}{\partial t(z;\hbar)}
\end{align}
is generated by the same quantum operators
\begin{align}\label{eq:q_mirror_map_tD}
    t_D(z;\hbar)=[1+\sum_{i=1}^{\infty}\hbar^{2i}\mathcal{D}_{2i}]t_D(z).
\end{align}
Here $C$ is entry of the $1\times 1$ intersection matrix which is equal to $-2$ for local $\mathbb{P}^1\times \mathbb{P}^1$ and is equal to $-3$ for local $\mathbb{P}^2$. $F^{\mathrm{NS}}$ is the NS free energy
\begin{align}
    F^{\mathrm{NS}}(t;\hbar)=\lim_{\substack{\epsilon_1\rightarrow \hbar,\\\epsilon_2\rightarrow 0}}\epsilon_1\epsilon_2 \F,\quad t_D(z)=\lim_{\hbar\rightarrow 0}C\partial_t F^{\mathrm{NS}}(t(z);\hbar).
\end{align}
See \cite{Huang:2013yta,Huang:2014nwa,Fischbach:2018yiu,Huang:2020neq} for more details about the derivation of the quantum operators $\mathcal{D}_{2j}$ from quantum curves. 

As we have discussed in Section \ref{sec:3.2}, the complex structure parameter $a_i$ at classical level is equal to the classical Wilson loop expectation value. This correspondence leads to the mirror dual of A-model Wilson loop expectation values in terms with K\"ahler parameters to the complex structure parameters $a_i$ in the B-model. 
This correspondence can be enhanced to quantum level. One key observation is that the quantum Seiberg-Witten curve \eqref{eq:q_SW} has the same form as the quantum mirror curve if one identifies the Wilson loop expectation values in the NS limit, with the same complex structure parameters $a_i$. At the level of amplitude, we have 
\begin{align}\label{eq:Wr_Q_op}
    \W^{(0,1)}_{\mathbf{r}}\left(t(z)\right)&=\sum_{n=0}^{\infty}\W^{(n,1)}_{\mathbf{r}}\left(t(z;\hbar)\right)\hbar^{2n}.
\end{align}
Notice that on the left hand side of \eqref{eq:Wr_Q_op} we use the classical mirror map $t(z)$ such that it maps to the original complex structure parameter $a_i$, while on the right hand side of \eqref{eq:Wr_Q_op}, the K\"ahler parameter is replaced by the quantum mirror map \eqref{eq:q_mirror_map}.

Substitute \eqref{eq:q_mirror_map} to \eqref{eq:Wr_Q_op} and expand both side of the equation \eqref{eq:Wr_Q_op} with small $\hbar$, we can solve that for the single parameter case
\begin{align}
    \mathcal{D}_2t(z)=&-\frac{\W_{\mathbf{r}}^{(1,1)}}{\partial_z \W_{\mathbf{r}}^{(0,1)}}\partial_z t,\nonumber\\
    \mathcal{D}_4t(z)=&-\frac{\W_{\mathbf{r}}^{(2,1)}}{\partial_z \W_{\mathbf{r}}^{(0,1)}}\partial_z t+\frac{2\W_{\mathbf{r}}^{(1,1)}\partial_z\W_{\mathbf{r}}^{(1,1)}\partial_z t+\left(\W_{\mathbf{r}}^{(1,1)}\right)^2\partial_z^2t}{2\left(\partial_z \W_{\mathbf{r}}^{(0,1)}\right)^2}-\frac{\left(\W_{\mathbf{r}}^{(1,1)}\right)^2\partial_z^2\W_{\mathbf{r}}^{(0,1)}\partial_zt}{2\left(\partial_z \W_{\mathbf{r}}^{(0,1)}\right)^3},\nonumber\\
    &\vdots\nonumber
\end{align}
Because of the additive property \eqref{eq:additiveruleinNS} in the NS limit, the quantum operators $\mathcal{D}_{2i}$ are the same for any representation $\mathbf{r}^{\otimes \n}$.  For local $\mathbb{P}^1\times \mathbb{P}^1$ model, by substituting the results from holomorphic anomaly equation and using the derivative rule \eqref{eq:P1P1_trule} to eliminate higher order derivatives, we observe that the quantum operators $\mathcal{D}_{2i}$ can be always written as degree two polynomials of the differential operator $\Theta= z\partial_z$, for example, 
\begin{align}\label{eq:Dop_P1P1}
    \mathcal{D}_2=\,&-\frac{  z}{3}\Theta +\frac{1-8z}{12} \Theta^2 ,\\
    \mathcal{D}_4=\,&\frac{1}{180\Delta^2}\left[\left(3584 z^4-320 z^3+34
   z^2\right)\Theta +\left(7168 z^4-864 z^3+154 z^2-z\right)\Theta ^2\right],\\
   &\vdots\nonumber
\end{align}
which agree with the results in \cite{Huang:2014nwa}. By using our expression of the Wilson loops, one can also analytically derive the quantum B-period, which is defined as
\begin{align}
   t_D(z;\hbar)&= \partial_{t(z;\hbar)}F^{NS}(t(z;\hbar);\hbar)= \sum_{n=0}^{\infty}t_D^{(n,0)}(t(z;\hbar))\hbar^{2n},\\
   &=[1+\hbar^2 \mathcal{D}_2^{\prime}+\hbar^4 \mathcal{D}_4^{\prime}+\cdots]t_D(z),
\end{align}
where
\begin{align}
    t_D^{(n,0)}=-2\partial_t \F^{(n,0)}(t).
\end{align}
Notice that from equation \eqref{eq:P1P1_Czzz}, 
\begin{align}
    C_{zzz}=\left(\frac{\partial t}{\partial z}\right)^3\cdot\frac{\partial^3 \F^{(0,0)}}{\partial t^3}=-\partial_z t_D(z) \partial_z^2t(z)+\partial_z^2t_D(z)\partial_z t(z),
\end{align}
we solve that
\begin{align}\label{eq:5.10}
    \partial_z^2t(z)=\frac{1}{\partial_z t_D(z)}\left(\partial_z^2t_D(z)\partial_z t(z)-C_{zzz}\right).
\end{align}
Taking derivative with respect to $z$ on both side of equation \eqref{eq:5.10}, and take use of equation \eqref{eq:P1P1_trule}, we have the replacement rule
\begin{align}
    \partial_z^3t_D(z)=\frac{1}{z^2\Delta}\left[(64z^2-3z)\partial_z^2+(36z-1)\partial_z\right]t_D(z),
\end{align}
which can be used to eliminate the higher order derivatives of $t_D(z)$, that we again solve the differential operator $\mathcal{D}_{2i}^{\prime}$, which is exactly the same as $\mathcal{D}_{2i}$ as illustrated in \eqref{eq:Dop_P1P1}. This fact agrees with the quantum curve approach from \cite{Huang:2014nwa}.

For local $\mathbb{P}^2$ model, by using the same techniques, we analytically derive the quantum differential operators for both quantum A- and B-periods
\begin{align}\label{eq:Dop_P2}
    \mathcal{D}_2=&\frac{1}{8}\Theta^2,\\
    \mathcal{D}_4=& \frac{1}{640\Delta^2}\left[  \left(1998 z^2-10 z\right)\Theta+\left(7857 z^2-87 z\right)\Theta^2\right],\\
    &\vdots \nonumber
\end{align}
which agree with the quantum differential operators computed in \cite{Huang:2014nwa}.
\paragraph{Refined A-periods}
In the gauge theory side, the quantum curve has a refined version contains both non-vanishing Omega deformed parameter $\epsilon_{1,2}$. The refined version of the curve is known as non-perturbative Dyson-Schwinger equation \cite{Nekrasov:2015wsu}, where the Wilson loop expectation values in the NS limit are lifted to their refined versions. Even though the refined curve does not have a quantum mechanics understanding, we can still identify the complex structure parameter with the Wilson loop expectation value
\begin{align}\label{eq:Wr_Q_op_refine}
    \W^{(0,1)}_{\mathbf{r}}\left(t(z)\right)&=\sum_{n,g=0}^{\infty}\W^{(n,g)}_{\mathbf{r}}\left(t(z;\epsilon_1,\epsilon_2)\right)(\epsilon_1+\epsilon_2)^{2n}(\epsilon_1\epsilon_2)^{g-1},
\end{align}
as a refined version of \eqref{eq:Wr_Q_op}, by proposing the refined A-period\footnote{There was another refined proposal \cite{Bourgine:2017jan}, it is interesting to check the connection between our proposals.}
\begin{align}
    t(z;\epsilon_1,\epsilon_2)=\left[1+\sum_{\substack{i,j=1,\\ i+j>0}}^{\infty}(\epsilon_1+\epsilon_2)^{2i}(\epsilon_1\epsilon_2)^j\mathcal{D}_{i,j}\right]t(z).
\end{align}
where $\mathcal{D}_{i,j}$ are expected to be differential operators with respect to $z$. Obviously, in the NS limit $\mathcal{D}_{i,0}=\mathcal{D}_{2i}$, the other operators are solved by expanding \eqref{eq:Wr_Q_op_refine} with small $\epsilon_{1,2}$ at both hand side, 
\begin{align}
    \mathcal{D}_{0,1}t(z)=&-\frac{\W_{\mathbf{r}}^{(0,2)}}{\partial_z \W_{\mathbf{r}}^{(0,1)}}\partial_z t,\nonumber\\
    \mathcal{D}_{0,2}t(z)=&-\frac{\W_{\mathbf{r}}^{(0,3)}}{\partial_z \W_{\mathbf{r}}^{(0,1)}}\partial_z t+
    \frac{1}{\partial_z \W_{\mathbf{r}}^{(0,1)}}\partial_z\left[\frac{\left(\W_{\mathbf{r}}^{(0,2)}\right)^2\partial_z t}{\partial_z \W_{\mathbf{r}}^{(0,1)}}\right],\nonumber\\
    \mathcal{D}_{1,1}t(z)=&-\frac{\W_{\mathbf{r}}^{(1,2)}}{\partial_z \W_{\mathbf{r}}^{(0,1)}}\partial_z t+
    \frac{1}{\partial_z \W_{\mathbf{r}}^{(0,1)}}\partial_z\left[\frac{\W_{\mathbf{r}}^{(0,2)}\W_{\mathbf{r}}^{(1,1)}\partial_z t}{\partial_z \W_{\mathbf{r}}^{(0,1)}}\right],\nonumber\\
    &\vdots\nonumber
\end{align}
By using the results of Wilson loop amplitudes, we observe that the operators $\mathcal{D}_{i,j}$ are still second order differential operators, but the coefficients depend on the propagator $S$. For examples, for local $\mathbb{P}^1\times \mathbb{P}^1$ in the fundamental representation,
\begin{align}\label{eq:Dop_P1P1_ref}
    \mathcal{D}_{0,1}=\,&\frac{1}{6
   z^2}[(-3 S+z^2-32 z^3) \Theta +(3 S+4 z^2-64 z^3) \Theta ^2],\\
    \mathcal{D}_{0,2}=\,&\frac{1}{12960 z^8 \Delta^2}[(-3240 S^4+1080 S^3 z^2+34560 S^3 z^3-2295 S^2 z^4+79920 S^2 z^5\nonumber\\
    &-1725 S z^6-1105920 S^2 z^6+125280 S z^7+830 z^8-3098880 S z^8-78920 z^9\nonumber\\
    &+25436160 S z^9+2891616 z^{10}-48084480 z^{11}+306118656 z^{12}) \Theta \nonumber\\
    &+(3240 S^4-1080 S^3 z^2-34560 S^3 z^3+2295 S^2 z^4-79920 S^2 z^5+1725 S z^6 \nonumber\\
    &+1105920 S^2 z^6-125280 S z^7+2545 z^8+3098880 S z^8-218944 z^9 \nonumber\\
    &-25436160 S z^9+7271616 z^{10}-108942336 z^{11}+612237312 z^{12}) \Theta ^2],\\
    \mathcal{D}_{1,1}=\,&\frac{1}{270 z^3 \Delta^2}[(45 S^2+720 S^2 z-30 S z^2-1560 S z^3+5 z^4-5760 S z^4+416 z^5 \nonumber\\
    &-12000 z^6-90624 z^7) \Theta +(-45 S^2-720 S^2 z+30 S z^2+1560 S z^3+z^4\nonumber\\
    &+5760 S z^4+1696 z^5-19776 z^6-181248 z^7) \Theta ^2],\\
   &\vdots\nonumber
\end{align}
For local $\mathbb{P}^2$ in the representation $[-1]$,
\begin{align}\label{eq:Dop_P2_ref}
    \mathcal{D}_{0,1}=\,&\frac{1}{4 z^2}[(-2 S+z^2+72 z^3) \Theta +(2 S+8 z^2+216 z^3) \Theta ^2],\\
    \mathcal{D}_{0,2}=\,&\frac{1}{8640 z^8 \Delta^2}[(-240 S^4-240 S^3 z^2-25920 S^3 z^3-1480 S^2 z^4-114480 S^2 z^5 \nonumber\\
    &-4820 S z^6-2799360 S^2 z^6-550800 S z^7+2825 z^8-22249080 S z^8+485460 z^9\nonumber\\
    &-302330880 S z^9+31913676 z^{10}+952132320 z^{11}+10883911680 z^{12}) \Theta \nonumber\\
    &+(240 S^4+240 S^3 z^2+25920 S^3 z^3+1480 S^2 z^4+114480 S^2 z^5+4820 S z^6\nonumber\\
    &+2799360 S^2 z^6+550800 S z^7+16975 z^8+22249080 S z^8+2459376 z^9 \nonumber\\
    &+302330880 S z^9+137471904 z^{10}+3461058720 z^{11}+32651735040 z^{12}) \Theta ^2],\\
    \mathcal{D}_{1,1}=\,&\frac{3}{320 z^3 \Delta^2}[ (-60 S^2+20 S z^2-7560 S z^3+5 z^4+4248 z^5+272160 z^6) \Theta \nonumber\\
    &\quad\quad\quad\quad+(60 S^2-20 S z^2+7560 S z^3+13 z^4+28242 z^5+816480 z^6) \Theta ^2],\\
   &\vdots\nonumber
\end{align}
It is interesting to find an explanation of these refined operators in the future.
\section{Conclusion}\label{sec:conclusion}
In this paper, we proposed the refined topological string correspondence to the half-BPS Wilson loop operators in 5d $\mathcal{N}=1$ quantum field theories in the Omega-deformed background $\mathbb{R}_{\epsilon_{1,2}}\times S^1$. Our proposal suggests a novel infinite class of generalization to the refined topological string amplitudes which are called Wilson loop amplitudes. The insertion of the Wilson loop operators can be understood in M-theory as the insertion of stationary curves. By using the refined topological vertex in the A-model, we explicitly calculated the expectation values of the Wilson loop operators in the representation $\mathbf{2}^{\otimes \n},\n=1,\cdots 5$ in the 5d pure $SU(2)$ gauge theory.

In the B-model, we proposed the refined holomorphic anomaly equation for the Wilson loop amplitudes. We calculated these amplitudes for local $\mathbb{P}^1\times \mathbb{P}^1$ model and local $\mathbb{P}^2$ models in various representations. The amplitudes solved from the refined holomorphic anomaly equation are polynomials of the propagator $S$, which are exact expressions and can be expanded at any point in the Calabi-Yau moduli space. Our results also indicate that the Wilson loop amplitudes are weight-zero quasi-modular forms. Finally, by taking use of the exact expressions of Wilson loop amplitudes, we derived the quantum operators studied in the quantum geometry.

Our work leads to many interesting problems. The first one is that why there is a refined holomorphic anomaly equation for Wilson loop amplitudes. In this paper, we focused on the BPS invariants of M2-branes winding on stationary curves, such that the partition function has a BPS expansion \eqref{eq:wilson_n}. The BPS invariants in \eqref{eq:wilson_n} are quite likely to be the descendant invariants as described in \cite{Losev:2003py,MR2512147}.  However, the existence of the holomorphic anomaly equation for the free energies suggests that there would be a suitable worldsheet description for the Wilson loop amplitudes, which gives a generalization of the Gromov-Witten invariants. It is interesting to find the worldsheet description and give a mathematical definition for the generalized invariants. 

In the meantime, we should check that if the refined holomorphic anomaly equation holds for more generic examples. In particular, the other local del Pezzo surfaces, including E-strings and try to compare with the quantum periods from quantum curves studied in  \cite{Furukawa:2019sfy,Moriyama:2020lyk,Chen:2021ivd}. It is also interesting to compute the Wilson loops/surfaces of generic 5d/6d SQFTs from more generic non-compact Calabi-Yau threefolds.

In Section \ref{sec:SU2}, we used the refined topological vertex to compute the Wilson loop expectation values for $SU(2)$ theory from $SU(2)$ theories with fundamental flavors. It is also interesting to recover our result in the $Sp(1)$ theory via the refined topological vertex with ON-planes \cite{Kim:2017jqn,Kim:2022dbr}. Even though we used the topological vertex to compute the partition function, the same partition function can also be computed from ADHM construction directly. It is worthy to use the ADHM construction to compute the Wilson loop expectation values for more generic 5d/6d SQFTs.

Finally, as we have done in Section \ref{sec:5}, the Wilson loop amplitudes calculated from holomorphic anomaly equations can help to derive the quantum operators that are derived from the quantum curves. It is also interesting to explore the connection between holomorphic anomaly equations and quantum curves more in the future.
\acknowledgments

We would like to thank Hee-Cheol Kim, Minsung Kim, Albrecht Klemm, Yuji Sugimoto, Kaiwen Sun, Longting Wu for valuable discussions. MH was supported in parts by the National Natural Science Foundation of China (Grants  No.11947301 and No.12047502). KL is supported by KIAS Individual Grant PG006904 and by the National Research Foundation of Korea (NRF) Grant funded by the Korea government (MSIT) (No.2017R1D1A1B06034369). XW is supported by KIAS Individual Grant QP079201.
\appendix

\section{Two-instanton results}\label{appendix:2-inst}
In this appendix, we list the two-instanton results of the Wilson loop expectation values for $SU(2)$ theory. They have the structure
\begin{align*}
    \langle W_{\mathbf{2}^{\otimes \n}}^{(2)}\rangle=\mathcal{N}^{(\n)}\left[(1-q_1q_2e^{-2\phi})(1-q_1^2q_2e^{-2\phi})(1-q_1q_2^2e^{-2\phi})\times(\phi\rightarrow -\phi)\right]^{-1},
\end{align*}
with
\begin{align*}
\mathcal{N}^{(1)}=\,&q_1^3 q_2^3 (1+q_1 q_2+q_1^2 q_2^2+q_1^3 q_2^3+q_1^4 q_2^4)(e^{\phi}+e^{-\phi})^{{3}}\nonumber\\
&-q_1^3 q_2^3 (1+q_1 q_2) (2+2 q_1 q_2+q_1^2 q_2+q_1 q_2^2+2 q_1^2 q_2^2+2 q_1^3 q_2^3)(e^{\phi}+e^{-\phi}),
\end{align*}
\begin{align*}
\mathcal{N}^{(2)}=\,q_1^5 q_2^5(e^{\phi}+e^{-\phi})^{{6}}
+q_1^3 q_2^3 (1+q_1^2 q_2+q_1 q_2^2-4 q_1^2 q_2^2+q_1^3 q_2^2+q_1^2 q_2^3+q_1^4 q_2^4)&(e^{\phi}+e^{-\phi})^{{4}}\nonumber\\
+q_1^2 q_2^2 (1+q_1 q_2)^2 (q_1+q_2-3 q_1 q_2-2 q_1^2 q_2-2 q_1 q_2^2+4 q_1^2 q_2^2-2 q_1^3 q_2^2&\nonumber\\
-2 q_1^2 q_2^3-3 q_1^3 q_2^3+q_1^4 q_2^3+q_1^3 q_2^4)&(e^{\phi}+e^{-\phi})^{{2}}\nonumber\\
+q_1^2q_2^2(1-q_1)(1-q_2)(1+q_1 q_2)^2 (1+q_1^2 q_2) (1+q_1 q_2^2)&,
\end{align*}
\begin{align*}
\mathcal{N}^{(3)}=3 q_1^5 q_2^5(e^{\phi}+e^{-\phi})^{{7}}
+q_1^3 q_2^3 (1-2 q_1+q_1^2-2 q_2-q_1 q_2+2 q_1^2 q_2
+q_1^3 q_2+q_2^2&+2 q_1 q_2^2-15 q_1^2 q_2^2\nonumber\\
+2 q_1^3 q_2^2+q_1^4 q_2^2+q_1 q_2^3+2 q_1^2 q_2^3-q_1^3 q_2^3-2 q_1^4 q_2^3+q_1^2 q_2^4-2 q_1^3 q_2^4+q_1^4 q_2^4)&(e^{\phi}+e^{-\phi})^{{5}}\nonumber\\
\quad +q_1^2 q_2^2 (2+2 q_1-q_1^2+2 q_2-3 q_1 q_2+6 q_1^2 q_2-5 q_1^3 q_2-q_2^2+6 q_1 q_2^2+2 q_1^2& q_2^2\nonumber\\
-9 q_1^3 q_2^2-4 q_1^4 q_2^2-5 q_1 q_2^3-9 q_1^2 q_2^3+22 q_1^3 q_2^3-9 q_1^4 q_2^3-5 q_1^5 q_2^3-4 q_1^2 q_2^4&\nonumber\\
-9 q_1^3 q_2^4+2 q_1^4 q_2^4+6 q_1^5 q_2^4-q_1^6 q_2^4-5 q_1^3 q_2^5+6 q_1^4 q_2^5-3 q_1^5 q_2^5&\nonumber\\+2 q_1^6 q_2^5-q_1^4 q_2^6+2 q_1^5 q_2^6+2 q_1^6 q_2^6)&(e^{\phi}+e^{-\phi})^{{3}}\nonumber\\
-q_1q_2(1-q_1) (1-q_2) (1+q_1 q_2) (2+q_1+q_2+2 q_1^2 q_2+2 q_1 q_2^2+2 q_1^2 q_2^2&\nonumber\\
-q_1^4 q_2^2+6 q_1^3 q_2^3-q_1^2 q_2^4+2 q_1^4 q_2^4+2 q_1^5 q_2^4+2 q_1^4 q_2^5+q_1^6 q_2^5+q_1^5 q_2^6+2 q_1^6 q_2^6)&(e^{\phi}+e^{-\phi}),
\end{align*}
\begin{align*}
\mathcal{N}^{(4)}=6 q_1^5 q_2^5(e^{\phi}+e^{-\phi})^{{8}}
+q_1^3 q_2^3 (1-5 q_1+q_1^2+q_1^3-5 q_2-6 q_1 q_2+6 q_1^2 q_2+2 q_1^3 q_2&\nonumber\\
+q_1^4 q_2+q_2^2+6 q_1 q_2^2-34 q_1^2 q_2^2+6 q_1^3 q_2^2+q_1^4 q_2^2+q_2^3+2 q_1 q_2^3&\nonumber\\
+6 q_1^2 q_2^3-6 q_1^3 q_2^3-5 q_1^4 q_2^3+q_1 q_2^4+q_1^2 q_2^4-5 q_1^3 q_2^4+q_1^4 q_2^4)&(e^{\phi}+e^{-\phi})^{{6}}\nonumber\\
+q_1^2 q_2^2 (5+7 q_1-5 q_1^2-q_1^3+7 q_2-8 q_1 q_2+18 q_1^2 q_2-9 q_1^3 q_2-5 q_1^4 q_2+q_1^5 q_2&\nonumber\\
-5 q_2^2+18 q_1 q_2^2+21 q_1^2 q_2^2-27 q_1^3 q_2^2-4 q_1^4 q_2^2-5 q_1^5 q_2^2-q_2^3-9 q_1 q_2^3&\nonumber\\
-27 q_1^2 q_2^3+74 q_1^3 q_2^3-27 q_1^4 q_2^3-9 q_1^5 q_2^3-q_1^6 q_2^3-5 q_1 q_2^4-4 q_1^2 q_2^4&\nonumber\\
-27 q_1^3 q_2^4+21 q_1^4 q_2^4+18 q_1^5 q_2^4-5 q_1^6 q_2^4+q_1 q_2^5-5 q_1^2 q_2^5-9 q_1^3 q_2^5&\nonumber\\
+18 q_1^4 q_2^5-8 q_1^5 q_2^5+7 q_1^6 q_2^5-q_1^3 q_2^6-5 q_1^4 q_2^6+7 q_1^5 q_2^6+5 q_1^6 q_2^6)&(e^{\phi}+e^{-\phi})^{{4}}\nonumber\\
-q_1q_2(1-q_1)(1-q_2)(1+q_1 q_2) (7+5 q_1+5 q_2+2 q_1 q_2+14 q_1^2 q_2-q_1^3 q_2&\nonumber\\
-2 q_1^4 q_2+14 q_1 q_2^2+9 q_1^2 q_2^2+q_1^3 q_2^2+2 q_1^4 q_2^2-2 q_1^5 q_2^2-q_1 q_2^3+q_1^2 q_2^3&\nonumber\\
+36 q_1^3 q_2^3+q_1^4 q_2^3-q_1^5 q_2^3-2 q_1 q_2^4+2 q_1^2 q_2^4+q_1^3 q_2^4+9 q_1^4 q_2^4+14 q_1^5 q_2^4&\nonumber\\
-2 q_1^2 q_2^5-q_1^3 q_2^5+14 q_1^4 q_2^5+2 q_1^5 q_2^5+5 q_1^6 q_2^5+5 q_1^5 q_2^6+7 q_1^6 q_2^6)&(e^{\phi}+e^{-\phi})^{{2}}\nonumber\\
+(1-q_1)^2 (1-q_2)^2 (1+q_1 q_2)^2 (1+q_1^2 q_2) (1+q_1 q_2^2) (2+q_1+q_2+q_1 q_2&\nonumber\\
+q_1^2 q_2+q_1 q_2^2+q_1^2 q_2^2+q_1^3 q_2^2+q_1^2 q_2^3+2 q_1^3 q_2^3)&,
\end{align*}
\begin{align*}
\mathcal{N}^{(5)}=10 q_1^5 q_2^5(e^{\phi}+e^{-\phi})^{{9}}
+q_1^3 q_2^3 (1-9 q_1+q_1^2+q_1^3+q_1^4-9 q_2-9 q_1 q_2-4 q_1^2 q_2&\nonumber\\
+16 q_1^3 q_2+q_1^4 q_2+q_2^2-4 q_1 q_2^2-29 q_1^2 q_2^2-4 q_1^3 q_2^2+q_1^4 q_2^2+q_2^3
+16 q_1 q_2^3&\nonumber\\
-4 q_1^2 q_2^3-9 q_1^3 q_2^3-9 q_1^4 q_2^3+q_2^4+q_1 q_2^4+q_1^2 q_2^4-9 q_1^3 q_2^4+q_1^4 q_2^4)&(e^{\phi}+e^{-\phi})^{{7}}\nonumber\\
+q_1^2 q_2^2 (9+10 q_1+5 q_1^2-15 q_1^3+q_1^5+10 q_2+11 q_1 q_2+2 q_1^2 q_2-8 q_1^3 q_2&\nonumber\\
-3 q_1^4 q_2-3 q_1^5 q_2+q_1^6 q_2+5 q_2^2+2 q_1 q_2^2+27 q_1^2 q_2^2+52 q_1^3 q_2^2-78 &q_1^4 q_2^2\nonumber\\
-3 q_1^5 q_2^2-15 q_2^3-8 q_1 q_2^3+52 q_1^2 q_2^3-48 q_1^3 q_2^3+52 q_1^4 q_2^3-8 q_1^5 q_2^3&\nonumber\\
-15 q_1^6 q_2^3-3 q_1 q_2^4-78 q_1^2 q_2^4+52 q_1^3 q_2^4+27 q_1^4 q_2^4+2 q_1^5 q_2^4+5 q_1^6& q_2^4\nonumber\\
+q_2^5-3 q_1 q_2^5-3 q_1^2 q_2^5-8 q_1^3 q_2^5+2 q_1^4 q_2^5+11 q_1^5 q_2^5+10 q_1^6 q_2^5&\nonumber\\
+q_1 q_2^6-15 q_1^3 q_2^6+5 q_1^4 q_2^6+10 q_1^5 q_2^6+9 q_1^6 q_2^6)&(e^{\phi}+e^{-\phi})^{{5}}\nonumber\\
-q_1q_2(1-q_1) (1-q_2) (10+17 q_1+9 q_1^2-4 q_1^3-2 q_1^4+17 q_2+30 q_1 q_2&\nonumber\\
+14 q_1^2 q_2+38 q_1^3 q_2-13 q_1^4 q_2-6 q_1^5 q_2+9 q_2^2+14 q_1 q_2^2+84 q_1^2 q_2^2+30 q_1^3 q_2^2&\nonumber\\
-9 q_1^4 q_2^2-2 q_1^5 q_2^2-6 q_1^6 q_2^2-4 q_2^3+38 q_1 q_2^3+30 q_1^2 q_2^3+16 q_1^3 q_2^3+114 q_1^4 q_2^3&\nonumber\\
-9 q_1^5 q_2^3-13 q_1^6 q_2^3-2 q_1^7 q_2^3-2 q_2^4-13 q_1 q_2^4-9 q_1^2 q_2^4+114 q_1^3 q_2^4+16 q_1^4 q_2^4&\nonumber\\
+30 q_1^5 q_2^4+38 q_1^6 q_2^4-4 q_1^7 q_2^4-6 q_1 q_2^5-2 q_1^2 q_2^5-9 q_1^3 q_2^5+30 q_1^4 q_2^5+84 q_1^5 q_2^5&\nonumber\\
+14 q_1^6 q_2^5+9 q_1^7 q_2^5-6 q_1^2 q_2^6-13 q_1^3 q_2^6+38 q_1^4 q_2^6+14 q_1^5 q_2^6+30 q_1^6 q_2^6&\nonumber\\
+17 q_1^7 q_2^6-2 q_1^3 q_2^7-4 q_1^4 q_2^7+9 q_1^5 q_2^7+17 q_1^6 q_2^7+10 q_1^7 q_2^7)&(e^{\phi}+e^{-\phi})^{{3}}\nonumber\\
+(-1+q_1)^2 (-1+q_2)^2 (1+q_1 q_2) (1+9 q_1+4 q_1^2+q_1^3+9 q_2+10 q_1 q_2&\nonumber\\
+7 q_1^2 q_2+15 q_1^3 q_2+4 q_1^4 q_2+4 q_2^2+7 q_1 q_2^2+30 q_1^2 q_2^2+18 q_1^3 q_2^2+11 q_1^4 q_2^2&\nonumber\\
+5 q_1^5 q_2^2+q_2^3+15 q_1 q_2^3+18 q_1^2 q_2^3+19 q_1^3 q_2^3+37 q_1^4 q_2^3+11 q_1^5 q_2^3+4 q_1^6 q_2^3&\nonumber\\
+4 q_1 q_2^4+11 q_1^2 q_2^4+37 q_1^3 q_2^4+19 q_1^4 q_2^4+18 q_1^5 q_2^4+15 q_1^6 q_2^4+q_1^7 q_2^4+5 q_1^2 q_2^5&\nonumber\\
+11 q_1^3 q_2^5+18 q_1^4 q_2^5+30 q_1^5 q_2^5+7 q_1^6 q_2^5+4 q_1^7 q_2^5+4 q_1^3 q_2^6+15 q_1^4 q_2^6+7 q_1^5 q_2^6&\nonumber\\
+10 q_1^6 q_2^6+9 q_1^7 q_2^6+q_1^4 q_2^7+4 q_1^5 q_2^7+9 q_1^6 q_2^7+q_1^7 q_2^7)&(e^{\phi}+e^{-\phi}).
\end{align*}

\section{Refined invariants}\label{sec:BPS}
We list the (pseudo)-BPS invariants $\widetilde{N}_{j_L,j_R}^{C}$ of the Wilson loop expectation values computed from the refined holomorphic anomaly equation, in the expansion
\begin{align}\label{eq:appbps}
    \langle W_{\mathbf{r}}\rangle =\exp\left(\mathcal{W}_{\mathbf{r}}\right)=\sum_{d=1}^{\infty}\sum_{j_L,j_R}(-1)^{2j_L+2j_R}\widetilde{N}_{j_L,j_R}^{C}{\chi_{j_L}(\epsilon_-)\chi_{j_R}(\epsilon_+)}e^{-d t},
\end{align}
As we explained in Section \ref{sec:Wilson_general}, the expansion \eqref{eq:appbps} is the most convenient but is not the proper BPS expansion for decomposable representations, so the pseudo-BPS invariant $\widetilde{N}_{j_L,j_R}^{C}$ may not be always positive integer. Negative invariants can be found in Table \ref{tabel:9} and Table \ref{tabel:10}. 
\begin{table}[!h]
\begin{center} 
\footnotesize
{\footnotesize\begin{tabular}{|c|c|}
 \hline $2j_L \backslash 2j_R$ & 0 \\ \hline
 0 & 1 \\ \hline
 \noalign{\vskip 2mm} \multispan{2} $d$\,=\,$-\frac{1}{2}$ \\
\end{tabular}}\hspace{0.5cm}
{\footnotesize\begin{tabular}{|c|c|}
 \hline $2j_L \backslash 2j_R$ & 0 \\ \hline
 0 & 2 \\ \hline
  \noalign{\vskip 2mm} \multispan{2} $d$\,=\,$\frac{1}{2}$ \\
\end{tabular}} \hspace{0.5cm}
{\footnotesize\begin{tabular}{|c|ccc|}
 \hline $2j_L \backslash 2j_R$ & 0 & 1 & 2 \\ \hline
 0 &  &  & 1 \\ \hline
  \noalign{\vskip 2mm} \multispan{4} $d$\,=\,$\frac{3}{2}$ \\
\end{tabular}}\hspace{0.5cm}
{\footnotesize\begin{tabular}{|c|ccccc|}
 \hline $2j_L \backslash 2j_R$ & 0 & 1 & 2 & 3 & 4 \\ \hline
 0 &  &  &  &  & 2 \\ \hline
 \noalign{\vskip 2mm} \multispan{6} $d$\,=\,$\frac{5}{2}$ \\
\end{tabular}} \vspace{10pt}

{\footnotesize\begin{tabular}{|c|cccccccc|}
 \hline $2j_L \backslash 2j_R$ & 0 & 1 & 2 & 3 & 4 & 5 & 6 & 7 \\ \hline
 0 &  &  &  &  & 1 &  & 4 &  \\
 1 &  &  &  &  &  &  &  & 1 \\ \hline
\end{tabular}} \vskip 3pt  $d=\frac{7}{2}$ \vskip 10pt

{\footnotesize\begin{tabular}{|c|ccccccccccc|}
 \hline $2j_L \backslash 2j_R$ & 0 & 1 & 2 & 3 & 4 & 5 & 6 & 7 & 8 & 9 & 10 \\ \hline
 0 &  &  &  &  & 2 &  & 4 &  & 8 &  &  \\
 1 &  &  &  &  &  &  &  & 2 &  & 4 &  \\
 2 &  &  &  &  &  &  &  &  &  &  & 2 \\ \hline
\end{tabular}} \vskip 3pt  $d=\frac{9}{2}$ \vskip 10pt

{\footnotesize\begin{tabular}{|c|ccccccccccccccc|}
 \hline $2j_L \backslash 2j_R$ & 0 & 1 & 2 & 3 & 4 & 5 & 6 & 7 & 8 & 9 & 10 & 11 & 12 & 13 & 14 \\ \hline
 0 &  &  & 1 &  & 4 &  & 9 &  & 12 &  & 17 &  & 1 &  &  \\
 1 &  &  &  &  &  & 1 &  & 5 &  & 10 &  & 12 &  & 1 &  \\
 2 &  &  &  &  &  &  &  &  & 1 &  & 5 &  & 9 &  &  \\
 3 &  &  &  &  &  &  &  &  &  &  &  & 1 &  & 4 &  \\
 4 &  &  &  &  &  &  &  &  &  &  &  &  &  &  & 1 \\ \hline
\end{tabular}} \vskip 3pt  $d=\frac{11}{2}$ \vskip 10pt

{\footnotesize\begin{tabular}{|c|ccccccccccccccccccc|}
 \hline $2j_L \backslash 2j_R$ & 0 & 1 & 2 & 3 & 4 & 5 & 6 & 7 & 8 & 9 & 10 & 11 & 12 & 13 & 14 & 15 & 16 & 17 & 18 \\ \hline
 0 & 2 &  & 4 &  & 12 &  & 20 &  & 32 &  & 36 &  & 40 &  & 6 &  & 2 &  &  \\
 1 &  &  &  & 2 &  & 6 &  & 16 &  & 28 &  & 38 &  & 38 &  & 8 &  &  &  \\
 2 &  &  &  &  &  &  & 2 &  & 6 &  & 18 &  & 28 &  & 32 &  & 4 &  &  \\
 3 &  &  &  &  &  &  &  &  &  & 2 &  & 6 &  & 16 &  & 20 &  & 2 &  \\
 4 &  &  &  &  &  &  &  &  &  &  &  &  & 2 &  & 6 &  & 12 &  &  \\
 5 &  &  &  &  &  &  &  &  &  &  &  &  &  &  &  & 2 &  & 4 &  \\
 6 &  &  &  &  &  &  &  &  &  &  &  &  &  &  &  &  &  &  & 2 \\ \hline
\end{tabular}} \vskip 3pt  $d=\frac{13}{2}$ \vskip 10pt
 \end{center}
 \caption{BPS spectrum of Wilson loop for local $\mathbb{P}^1\times\mathbb{P}^1$ in the representation $\mathbf{2}$ of $SU(2)$ for the curve class $dt=2d\phi$.}\label{tabel:6}
\end{table}

\begin{table}[h]
\begin{center} 
\footnotesize
{\footnotesize\begin{tabular}{|c|c|}
 \hline $2j_L \backslash 2j_R$ & 0 \\ \hline
 0 & 1 \\ \hline
\noalign{\vskip 2mm} \multispan{2} $d$\,=\,$-1$ \\
\end{tabular}}\hspace{0.5cm}
{\footnotesize\begin{tabular}{|c|c|}
 \hline $2j_L \backslash 2j_R$ & 0 \\ \hline
 0 & 4 \\ \hline
\noalign{\vskip 2mm} \multispan{2} $d$\,=\,$0$ \\
\end{tabular}}\hspace{0.5cm}
{\footnotesize\begin{tabular}{|c|ccc|}
 \hline $2j_L \backslash 2j_R$ & 0 & 1 & 2 \\ \hline
 0 & 3 &  & 1 \\
 1 &  & 1 &  \\ \hline
 \noalign{\vskip 2mm} \multispan{4} $d$\,=\,$1$ \\
\end{tabular}} \hspace{0.5cm}
{\footnotesize\begin{tabular}{|c|ccccc|}
 \hline $2j_L \backslash 2j_R$ & 0 & 1 & 2 & 3 & 4 \\ \hline
 0 &  &  & 2 &  & 2 \\
 1 &  &  &  & 2 &  \\ \hline
 \noalign{\vskip 2mm} \multispan{6} $d$\,=\,$2$ \\
\end{tabular}} \vspace{10pt}

{\footnotesize\begin{tabular}{|c|cccccccc|}
 \hline $2j_L \backslash 2j_R$ & 0 & 1 & 2 & 3 & 4 & 5 & 6 & 7 \\ \hline
 0 & 1 &  &  &  & 5 &  & 4 &  \\
 1 &  &  &  & 1 &  & 4 &  & 1 \\
 2 &  &  &  &  &  &  & 1 &  \\ \hline
\end{tabular}} \vskip 3pt  $d=3$ \vskip 10pt

{\footnotesize\begin{tabular}{|c|ccccccccccc|}
 \hline $2j_L \backslash 2j_R$ & 0 & 1 & 2 & 3 & 4 & 5 & 6 & 7 & 8 & 9 & 10 \\ \hline
 0 &  &  & 2 &  & 4 &  & 12 &  & 10 &  &  \\
 1 &  &  &  & 2 &  & 4 &  & 12 &  & 4 &  \\
 2 &  &  &  &  &  &  & 2 &  & 4 &  & 2 \\
 3 &  &  &  &  &  &  &  &  &  & 2 &  \\ \hline
\end{tabular}} \vskip 3pt  $d=4$ \vskip 10pt

{\footnotesize\begin{tabular}{|c|ccccccccccccccc|}
 \hline $2j_L \backslash 2j_R$ & 0 & 1 & 2 & 3 & 4 & 5 & 6 & 7 & 8 & 9 & 10 & 11 & 12 & 13 & 14 \\ \hline
 0 & 3 &  & 2 &  & 12 &  & 18 &  & 33 &  & 25 &  & 1 &  &  \\
 1 &  & 1 &  & 4 &  & 11 &  & 21 &  & 35 &  & 17 &  & 1 &  \\
 2 &  &  &  &  & 1 &  & 5 &  & 11 &  & 21 &  & 10 &  &  \\
 3 &  &  &  &  &  &  &  & 1 &  & 5 &  & 10 &  & 4 &  \\
 4 &  &  &  &  &  &  &  &  &  &  & 1 &  & 4 &  & 1 \\
 5 &  &  &  &  &  &  &  &  &  &  &  &  &  & 1 &  \\ \hline
\end{tabular}} \vskip 3pt  $d=5$ \vskip 10pt

{\footnotesize\begin{tabular}{|c|ccccccccccccccccccc|}
 \hline $2j_L \backslash 2j_R$ & 0 & 1 & 2 & 3 & 4 & 5 & 6 & 7 & 8 & 9 & 10 & 11 & 12 & 13 & 14 & 15 & 16 & 17 & 18 \\ \hline
 0 & 2 &  & 14 &  & 26 &  & 52 &  & 72 &  & 94 &  & 68 &  & 10 &  & 2 &  &  \\
 1 &  & 4 &  & 16 &  & 30 &  & 62 &  & 88 &  & 114 &  & 64 &  & 10 &  &  &  \\
 2 &  &  & 2 &  & 6 &  & 18 &  & 38 &  & 66 &  & 86 &  & 48 &  & 4 &  &  \\
 3 &  &  &  &  &  & 2 &  & 6 &  & 20 &  & 36 &  & 56 &  & 26 &  & 2 &  \\
 4 &  &  &  &  &  &  &  &  & 2 &  & 6 &  & 18 &  & 28 &  & 14 &  &  \\
 5 &  &  &  &  &  &  &  &  &  &  &  & 2 &  & 6 &  & 14 &  & 4 &  \\
 6 &  &  &  &  &  &  &  &  &  &  &  &  &  &  & 2 &  & 4 &  & 2 \\
 7 &  &  &  &  &  &  &  &  &  &  &  &  &  &  &  &  &  & 2 &  \\ \hline
\end{tabular}} \vskip 3pt  $d=6$ \vskip 10pt
\end{center}
 \caption{Pseudo-BPS spectrum of Wilson loop for local $\mathbb{P}^1\times\mathbb{P}^1$ in the representation $\mathbf{2}\otimes\mathbf{2}$ of $SU(2)$ for the curve class $dt=2d\phi$.}\label{tabel:7}
\end{table}

\begin{table}[h]
\begin{center} 
\footnotesize
{\footnotesize\begin{tabular}{|c|c|}
 \hline $2j_L \backslash 2j_R$ & 0 \\ \hline
 0 & 1 \\ \hline
  \noalign{\vskip 2mm} \multispan{2} $d$\,=\,$-\frac{3}{2}$ \\
\end{tabular}} \hspace{0.5cm}
{\footnotesize\begin{tabular}{|c|c|}
 \hline $2j_L \backslash 2j_R$ & 0 \\ \hline
 0 & 6 \\ \hline
  \noalign{\vskip 2mm} \multispan{2} $d$\,=\,$-\frac{1}{2}$ \\
\end{tabular}} \hspace{0.5cm}
{\footnotesize\begin{tabular}{|c|ccc|}
 \hline $2j_L \backslash 2j_R$ & 0 & 1 & 2 \\ \hline
 0 & 11 &  & 1 \\
 1 &  & 1 &  \\
 2 & 1 &  &  \\ \hline
  \noalign{\vskip 2mm} \multispan{4} $d$\,=\,$\frac{1}{2}$ \\
\end{tabular}}\vspace{10pt} 

{\footnotesize\begin{tabular}{|c|ccccc|}
 \hline $2j_L \backslash 2j_R$ & 0 & 1 & 2 & 3 & 4 \\ \hline
 0 & 4 &  & 6 &  & 2 \\
 1 &  & 2 &  & 2 &  \\
 2 &  &  & 2 &  &  \\ \hline
\end{tabular}} \vskip 3pt  $d=\frac{3}{2}$ \vskip 10pt

{\footnotesize\begin{tabular}{|c|cccccccc|}
 \hline $2j_L \backslash 2j_R$ & 0 & 1 & 2 & 3 & 4 & 5 & 6 & 7 \\ \hline
 0 & 1 &  & 3 &  & 14 &  & 4 &  \\
 1 &  & 1 &  & 5 &  & 4 &  & 1 \\
 2 &  &  & 1 &  & 4 &  & 1 &  \\
 3 &  &  &  &  &  & 1 &  &  \\ \hline
\end{tabular}} \vskip 3pt  $d=\frac{5}{2}$ \vskip 10pt

{\footnotesize\begin{tabular}{|c|ccccccccccc|}
 \hline $2j_L \backslash 2j_R$ & 0 & 1 & 2 & 3 & 4 & 5 & 6 & 7 & 8 & 9 & 10 \\ \hline
 0 & 2 &  & 2 &  & 16 &  & 30 &  & 10 &  &  \\
 1 &  & 2 &  & 6 &  & 12 &  & 18 &  & 4 &  \\
 2 &  &  & 2 &  & 4 &  & 12 &  & 4 &  & 2 \\
 3 &  &  &  &  &  & 2 &  & 4 &  & 2 &  \\
 4 &  &  &  &  &  &  &  &  & 2 &  &  \\ \hline
\end{tabular}} \vskip 3pt  $d=\frac{7}{2}$ \vskip 10pt

{\footnotesize\begin{tabular}{|c|ccccccccccccccc|}
 \hline $2j_L \backslash 2j_R$ & 0 & 1 & 2 & 3 & 4 & 5 & 6 & 7 & 8 & 9 & 10 & 11 & 12 & 13 & 14 \\ \hline
 0 & 3 &  & 7 &  & 27 &  & 52 &  & 73 &  & 26 &  & 1 &  &  \\
 1 &  & 5 &  & 13 &  & 24 &  & 54 &  & 60 &  & 17 &  & 1 &  \\
 2 & 1 &  & 4 &  & 11 &  & 22 &  & 36 &  & 34 &  & 10 &  &  \\
 3 &  &  &  & 1 &  & 5 &  & 11 &  & 21 &  & 11 &  & 4 &  \\
 4 &  &  &  &  &  &  & 1 &  & 5 &  & 10 &  & 4 &  & 1 \\
 5 &  &  &  &  &  &  &  &  &  & 1 &  & 4 &  & 1 &  \\
 6 &  &  &  &  &  &  &  &  &  &  &  &  & 1 &  &  \\ \hline
\end{tabular}} \vskip 3pt  $d=\frac{9}{2}$ \vskip 10pt

{\footnotesize\begin{tabular}{|c|ccccccccccccccccccc|}
 \hline $2j_L \backslash 2j_R$ & 0 & 1 & 2 & 3 & 4 & 5 & 6 & 7 & 8 & 9 & 10 & 11 & 12 & 13 & 14 & 15 & 16 & 17 & 18 \\ \hline
 0 & 10 &  & 24 &  & 68 &  & 118 &  & 172 &  & 192 &  & 84 &  & 10 &  & 2 &  &  \\
 1 &  & 14 &  & 38 &  & 74 &  & 140 &  & 206 &  & 202 &  & 74 &  & 10 &  &  &  \\
 2 & 2 &  & 16 &  & 32 &  & 66 &  & 106 &  & 170 &  & 150 &  & 50 &  & 4 &  &  \\
 3 &  & 2 &  & 6 &  & 18 &  & 38 &  & 70 &  & 98 &  & 88 &  & 26 &  & 2 &  \\
 4 &  &  &  &  & 2 &  & 6 &  & 20 &  & 36 &  & 58 &  & 38 &  & 14 &  &  \\
 5 &  &  &  &  &  &  &  & 2 &  & 6 &  & 18 &  & 28 &  & 16 &  & 4 &  \\
 6 &  &  &  &  &  &  &  &  &  &  & 2 &  & 6 &  & 14 &  & 4 &  & 2 \\
 7 &  &  &  &  &  &  &  &  &  &  &  &  &  & 2 &  & 4 &  & 2 &  \\
 8 &  &  &  &  &  &  &  &  &  &  &  &  &  &  &  &  & 2 &  &  \\ \hline
\end{tabular}} \vskip 3pt  $d=\frac{11}{2}$ \vskip 10pt
\end{center}
 \caption{Pseudo-BPS spectrum of Wilson loop for local $\mathbb{P}^1\times\mathbb{P}^1$ in the representation $\mathbf{2}\otimes\mathbf{2}\otimes\mathbf{2}$ of $SU(2)$ for the curve class $dt=2d\phi$.}\label{tabel:8}
\end{table}

\begin{table}[h]
\begin{center} 
{\footnotesize\begin{tabular}{|c|c|}
 \hline $2j_L \backslash 2j_R$ & 0 \\ \hline
 0 & 1 \\ \hline
\noalign{\vskip 2mm} \multispan{2} $d$\,=\,$-2$ \\
\end{tabular}}\hspace{0.5cm}
{\footnotesize\begin{tabular}{|c|c|}
 \hline $2j_L \backslash 2j_R$ & 0 \\ \hline
 0 & 8 \\ \hline
 \noalign{\vskip 2mm} \multispan{2} $d$\,=\,$-1$ \\
\end{tabular}}\vspace{10pt} 

{\footnotesize\begin{tabular}{|c|ccc|}
 \hline $2j_L \backslash 2j_R$ & 0 & 1 & 2 \\ \hline
 0 & 26 &  & 2 \\
 1 &  & -2 &  \\
 2 & 4 &  &  \\ \hline
  \noalign{\vskip 2mm} \multispan{2} $d$\,=\,$0$ \\
\end{tabular}}\hspace{0.5cm}
{\footnotesize\begin{tabular}{|c|ccccc|}
 \hline $2j_L \backslash 2j_R$ & 0 & 1 & 2 & 3 & 4 \\ \hline
 0 & 22 &  & 8 &  & 2 \\
 1 &  & 8 &  & 2 &  \\
 2 & 2 &  & 2 &  &  \\
 3 &  & 2 &  &  &  \\ \hline
\noalign{\vskip 2mm} \multispan{2} $d$\,=\,$1$ \\
\end{tabular}}\vspace{10pt}

{\footnotesize\begin{tabular}{|c|cccccccc|}
 \hline $2j_L \backslash 2j_R$ & 0 & 1 & 2 & 3 & 4 & 5 & 6 & 7 \\ \hline
 0 & 7 &  & 15 &  & 18 &  & 4 &  \\
 1 &  & 4 &  & 18 &  & 4 &  & 1 \\
 2 &  &  & 6 &  & 4 &  & 1 &  \\
 3 &  & 1 &  & 4 &  & 1 &  &  \\
 4 &  &  &  &  & 1 &  &  &  \\ \hline
\end{tabular}} \vskip 3pt  $d=2$ \vskip 10pt

{\footnotesize\begin{tabular}{|c|ccccccccccc|}
 \hline $2j_L \backslash 2j_R$ & 0 & 1 & 2 & 3 & 4 & 5 & 6 & 7 & 8 & 9 & 10 \\ \hline
 0 & 8 &  & 8 &  & 44 &  & 38 &  & 10 &  &  \\
 1 &  & 4 &  & 20 &  & 40 &  & 20 &  & 4 &  \\
 2 & 2 &  & 6 &  & 14 &  & 20 &  & 4 &  & 2 \\
 3 &  & 2 &  & 4 &  & 12 &  & 4 &  & 2 &  \\
 4 &  &  &  &  & 2 &  & 4 &  & 2 &  &  \\
 5 &  &  &  &  &  &  &  & 2 &  &  &  \\ \hline
\end{tabular}} \vskip 3pt  $d=3$ \vskip 10pt

{\footnotesize\begin{tabular}{|c|ccccccccccccccc|}
 \hline $2j_L \backslash 2j_R$ & 0 & 1 & 2 & 3 & 4 & 5 & 6 & 7 & 8 & 9 & 10 & 11 & 12 & 13 & 14 \\ \hline
 0 & 9 &  & 21 &  & 59 &  & 113 &  & 105 &  & 26 &  & 1 &  &  \\
 1 &  & 10 &  & 37 &  & 70 &  & 131 &  & 69 &  & 17 &  & 1 &  \\
 2 & 2 &  & 16 &  & 25 &  & 63 &  & 70 &  & 38 &  & 10 &  &  \\
 3 &  & 4 &  & 11 &  & 22 &  & 37 &  & 38 &  & 11 &  & 4 &  \\
 4 &  &  & 1 &  & 5 &  & 11 &  & 21 &  & 11 &  & 4 &  & 1 \\
 5 &  &  &  &  &  & 1 &  & 5 &  & 10 &  & 4 &  & 1 &  \\
 6 &  &  &  &  &  &  &  &  & 1 &  & 4 &  & 1 &  &  \\
 7 &  &  &  &  &  &  &  &  &  &  &  & 1 &  &  &  \\ \hline
\end{tabular}} \vskip 3pt  $d=4$ 
\end{center}
 \caption{Pseudo-BPS spectrum of Wilson loop for local $\mathbb{P}^1\times\mathbb{P}^1$ in the representation $\mathbf{2}^{\otimes 4}$ of $SU(2)$ for the curve class $dt=2d\phi$.}\label{tabel:9}
\end{table}

\begin{table}[h]
\begin{center} 
{\footnotesize\begin{tabular}{|c|c|}
 \hline $2j_L \backslash 2j_R$ & 0 \\ \hline
 0 & 1 \\ \hline
 \noalign{\vskip 2mm} \multispan{2} $d$\,=\,$-\frac{5}{2}$ \\
\end{tabular}}
\hspace{0.5cm}
{\footnotesize\begin{tabular}{|c|c|}
 \hline $2j_L \backslash 2j_R$ & 0 \\ \hline
 0 & 10 \\ \hline
  \noalign{\vskip 2mm} \multispan{2} $d$\,=\,$-\frac{3}{2}$ \\
\end{tabular}}
\hspace{0.5cm}
{\footnotesize\begin{tabular}{|c|ccc|}
 \hline $2j_L \backslash 2j_R$ & 0 & 1 & 2 \\ \hline
 0 & 50 &  & 5 \\
 1 &  & -10 &  \\
 2 & 10 &  &  \\ \hline
  \noalign{\vskip 2mm} \multispan{2} $d$\,=\,$-\frac{1}{2}$ \\
\end{tabular}}\vspace{10pt}

{\footnotesize\begin{tabular}{|c|ccccc|}
 \hline $2j_L \backslash 2j_R$ & 0 & 1 & 2 & 3 & 4 \\ \hline
 0 & 90 &  & 18 &  & 2 \\
 1 &  & -12 &  & 2 &  \\
 2 & 28 &  & 2 &  &  \\
 3 &  & 2 &  &  &  \\
 4 & 2 &  &  &  &  \\ \hline
\end{tabular}} \vskip 3pt  $d=\frac{1}{2}$ \vskip 10pt

{\footnotesize\begin{tabular}{|c|cccccccc|}
 \hline $2j_L \backslash 2j_R$ & 0 & 1 & 2 & 3 & 4 & 5 & 6 & 7 \\ \hline
 0 & 43 &  & 54 &  & 26 &  & 4 &  \\
 1 &  & 10 &  & 11 &  & 4 &  & 1 \\
 2 & 3 &  & 32 &  & 4 &  & 1 &  \\
 3 &  & 5 &  & 4 &  & 1 &  &  \\
 4 & 1 &  & 4 &  & 1 &  &  &  \\
 5 &  &  &  & 1 &  &  &  &  \\ \hline
\end{tabular}} \vskip 3pt  $d=\frac{3}{2}$ \vskip 10pt

{\footnotesize\begin{tabular}{|c|ccccccccccc|}
 \hline $2j_L \backslash 2j_R$ & 0 & 1 & 2 & 3 & 4 & 5 & 6 & 7 & 8 & 9 & 10 \\ \hline
 0 & 18 &  & 40 &  & 134 &  & 54 &  & 10 &  &  \\
 1 &  & 18 &  & 36 &  & 26 &  & 22 &  & 4 &  \\
 2 & 2 &  & 24 &  & 68 &  & 22 &  & 4 &  & 2 \\
 3 &  & 6 &  & 14 &  & 22 &  & 4 &  & 2 &  \\
 4 & 2 &  & 4 &  & 12 &  & 4 &  & 2 &  &  \\
 5 &  &  &  & 2 &  & 4 &  & 2 &  &  &  \\
 6 &  &  &  &  &  &  & 2 &  &  &  &  \\ \hline
\end{tabular}} \vskip 3pt  $d=\frac{5}{2}$ \vskip 10pt

{\footnotesize\begin{tabular}{|c|ccccccccccccccc|}
 \hline $2j_L \backslash 2j_R$ & 0 & 1 & 2 & 3 & 4 & 5 & 6 & 7 & 8 & 9 & 10 & 11 & 12 & 13 & 14 \\ \hline
 0 & 25 &  & 39 &  & 189 &  & 292 &  & 130 &  & 26 &  & 1 &  &  \\
 1 &  & 34 &  & 80 &  & 104 &  & 176 &  & 81 &  & 17 &  & 1 &  \\
 2 & 8 &  & 44 &  & 93 &  & 188 &  & 68 &  & 42 &  & 10 &  &  \\
 3 &  & 13 &  & 28 &  & 67 &  & 88 &  & 42 &  & 11 &  & 4 &  \\
 4 & 3 &  & 11 &  & 22 &  & 37 &  & 42 &  & 11 &  & 4 &  & 1 \\
 5 &  & 1 &  & 5 &  & 11 &  & 21 &  & 11 &  & 4 &  & 1 &  \\
 6 &  &  &  &  & 1 &  & 5 &  & 10 &  & 4 &  & 1 &  &  \\
 7 &  &  &  &  &  &  &  & 1 &  & 4 &  & 1 &  &  &  \\
 8 &  &  &  &  &  &  &  &  &  &  & 1 &  &  &  &  \\ \hline
\end{tabular}} \vskip 3pt  $d=\frac{7}{2}$
\end{center}
 \caption{Pseudo-BPS spectrum of Wilson loop for local $\mathbb{P}^1\times\mathbb{P}^1$ in the representation $\mathbf{2}^{\otimes 5}$ of $SU(2)$ for the curve class $dt=2d\phi$.}\label{tabel:10}
\end{table}
\begin{table}[!h]
\begin{center} 
\footnotesize
{\footnotesize\begin{tabular}{|c|c|}
 \hline $2j_L \backslash 2j_R$ & 0 \\ \hline
 0 & 1 \\ \hline
\end{tabular}} \vskip 3pt  $d=-\frac{1}{3}$ \vskip 10pt

{\footnotesize\begin{tabular}{|c|cc|}
 \hline $2j_L \backslash 2j_R$ & 0 & 1 \\ \hline
 0 &  & 1 \\ \hline
\end{tabular}} \vskip 3pt  $d=\frac{2}{3}$ \vskip 10pt

{\footnotesize\begin{tabular}{|c|ccccc|}
 \hline $2j_L \backslash 2j_R$ & 0 & 1 & 2 & 3 & 4 \\ \hline
 0 &  &  &  &  & 1 \\ \hline
\end{tabular}} \vskip 3pt  $d=\frac{5}{3}$ \vskip 10pt

{\footnotesize\begin{tabular}{|c|ccccccccc|}
 \hline $2j_L \backslash 2j_R$ & 0 & 1 & 2 & 3 & 4 & 5 & 6 & 7 & 8 \\ \hline
 0 &  &  &  &  &  & 1 &  & 1 &  \\
 1 &  &  &  &  &  &  &  &  & 1 \\ \hline
\end{tabular}} \vskip 3pt  $d=\frac{8}{3}$ \vskip 10pt

{\footnotesize\begin{tabular}{|c|cccccccccccccc|}
 \hline $2j_L \backslash 2j_R$ & 0 & 1 & 2 & 3 & 4 & 5 & 6 & 7 & 8 & 9 & 10 & 11 & 12 & 13 \\ \hline
 0 &  &  &  &  & 1 &  & 1 &  & 2 &  & 1 &  & 1 &  \\
 1 &  &  &  &  &  &  &  & 1 &  & 2 &  & 2 &  &  \\
 2 &  &  &  &  &  &  &  &  &  &  & 1 &  & 1 &  \\
 3 &  &  &  &  &  &  &  &  &  &  &  &  &  & 1 \\ \hline
\end{tabular}} \vskip 3pt  $d=\frac{11}{3}$ \vskip 10pt

{\footnotesize\begin{tabular}{|c|cccccccccccccccccccc|}
 \hline $2j_L \backslash 2j_R$ & 0 & 1 & 2 & 3 & 4 & 5 & 6 & 7 & 8 & 9 & 10 & 11 & 12 & 13 & 14 & 15 & 16 & 17 & 18 & 19 \\ \hline
 0 &  & 1 &  & 1 &  & 2 &  & 3 &  & 4 &  & 5 &  & 4 &  & 3 &  &  &  &  \\
 1 &  &  &  &  & 1 &  & 2 &  & 4 &  & 5 &  & 7 &  & 5 &  & 2 &  &  &  \\
 2 &  &  &  &  &  &  &  & 1 &  & 2 &  & 4 &  & 5 &  & 4 &  & 2 &  &  \\
 3 &  &  &  &  &  &  &  &  &  &  & 1 &  & 2 &  & 4 &  & 3 &  & 1 &  \\
 4 &  &  &  &  &  &  &  &  &  &  &  &  &  & 1 &  & 2 &  & 2 &  &  \\
 5 &  &  &  &  &  &  &  &  &  &  &  &  &  &  &  &  & 1 &  & 1 &  \\
 6 &  &  &  &  &  &  &  &  &  &  &  &  &  &  &  &  &  &  &  & 1 \\ \hline
\end{tabular}} \vskip 3pt  $d=\frac{14}{3}$ \vskip 10pt
\end{center}
 \caption{BPS spectrum of Wilson loop for local $\mathbb{P}^2$ in the representation $[-1]$ for the curve class $dt$.}
 \label{tab:P2_z1}
\end{table}

\begin{table}[h]
\begin{center} 
\footnotesize
{\footnotesize\begin{tabular}{|c|c|}
 \hline $2j_L \backslash 2j_R$ & 0 \\ \hline
 0 & 1 \\ \hline
\end{tabular}} \vskip 3pt  $d=-\frac{2}{3}$ \vskip 10pt

{\footnotesize\begin{tabular}{|c|cc|}
 \hline $2j_L \backslash 2j_R$ & 0 & 1 \\ \hline
 0 &  & 1 \\
 1 & 1 &  \\ \hline
\end{tabular}} \vskip 3pt  $d=\frac{1}{3}$ \vskip 10pt

{\footnotesize\begin{tabular}{|c|ccccc|}
 \hline $2j_L \backslash 2j_R$ & 0 & 1 & 2 & 3 & 4 \\ \hline
 0 & 1 &  &  &  & 1 \\
 1 &  &  &  & 1 &  \\ \hline
\end{tabular}} \vskip 3pt  $d=\frac{4}{3}$ \vskip 10pt

{\footnotesize\begin{tabular}{|c|ccccccccc|}
 \hline $2j_L \backslash 2j_R$ & 0 & 1 & 2 & 3 & 4 & 5 & 6 & 7 & 8 \\ \hline
 0 &  &  &  & 1 &  & 1 &  & 1 &  \\
 1 &  &  &  &  & 1 &  & 1 &  & 1 \\
 2 &  &  &  &  &  &  &  & 1 &  \\ \hline
\end{tabular}} \vskip 3pt  $d=\frac{7}{3}$ \vskip 10pt

{\footnotesize\begin{tabular}{|c|cccccccccccccc|}
 \hline $2j_L \backslash 2j_R$ & 0 & 1 & 2 & 3 & 4 & 5 & 6 & 7 & 8 & 9 & 10 & 11 & 12 & 13 \\ \hline
 0 & 1 &  &  &  & 2 &  & 2 &  & 3 &  & 2 &  & 1 &  \\
 1 &  &  &  & 1 &  & 1 &  & 4 &  & 3 &  & 3 &  &  \\
 2 &  &  &  &  &  &  & 1 &  & 2 &  & 3 &  & 1 &  \\
 3 &  &  &  &  &  &  &  &  &  & 1 &  & 1 &  & 1 \\
 4 &  &  &  &  &  &  &  &  &  &  &  &  & 1 &  \\ \hline
\end{tabular}} \vskip 3pt  $d=\frac{10}{3}$ \vskip 10pt

{\footnotesize\begin{tabular}{|c|cccccccccccccccccccc|}
 \hline $2j_L \backslash 2j_R$ & 0 & 1 & 2 & 3 & 4 & 5 & 6 & 7 & 8 & 9 & 10 & 11 & 12 & 13 & 14 & 15 & 16 & 17 & 18 & 19 \\ \hline
 0 &  & 1 &  & 3 &  & 4 &  & 7 &  & 8 &  & 10 &  & 7 &  & 4 &  &  &  &  \\
 1 & 1 &  & 1 &  & 4 &  & 6 &  & 10 &  & 13 &  & 13 &  & 9 &  & 3 &  &  &  \\
 2 &  &  &  & 1 &  & 2 &  & 5 &  & 8 &  & 12 &  & 11 &  & 7 &  & 2 &  &  \\
 3 &  &  &  &  &  &  & 1 &  & 2 &  & 5 &  & 8 &  & 8 &  & 5 &  & 1 &  \\
 4 &  &  &  &  &  &  &  &  &  & 1 &  & 2 &  & 5 &  & 5 &  & 3 &  &  \\
 5 &  &  &  &  &  &  &  &  &  &  &  &  & 1 &  & 2 &  & 3 &  & 1 &  \\
 6 &  &  &  &  &  &  &  &  &  &  &  &  &  &  &  & 1 &  & 1 &  & 1 \\
 7 &  &  &  &  &  &  &  &  &  &  &  &  &  &  &  &  &  &  & 1 &  \\ \hline
\end{tabular}} \vskip 3pt  $d=\frac{13}{3}$ \vskip 10pt
\end{center}
 \caption{Pseudo-BPS spectrum of Wilson loop for local $\mathbb{P}^2$ in the representation $[-1]\otimes [-1]$  for the curve class $dt$.}
 \label{tab:P2_z2}
\end{table}

\clearpage
\section{Wilson loop amplitudes}
\subsection{$\mathbb{P}^1\times\mathbb{P}^1$}\label{sec:C}
\paragraph{$(n,g)=(2,1)$}
\begin{dmath*}
\mathcal{W}^{(2,1)}_{\mathbf{2}^{\otimes \n}}=\frac{\n}{77760 z^6(1-16z)^3}(135 S^3-2160 S^3 z-135 S^2 z^2+8640 S^3 z^2+3240 S^2 z^3+45 S z^4-43200 S^2 z^4-1656 S
   z^5+207360 S^2 z^5-5 z^6+48384 S z^6+272 z^7-324864 S z^7-6032 z^8+1548288 S
   z^8+112896 z^9-129024 z^{10}).
\end{dmath*}
\paragraph{$(n,g)=(1,2)$}
\begin{dmath*}
\mathcal{W}^{(1,2)}_{\mathbf{2}}=\frac{1}{9720 z^8(1-16 z)^3}(-3240 S^4 z+405 S^4-56160 S^3 z^4+7020 S^3 z^3-405 S^3 z^2-311040 S^2 z^7+60480 S^2 z^6-2970 S^2 z^5+135 S^2 z^4-774144 S z^{10}+196992 S z^9-6912 S z^8+228 S z^7-15 S z^6+64512 z^{12}+6912 z^{11}-1904 z^{10}+34 z^9).
\end{dmath*}
\begin{dmath*}
\mathcal{W}^{(1,2)}_{\mathbf{2}\otimes\mathbf{2}}=\frac{1}{19440 z^8(1-16 z)^3}(-12960 S^4 z+1620 S^4-172800 S^3 z^4+18360 S^3 z^3-1215 S^3 z^2+8640 S^2 z^6+4320 S^2 z^5+135 S^2 z^4+221184 S z^{10}-248832 S z^9+110592 S z^8-6648 S z^7+75 S z^6-847872 z^{12}+338688 z^{11}-36176 z^{10}+1216 z^9-15 z^8).
\end{dmath*}
\begin{dmath*}
\mathcal{W}^{(1,2)}_{\mathbf{2}\otimes\mathbf{2}\otimes\mathbf{2}}=\frac{1}{6480 z^8(1-16 z)^3}(-6480 S^4 z+810 S^4-60480 S^3 z^4+4320 S^3 z^3-405 S^3 z^2+622080 S^2 z^7-112320 S^2 z^6+10260 S^2 z^5-135 S^2 z^4+1769472 S z^{10}-642816 S z^9+124416 S z^8-7104 S z^7+105 S z^6-976896 z^{12}+324864 z^{11}-32368 z^{10}+1148 z^9-15 z^8).
\end{dmath*}
\begin{dmath*}
\mathcal{W}^{(1,2)}_{\mathbf{2}^{\otimes 4}}=\frac{1}{9720 z^8(1-16 z)^3}(-12960 S^4 z+1620 S^4-69120 S^3 z^4-1080 S^3 z^3-405 S^3 z^2+2488320 S^2 z^7-457920 S^2 z^6+36720 S^2 z^5-675 S^2 z^4+6856704 S z^{10}-2322432 S z^9+387072 S z^8-21768 S z^7+345 S z^6-3059712 z^{12}+960768 z^{11}-93296 z^{10}+3376 z^9-45 z^8).
\end{dmath*}
\begin{dmath*}
\mathcal{W}^{(1,2)}_{\mathbf{2}^{\otimes 5}}=\frac{1}{1944 z^8(1-16 z)^3}(-3240 S^4 z+405 S^4-4320 S^3 z^4-2700 S^3 z^3+933120 S^2 z^7-172800 S^2 z^6+13230 S^2 z^5-270 S^2 z^4+2543616 S z^{10}-839808 S z^9+131328 S z^8-7332 S z^7+120 S z^6-1041408 z^{12}+317952 z^{11}-30464 z^{10}+1114 z^9-15 z^8).
\end{dmath*}
\paragraph{$(n,g)=(0,3)$}
\begin{dmath*}
\mathcal{W}^{(0,3)}_{\mathbf{2}}=\frac{1}{77760 z^{10}(1-16 z)^3}(24300 S^5+298080 S^4 z^3-27540 S^4 z^2+1486080 S^3 z^6-257040 S^3 z^5+12285 S^3 z^4+4147200 S^2 z^9-950400 S^2 z^8+91800 S^2 z^7-2925 S^2 z^6+5308416 S z^{12}-1686528 S z^{11}+260928 S z^{10}-19032 S z^9+435 S z^8-442368 z^{14}+225792 z^{13}-33824 z^{12}+1984 z^{11}-35 z^{10}).
\end{dmath*}
\begin{dmath*}
\mathcal{W}^{(0,3)}_{\mathbf{2}\otimes\mathbf{2}}=\frac{1}{38880 z^{10}(1-16 z)^3}(24300 S^5+142560 S^4 z^3-17820 S^4 z^2-587520 S^3 z^6-10800 S^3 z^5+4995 S^3 z^4-829440 S^2 z^9-17280 S^2 z^8+39960 S^2 z^7-2115 S^2 z^6+31850496 S z^{12}-7492608 S z^{11}+796608 S z^{10}-43512 S z^9+885 S z^8+79626240 z^{15}-24219648 z^{14}+3232512 z^{13}-232784 z^{12}+8704 z^{11}-125 z^{10}).
\end{dmath*}
\begin{dmath*}
\mathcal{W}^{(0,3)}_{\mathbf{2}\otimes\mathbf{2}\otimes\mathbf{2}}=\frac{1}{25920 z^{10}(1-16 z)^3}(24300 S^5-12960 S^4 z^3-8100 S^4 z^2-1831680 S^3 z^6+131760 S^3 z^5+945 S^3 z^4+14100480 S^2 z^9-2401920 S^2 z^8+169560 S^2 z^7-4545 S^2 z^6+217645056 S z^{12}-46476288 S z^{11}+3889728 S z^{10}-154392 S z^9+2415 S z^8+477757440 z^{15}-134258688 z^{14}+15501312 z^{13}-920384 z^{12}+27904 z^{11}-335 z^{10}).
\end{dmath*}
\begin{dmath*}
\mathcal{W}^{(0,3)}_{\mathbf{2}^{\otimes 4}}=\frac{1}{19440 z^{10}(1-16 z)^3}(24300 S^5-168480 S^4 z^3+1620 S^4 z^2-2246400 S^3 z^6+170640 S^3 z^5+135 S^3 z^4+48936960 S^2 z^9-8104320 S^2 z^8+480600 S^2 z^7-10215 S^2 z^6+562692096 S z^{12}-118637568 S z^{11}+9540288 S z^{10}-351672 S z^9+5025 S z^8+1194393600 z^{15}-330559488
   z^{14}+37032192 z^{13}-2096624 z^{12}+59584 z^{11}-665 z^{10}).
\end{dmath*}
\begin{dmath*}
\mathcal{W}^{(0,3)}_{\mathbf{2}^{\otimes 5}}=\frac{1}{15552 z^{10}(1-16 z)^3}(24300 S^5-324000 S^4 z^3+11340 S^4 z^2-1831680 S^3 z^6+105840 S^3 z^5+2565 S^3 z^4+103680000 S^2 z^9-17124480 S^2 z^8+973080 S^2 z^7-19125 S^2 z^6+1066991616 S z^{12}-223976448 S z^{11}+17748288 S z^{10}-635352 S z^9+8715 S z^8+2229534720 z^{15}-613122048 z^{14}+67825152 z^{13}-3761504 z^{12}+103744 z^{11}-1115 z^{10}).
\end{dmath*}
\subsection{$\mathbb{P}^2$}
\paragraph{$(n,g)=(2,1)$}
\begin{dmath*}
\mathcal{W}^{(2,1)}_{[-1]}=\frac{1}{414720 z^6 (1+27 z)^3}(40 S^3-60 S^2 z^2+3240 S^2 z^3+30 S z^4-5184 S z^5-5 z^6+419904 S z^6-378z^7+215784z^8-2834352 z^9).
\end{dmath*}
\begin{dmath*}
\mathcal{W}^{(2,1)}_{[-1]\otimes[-1]}=\frac{1}{207360 z^6 (1+27 z)^3}(40 S^3-60 S^2 z^2+3240 S^2 z^3+30 S z^4-5184 S z^5-5 z^6+419904 S z^6-378 z^7+215784 z^8-2834352 z^9).
\end{dmath*}

\paragraph{$(n,g)=(1,2)$}
\begin{dmath*}
\mathcal{W}^{(1,2)}_{[-1]}=\frac{1}{103680 z^8 (1+27 z)^3}(160 S^4-200 S^3 z^2+6480 S^3 z^3+60 S^2 z^4-15120 S^2 z^5+10 S z^6+116640 S^2 z^6+5724 S z^7-5 z^8+279936 S z^8+108 z^9-128304 z^{10}-314928 z^{11}).
\end{dmath*}
\begin{dmath*}
\mathcal{W}^{(1,2)}_{[-1]\otimes[-1]}=\frac{1}{51840 z^8 (1+27 z)^3}(160 S^4-120 S^3 z^2+8640 S^3 z^3-60 S^2 z^4-18360 S^2 z^5+70 S z^6+116640 S^2 z^6+13824 S z^7-15 z^8+454896 S z^8-1242 z^9-215784 z^{10}-1889568 z^{11}).
\end{dmath*}

\paragraph{$(n,g)=(0,3)$}
\begin{dmath*}
\mathcal{W}^{(0,3)}_{[-1]}=\frac{1}{155520 z^{10} (1+27 z)^3}(1200 S^5-1560 S^4 z^2-6480 S^4 z^3+740 S^3 z^4-4320 S^3 z^5-270 S^2 z^6-4860 S^2 z^7+120 S z^8+174960 S^2 z^8+10692 S z^9-25 z^{10}+8748 S z^{10}-3186 z^{11}-42282 z^{12}-551124 z^{13}).
\end{dmath*}
\begin{dmath*}
\mathcal{W}^{(0,3)}_{[-1]\otimes[-1]}=\frac{1}{77760 z^{10} (1+27 z)^3}(1200 S^5-600 S^4 z^2+19440 S^4 z^3+20 S^3 z^4-49680 S^3 z^5-630 S^2 z^6-699840 S^3 z^6-27540 S^2 z^7+540 S z^8-524880 S^2 z^8+34992 S z^9-9447840 S^2 z^9-115 z^{10}+358668 S z^{10}-12096 z^{11}-304722 z^{12}-2913084 z^{13}).
\end{dmath*}

\clearpage
\bibliographystyle{JHEP}     
{\small{\bibliography{reference}}}

\providecommand{\href}[2]{#2}\begingroup\raggedright\begin{thebibliography}{10}

\bibitem{Witten:1988xj}
E.~Witten, {\it {Topological Sigma Models}},  {\em Commun. Math. Phys.} {\bf
  118} (1988) 411.

\bibitem{Katz:1996fh}
S.~H. Katz, A.~Klemm, and C.~Vafa, {\it {Geometric engineering of quantum field
  theories}},  {\em Nucl. Phys. B} {\bf 497} (1997) 173--195,
  [\href{http://arxiv.org/abs/hep-th/9609239}{{\tt hep-th/9609239}}].

\bibitem{Katz:1997eq}
S.~Katz, P.~Mayr, and C.~Vafa, {\it {Mirror symmetry and exact solution of 4-D
  N=2 gauge theories: 1.}},  {\em Adv. Theor. Math. Phys.} {\bf 1} (1998)
  53--114, [\href{http://arxiv.org/abs/hep-th/9706110}{{\tt hep-th/9706110}}].

\bibitem{Gopakumar:1998ii}
R.~Gopakumar and C.~Vafa, {\it {M theory and topological strings. 1.}},
  \href{http://arxiv.org/abs/hep-th/9809187}{{\tt hep-th/9809187}}.

\bibitem{Gopakumar:1998jq}
R.~Gopakumar and C.~Vafa, {\it {M theory and topological strings. 2.}},
  \href{http://arxiv.org/abs/hep-th/9812127}{{\tt hep-th/9812127}}.

\bibitem{Nekrasov:2002qd}
N.~A. Nekrasov, {\it {Seiberg-Witten prepotential from instanton counting}},
  {\em Adv. Theor. Math. Phys.} {\bf 7} (2003), no.~5 831--864,
  [\href{http://arxiv.org/abs/hep-th/0206161}{{\tt hep-th/0206161}}].

\bibitem{Nekrasov:2003rj}
N.~Nekrasov and A.~Okounkov, {\it {Seiberg-Witten theory and random
  partitions}},  {\em Prog. Math.} {\bf 244} (2006) 525--596,
  [\href{http://arxiv.org/abs/hep-th/0306238}{{\tt hep-th/0306238}}].

\bibitem{Kim:2011mv}
H.-C. Kim, S.~Kim, E.~Koh, K.~Lee, and S.~Lee, {\it {On instantons as
  Kaluza-Klein modes of M5-branes}},  {\em JHEP} {\bf 12} (2011) 031,
  [\href{http://arxiv.org/abs/1110.2175}{{\tt arXiv:1110.2175}}].

\bibitem{Rodriguez-Gomez:2013dpa}
D.~Rodr\'\i{}guez-G\'omez and G.~Zafrir, {\it {On the 5d instanton index as a
  Hilbert series}},  {\em Nucl. Phys. B} {\bf 878} (2014) 1--11,
  [\href{http://arxiv.org/abs/1305.5684}{{\tt arXiv:1305.5684}}].

\bibitem{Bergman:2013ala}
O.~Bergman, D.~Rodr\'\i{}guez-G\'omez, and G.~Zafrir, {\it {Discrete $\theta$
  and the 5d superconformal index}},  {\em JHEP} {\bf 01} (2014) 079,
  [\href{http://arxiv.org/abs/1310.2150}{{\tt arXiv:1310.2150}}].

\bibitem{Hwang:2014uwa}
C.~Hwang, J.~Kim, S.~Kim, and J.~Park, {\it {General instanton counting and 5d
  SCFT}},  {\em JHEP} {\bf 07} (2015) 063,
  [\href{http://arxiv.org/abs/1406.6793}{{\tt arXiv:1406.6793}}]. [Addendum:
  JHEP 04, 094 (2016)].

\bibitem{Haghighat:2013gba}
B.~Haghighat, A.~Iqbal, C.~Koz\c{c}az, G.~Lockhart, and C.~Vafa, {\it
  {M-Strings}},  {\em Commun. Math. Phys.} {\bf 334} (2015), no.~2 779--842,
  [\href{http://arxiv.org/abs/1305.6322}{{\tt arXiv:1305.6322}}].

\bibitem{Kim:2014dza}
J.~Kim, S.~Kim, K.~Lee, J.~Park, and C.~Vafa, {\it {Elliptic Genus of
  E-strings}},  {\em JHEP} {\bf 09} (2017) 098,
  [\href{http://arxiv.org/abs/1411.2324}{{\tt arXiv:1411.2324}}].

\bibitem{Haghighat:2014vxa}
B.~Haghighat, A.~Klemm, G.~Lockhart, and C.~Vafa, {\it {Strings of Minimal 6d
  SCFTs}},  {\em Fortsch. Phys.} {\bf 63} (2015) 294--322,
  [\href{http://arxiv.org/abs/1412.3152}{{\tt arXiv:1412.3152}}].

\bibitem{Gadde:2015tra}
A.~Gadde, B.~Haghighat, J.~Kim, S.~Kim, G.~Lockhart, and C.~Vafa, {\it {6d
  String Chains}},  {\em JHEP} {\bf 02} (2018) 143,
  [\href{http://arxiv.org/abs/1504.04614}{{\tt arXiv:1504.04614}}].

\bibitem{Kim:2015fxa}
J.~Kim, S.~Kim, and K.~Lee, {\it {Higgsing towards E-strings}},  {\em JHEP}
  {\bf 01} (2021) 110, [\href{http://arxiv.org/abs/1510.03128}{{\tt
  arXiv:1510.03128}}].

\bibitem{Kim:2016foj}
H.-C. Kim, S.~Kim, and J.~Park, {\it {6d strings from new chiral gauge
  theories}},  \href{http://arxiv.org/abs/1608.03919}{{\tt arXiv:1608.03919}}.

\bibitem{Kim:2018gjo}
H.-C. Kim, J.~Kim, S.~Kim, K.-H. Lee, and J.~Park, {\it {6d strings and
  exceptional instantons}},  {\em Phys. Rev. D} {\bf 103} (2021), no.~2 025012,
  [\href{http://arxiv.org/abs/1801.03579}{{\tt arXiv:1801.03579}}].

\bibitem{Gomis:2006sb}
J.~Gomis and F.~Passerini, {\it {Holographic Wilson Loops}},  {\em JHEP} {\bf
  08} (2006) 074, [\href{http://arxiv.org/abs/hep-th/0604007}{{\tt
  hep-th/0604007}}].

\bibitem{Tong:2014cha}
D.~Tong and K.~Wong, {\it {Instantons, Wilson lines, and D-branes}},  {\em
  Phys. Rev. D} {\bf 91} (2015), no.~2 026007,
  [\href{http://arxiv.org/abs/1410.8523}{{\tt arXiv:1410.8523}}].

\bibitem{Gaiotto:2014ina}
D.~Gaiotto and H.-C. Kim, {\it {Surface defects and instanton partition
  functions}},  {\em JHEP} {\bf 10} (2016) 012,
  [\href{http://arxiv.org/abs/1412.2781}{{\tt arXiv:1412.2781}}].

\bibitem{Nekrasov:2015wsu}
N.~Nekrasov, {\it {BPS/CFT correspondence: non-perturbative Dyson-Schwinger
  equations and qq-characters}},  {\em JHEP} {\bf 03} (2016) 181,
  [\href{http://arxiv.org/abs/1512.05388}{{\tt arXiv:1512.05388}}].

\bibitem{Kim:2016qqs}
H.-C. Kim, {\it {Line defects and 5d instanton partition functions}},  {\em
  JHEP} {\bf 03} (2016) 199, [\href{http://arxiv.org/abs/1601.06841}{{\tt
  arXiv:1601.06841}}].

\bibitem{Assel:2018rcw}
B.~Assel and A.~Sciarappa, {\it {Wilson loops in 5d $\mathcal{N}=1$ theories
  and S-duality}},  {\em JHEP} {\bf 10} (2018) 082,
  [\href{http://arxiv.org/abs/1806.09636}{{\tt arXiv:1806.09636}}].

\bibitem{Gaiotto:2015una}
D.~Gaiotto and H.-C. Kim, {\it {Duality walls and defects in 5d $ \mathcal{N}=1
  $ theories}},  {\em JHEP} {\bf 01} (2017) 019,
  [\href{http://arxiv.org/abs/1506.03871}{{\tt arXiv:1506.03871}}].

\bibitem{Haouzi:2020yxy}
N.~Haouzi and J.~Oh, {\it {On the Quantization of Seiberg-Witten Geometry}},
  {\em JHEP} {\bf 01} (2021) 184, [\href{http://arxiv.org/abs/2004.00654}{{\tt
  arXiv:2004.00654}}].

\bibitem{Chen:2007ir}
B.~Chen, W.~He, J.-B. Wu, and L.~Zhang, {\it {M5-branes and Wilson Surfaces}},
  {\em JHEP} {\bf 08} (2007) 067, [\href{http://arxiv.org/abs/0707.3978}{{\tt
  arXiv:0707.3978}}].

\bibitem{Agarwal:2018tso}
P.~Agarwal, J.~Kim, S.~Kim, and A.~Sciarappa, {\it {Wilson surfaces in
  M5-branes}},  {\em JHEP} {\bf 08} (2018) 119,
  [\href{http://arxiv.org/abs/1804.09932}{{\tt arXiv:1804.09932}}].

\bibitem{Chen:2020jla}
J.~Chen, B.~Haghighat, H.-C. Kim, and M.~Sperling, {\it {Elliptic quantum
  curves of class $ {\mathcal{S}}_k $}},  {\em JHEP} {\bf 03} (2021) 028,
  [\href{http://arxiv.org/abs/2008.05155}{{\tt arXiv:2008.05155}}].

\bibitem{Chen:2021ivd}
J.~Chen, B.~Haghighat, H.-C. Kim, M.~Sperling, and X.~Wang, {\it {E-string
  quantum curve}},  {\em Nucl. Phys. B} {\bf 973} (2021) 115602,
  [\href{http://arxiv.org/abs/2103.16996}{{\tt arXiv:2103.16996}}].

\bibitem{Chen:2021rek}
J.~Chen, B.~Haghighat, H.-C. Kim, K.~Lee, M.~Sperling, and X.~Wang, {\it
  {Elliptic Quantum Curves of 6d SO(N) theories}},
  \href{http://arxiv.org/abs/2110.13487}{{\tt arXiv:2110.13487}}.

\bibitem{Kim:2021gyj}
H.-C. Kim, M.~Kim, and S.-S. Kim, {\it {5d/6d Wilson loops from blowups}},
  {\em JHEP} {\bf 08} (2021) 131, [\href{http://arxiv.org/abs/2106.04731}{{\tt
  arXiv:2106.04731}}].

\bibitem{Witten:1992fb}
E.~Witten, {\it {Chern-Simons gauge theory as a string theory}},  {\em Prog.
  Math.} {\bf 133} (1995) 637--678,
  [\href{http://arxiv.org/abs/hep-th/9207094}{{\tt hep-th/9207094}}].

\bibitem{Iqbal:2007ii}
A.~Iqbal, C.~Kozcaz, and C.~Vafa, {\it {The Refined topological vertex}},  {\em
  JHEP} {\bf 10} (2009) 069, [\href{http://arxiv.org/abs/hep-th/0701156}{{\tt
  hep-th/0701156}}].

\bibitem{Aganagic:2012hs}
M.~Aganagic and S.~Shakirov, {\it {Refined Chern-Simons Theory and Topological
  String}},  \href{http://arxiv.org/abs/1210.2733}{{\tt arXiv:1210.2733}}.

\bibitem{Taki:2007dh}
M.~Taki, {\it {Refined Topological Vertex and Instanton Counting}},  {\em JHEP}
  {\bf 03} (2008) 048, [\href{http://arxiv.org/abs/0710.1776}{{\tt
  arXiv:0710.1776}}].

\bibitem{Tachikawa:2004ur}
Y.~Tachikawa, {\it {Five-dimensional Chern-Simons terms and Nekrasov's
  instanton counting}},  {\em JHEP} {\bf 02} (2004) 050,
  [\href{http://arxiv.org/abs/hep-th/0401184}{{\tt hep-th/0401184}}].

\bibitem{Gottsche:2006bm}
L.~Gottsche, H.~Nakajima, and K.~Yoshioka, {\it {K-theoretic Donaldson
  invariants via instanton counting}},  {\em Pure Appl. Math. Quart.} {\bf 5}
  (2009) 1029--1111, [\href{http://arxiv.org/abs/math/0611945}{{\tt
  math/0611945}}].

\bibitem{Bershadsky:1993cx}
M.~Bershadsky, S.~Cecotti, H.~Ooguri, and C.~Vafa, {\it {Kodaira-Spencer theory
  of gravity and exact results for quantum string amplitudes}},  {\em Commun.
  Math. Phys.} {\bf 165} (1994) 311--428,
  [\href{http://arxiv.org/abs/hep-th/9309140}{{\tt hep-th/9309140}}].

\bibitem{Huang:2006si}
M.-x. Huang and A.~Klemm, {\it {Holomorphic Anomaly in Gauge Theories and
  Matrix Models}},  {\em JHEP} {\bf 09} (2007) 054,
  [\href{http://arxiv.org/abs/hep-th/0605195}{{\tt hep-th/0605195}}].

\bibitem{Huang:2006hq}
M.-x. Huang, A.~Klemm, and S.~Quackenbush, {\it {Topological string theory on
  compact Calabi-Yau: Modularity and boundary conditions}},  {\em Lect. Notes
  Phys.} {\bf 757} (2009) 45--102,
  [\href{http://arxiv.org/abs/hep-th/0612125}{{\tt hep-th/0612125}}].

\bibitem{Grimm:2007tm}
T.~W. Grimm, A.~Klemm, M.~Marino, and M.~Weiss, {\it {Direct Integration of the
  Topological String}},  {\em JHEP} {\bf 08} (2007) 058,
  [\href{http://arxiv.org/abs/hep-th/0702187}{{\tt hep-th/0702187}}].

\bibitem{Haghighat:2008gw}
B.~Haghighat, A.~Klemm, and M.~Rauch, {\it {Integrability of the holomorphic
  anomaly equations}},  {\em JHEP} {\bf 10} (2008) 097,
  [\href{http://arxiv.org/abs/0809.1674}{{\tt arXiv:0809.1674}}].

\bibitem{Huang:2010kf}
M.-x. Huang and A.~Klemm, {\it {Direct integration for general $\Omega$
  backgrounds}},  {\em Adv. Theor. Math. Phys.} {\bf 16} (2012), no.~3
  805--849, [\href{http://arxiv.org/abs/1009.1126}{{\tt arXiv:1009.1126}}].

\bibitem{Krefl:2010fm}
D.~Krefl and J.~Walcher, {\it {Extended Holomorphic Anomaly in Gauge Theory}},
  {\em Lett. Math. Phys.} {\bf 95} (2011) 67--88,
  [\href{http://arxiv.org/abs/1007.0263}{{\tt arXiv:1007.0263}}].

\bibitem{Huang:2011qx}
M.-x. Huang, A.-K. Kashani-Poor, and A.~Klemm, {\it {The $\Omega$ deformed
  B-model for rigid $\mathcal{N}=2$ theories}},  {\em Annales Henri Poincare}
  {\bf 14} (2013) 425--497, [\href{http://arxiv.org/abs/1109.5728}{{\tt
  arXiv:1109.5728}}].

\bibitem{Huang:2013yta}
M.-X. Huang, A.~Klemm, and M.~Poretschkin, {\it {Refined stable pair invariants
  for E-, M- and $[p, q]$-strings}},  {\em JHEP} {\bf 11} (2013) 112,
  [\href{http://arxiv.org/abs/1308.0619}{{\tt arXiv:1308.0619}}].

\bibitem{Klemm:2015iya}
A.~Klemm, M.~Poretschkin, T.~Schimannek, and M.~Westerholt-Raum, {\it {Direct
  Integration for Mirror Curves of Genus Two and an Almost Meromorphic Siegel
  Modular Form}},  \href{http://arxiv.org/abs/1502.00557}{{\tt
  arXiv:1502.00557}}.

\bibitem{Nakajima:2003pg}
H.~Nakajima and K.~Yoshioka, {\it {Instanton counting on blowup. 1.}},  {\em
  Invent. Math.} {\bf 162} (2005) 313--355,
  [\href{http://arxiv.org/abs/math/0306198}{{\tt math/0306198}}].

\bibitem{Nakajima:2003uh}
H.~Nakajima and K.~Yoshioka, {\it {Lectures on instanton counting}},  in {\em
  {CRM Workshop on Algebraic Structures and Moduli Spaces}}, 11, 2003.
\newblock \href{http://arxiv.org/abs/math/0311058}{{\tt math/0311058}}.

\bibitem{Nakajima:2005fg}
H.~Nakajima and K.~Yoshioka, {\it {Instanton counting on blowup. II.
  K-theoretic partition function}},
  \href{http://arxiv.org/abs/math/0505553}{{\tt math/0505553}}.

\bibitem{Nakajima:2009qjc}
H.~Nakajima and K.~Yoshioka, {\it {Perverse coherent sheaves on blowup, III:
  Blow-up formula from wall-crossing}},  {\em Kyoto J. Math.} {\bf 51} (2011),
  no.~2 263--335, [\href{http://arxiv.org/abs/0911.1773}{{\tt
  arXiv:0911.1773}}].

\bibitem{Huang:2017mis}
M.-x. Huang, K.~Sun, and X.~Wang, {\it {Blowup Equations for Refined
  Topological Strings}},  {\em JHEP} {\bf 10} (2018) 196,
  [\href{http://arxiv.org/abs/1711.09884}{{\tt arXiv:1711.09884}}].

\bibitem{Gu:2018gmy}
J.~Gu, B.~Haghighat, K.~Sun, and X.~Wang, {\it {Blowup Equations for 6d SCFTs.
  I}},  {\em JHEP} {\bf 03} (2019) 002,
  [\href{http://arxiv.org/abs/1811.02577}{{\tt arXiv:1811.02577}}].

\bibitem{Gu:2019dan}
J.~Gu, A.~Klemm, K.~Sun, and X.~Wang, {\it {Elliptic blowup equations for 6d
  SCFTs. Part II. Exceptional cases}},  {\em JHEP} {\bf 12} (2019) 039,
  [\href{http://arxiv.org/abs/1905.00864}{{\tt arXiv:1905.00864}}].

\bibitem{Gu:2019pqj}
J.~Gu, B.~Haghighat, A.~Klemm, K.~Sun, and X.~Wang, {\it {Elliptic blowup
  equations for 6d SCFTs. Part III. E-strings, M-strings and chains}},  {\em
  JHEP} {\bf 07} (2020) 135, [\href{http://arxiv.org/abs/1911.11724}{{\tt
  arXiv:1911.11724}}].

\bibitem{Gu:2020fem}
J.~Gu, B.~Haghighat, A.~Klemm, K.~Sun, and X.~Wang, {\it {Elliptic blowup
  equations for 6d SCFTs. Part IV. Matters}},  {\em JHEP} {\bf 11} (2021) 090,
  [\href{http://arxiv.org/abs/2006.03030}{{\tt arXiv:2006.03030}}].

\bibitem{Kim:2019uqw}
J.~Kim, S.-S. Kim, K.-H. Lee, K.~Lee, and J.~Song, {\it {Instantons from
  Blow-up}},  {\em JHEP} {\bf 11} (2019) 092,
  [\href{http://arxiv.org/abs/1908.11276}{{\tt arXiv:1908.11276}}]. [Erratum:
  JHEP 06, 124 (2020)].

\bibitem{Kim:2020hhh}
H.-C. Kim, M.~Kim, S.-S. Kim, and K.-H. Lee, {\it {Bootstrapping BPS spectra of
  5d/6d field theories}},  {\em JHEP} {\bf 04} (2021) 161,
  [\href{http://arxiv.org/abs/2101.00023}{{\tt arXiv:2101.00023}}].

\bibitem{Duan:2021ges}
Z.~Duan, K.~Lee, J.~Nahmgoong, and X.~Wang, {\it {Twisted 6d (2, 0) SCFTs on a
  circle}},  {\em JHEP} {\bf 07} (2021) 179,
  [\href{http://arxiv.org/abs/2103.06044}{{\tt arXiv:2103.06044}}].

\bibitem{Aganagic:2006wq}
M.~Aganagic, V.~Bouchard, and A.~Klemm, {\it {Topological Strings and (Almost)
  Modular Forms}},  {\em Commun. Math. Phys.} {\bf 277} (2008) 771--819,
  [\href{http://arxiv.org/abs/hep-th/0607100}{{\tt hep-th/0607100}}].

\bibitem{Bousseau:2020ckw}
P.~Bousseau, H.~Fan, S.~Guo, and L.~Wu, {\it {Holomorphic anomaly equation for
  $(\mathbb{P}^2,E)$ and the Nekrasov-Shatashvili limit of local
  $\mathbb{P}^2$}},  \href{http://arxiv.org/abs/2001.05347}{{\tt
  arXiv:2001.05347}}.

\bibitem{Choi:2012jz}
J.~Choi, S.~Katz, and A.~Klemm, {\it {The refined BPS index from stable pair
  invariants}},  {\em Commun. Math. Phys.} {\bf 328} (2014) 903--954,
  [\href{http://arxiv.org/abs/1210.4403}{{\tt arXiv:1210.4403}}].

\bibitem{Chiang:1999tz}
T.~M. Chiang, A.~Klemm, S.-T. Yau, and E.~Zaslow, {\it {Local mirror symmetry:
  Calculations and interpretations}},  {\em Adv. Theor. Math. Phys.} {\bf 3}
  (1999) 495--565, [\href{http://arxiv.org/abs/hep-th/9903053}{{\tt
  hep-th/9903053}}].

\bibitem{Aharony:1997bh}
O.~Aharony, A.~Hanany, and B.~Kol, {\it {Webs of (p,q) five-branes,
  five-dimensional field theories and grid diagrams}},  {\em JHEP} {\bf 01}
  (1998) 002, [\href{http://arxiv.org/abs/hep-th/9710116}{{\tt
  hep-th/9710116}}].

\bibitem{Gomis:2006im}
J.~Gomis and F.~Passerini, {\it {Wilson Loops as D3-Branes}},  {\em JHEP} {\bf
  01} (2007) 097, [\href{http://arxiv.org/abs/hep-th/0612022}{{\tt
  hep-th/0612022}}].

\bibitem{MR1318878}
L.~C. Jeffrey and F.~C. Kirwan, {\it Localization for nonabelian group
  actions},  {\em Topology} {\bf 34} (1995), no.~2 291--327.

\bibitem{Bao:2013pwa}
L.~Bao, V.~Mitev, E.~Pomoni, M.~Taki, and F.~Yagi, {\it {Non-Lagrangian
  Theories from Brane Junctions}},  {\em JHEP} {\bf 01} (2014) 175,
  [\href{http://arxiv.org/abs/1310.3841}{{\tt arXiv:1310.3841}}].

\bibitem{Losev:2003py}
A.~S. Losev, A.~Marshakov, and N.~A. Nekrasov, {\it {Small instantons, little
  strings and free fermions}},  in {\em {From Fields to Strings:
  Circumnavigating Theoretical Physics: A Conference in Tribute to Ian Kogan}},
  pp.~581--621, 2, 2003.
\newblock \href{http://arxiv.org/abs/hep-th/0302191}{{\tt hep-th/0302191}}.

\bibitem{Shadchin:2004yx}
S.~Shadchin, {\it {Saddle point equations in Seiberg-Witten theory}},  {\em
  JHEP} {\bf 10} (2004) 033, [\href{http://arxiv.org/abs/hep-th/0408066}{{\tt
  hep-th/0408066}}].

\bibitem{Hayashi:2013qwa}
H.~Hayashi, H.-C. Kim, and T.~Nishinaka, {\it {Topological strings and 5d $T_N$
  partition functions}},  {\em JHEP} {\bf 06} (2014) 014,
  [\href{http://arxiv.org/abs/1310.3854}{{\tt arXiv:1310.3854}}].

\bibitem{Hayashi:2014wfa}
H.~Hayashi and G.~Zoccarato, {\it {Exact partition functions of Higgsed 5d
  $T_N$ theories}},  {\em JHEP} {\bf 01} (2015) 093,
  [\href{http://arxiv.org/abs/1409.0571}{{\tt arXiv:1409.0571}}].

\bibitem{Nekrasov:1996cz}
N.~Nekrasov, {\it {Five dimensional gauge theories and relativistic integrable
  systems}},  {\em Nucl. Phys. B} {\bf 531} (1998) 323--344,
  [\href{http://arxiv.org/abs/hep-th/9609219}{{\tt hep-th/9609219}}].

\bibitem{Nekrasov:2009rc}
N.~A. Nekrasov and S.~L. Shatashvili, {\it {Quantization of Integrable Systems
  and Four Dimensional Gauge Theories}},  in {\em {16th International Congress
  on Mathematical Physics}}, pp.~265--289, 8, 2009.
\newblock \href{http://arxiv.org/abs/0908.4052}{{\tt arXiv:0908.4052}}.

\bibitem{Grassi:2017qee}
A.~Grassi and M.~Marino, {\it {The complex side of the TS/ST correspondence}},
  {\em J. Phys. A} {\bf 52} (2019), no.~5 055402,
  [\href{http://arxiv.org/abs/1708.08642}{{\tt arXiv:1708.08642}}].

\bibitem{Grassi:2018bci}
A.~Grassi and M.~Mari\~no, {\it {A Solvable Deformation of Quantum Mechanics}},
   {\em SIGMA} {\bf 15} (2019) 025,
  [\href{http://arxiv.org/abs/1806.01407}{{\tt arXiv:1806.01407}}].

\bibitem{Witten:1993ed}
E.~Witten, {\it {Quantum background independence in string theory}},  in {\em
  {Conference on Highlights of Particle and Condensed Matter Physics
  (SALAMFEST)}}, 6, 1993.
\newblock \href{http://arxiv.org/abs/hep-th/9306122}{{\tt hep-th/9306122}}.

\bibitem{Lian:1994zv}
B.~H. Lian and S.-T. Yau, {\it {Arithmetic properties of mirror map and quantum
  coupling}},  {\em Commun. Math. Phys.} {\bf 176} (1996) 163--192,
  [\href{http://arxiv.org/abs/hep-th/9411234}{{\tt hep-th/9411234}}].

\bibitem{Closset:2021lhd}
C.~Closset and H.~Magureanu, {\it {The $U$-plane of rank-one 4d $\mathcal{N}=2$
  KK theories}},  {\em SciPost Phys.} {\bf 12} (2022) 065,
  [\href{http://arxiv.org/abs/2107.03509}{{\tt arXiv:2107.03509}}].

\bibitem{Aganagic:2011mi}
M.~Aganagic, M.~C.~N. Cheng, R.~Dijkgraaf, D.~Krefl, and C.~Vafa, {\it {Quantum
  Geometry of Refined Topological Strings}},  {\em JHEP} {\bf 11} (2012) 019,
  [\href{http://arxiv.org/abs/1105.0630}{{\tt arXiv:1105.0630}}].

\bibitem{Huang:2014nwa}
M.-x. Huang, A.~Klemm, J.~Reuter, and M.~Schiereck, {\it {Quantum geometry of
  del Pezzo surfaces in the Nekrasov-Shatashvili limit}},  {\em JHEP} {\bf 02}
  (2015) 031, [\href{http://arxiv.org/abs/1401.4723}{{\tt arXiv:1401.4723}}].

\bibitem{Fischbach:2018yiu}
F.~Fischbach, A.~Klemm, and C.~Nega, {\it {WKB Method and Quantum Periods
  beyond Genus One}},  {\em J. Phys. A} {\bf 52} (2019), no.~7 075402,
  [\href{http://arxiv.org/abs/1803.11222}{{\tt arXiv:1803.11222}}].

\bibitem{Huang:2020neq}
M.-x. Huang, Y.~Sugimoto, and X.~Wang, {\it {Quantum periods and spectra in
  dimer models and Calabi-Yau geometries}},  {\em JHEP} {\bf 09} (2020) 168,
  [\href{http://arxiv.org/abs/2006.13482}{{\tt arXiv:2006.13482}}].

\bibitem{Bourgine:2017jan}
J.-E. Bourgine and D.~Fioravanti, {\it {Seiberg-Witten period relations in
  Omega background}},  {\em JHEP} {\bf 08} (2018) 124,
  [\href{http://arxiv.org/abs/1711.07570}{{\tt arXiv:1711.07570}}].

\bibitem{MR2512147}
N.~A. Nekrasov, {\it Two-dimensional topological strings revisited},  {\em
  Lett. Math. Phys.} {\bf 88} (2009), no.~1-3 207--253.

\bibitem{Furukawa:2019sfy}
T.~Furukawa, S.~Moriyama, and Y.~Sugimoto, {\it {Quantum Mirror Map for Del
  Pezzo Geometries}},  {\em J. Phys. A} {\bf 53} (2020), no.~38 38,
  [\href{http://arxiv.org/abs/1908.11396}{{\tt arXiv:1908.11396}}].

\bibitem{Moriyama:2020lyk}
S.~Moriyama, {\it {Spectral Theories and Topological Strings on del Pezzo
  Geometries}},  {\em JHEP} {\bf 10} (2020) 154,
  [\href{http://arxiv.org/abs/2007.05148}{{\tt arXiv:2007.05148}}].

\bibitem{Kim:2017jqn}
S.-S. Kim and F.~Yagi, {\it {Topological vertex formalism with O5-plane}},
  {\em Phys. Rev. D} {\bf 97} (2018), no.~2 026011,
  [\href{http://arxiv.org/abs/1709.01928}{{\tt arXiv:1709.01928}}].

\bibitem{Kim:2022dbr}
S.-S. Kim and X.-Y. Wei, {\it {Refined topological vertex with ON-planes}},
  \href{http://arxiv.org/abs/2201.12264}{{\tt arXiv:2201.12264}}.

\end{thebibliography}\endgroup

\end{document}